\documentclass[twocolumn,floatfix,showpacs,preprintnumbers,amsmath,amssymb,nofootinbib]{revtex4-1}

\usepackage{amsmath}
\usepackage{amssymb}
\usepackage{adjustbox}
\usepackage{mathrsfs}
\usepackage{bm}
\usepackage{graphicx}
\usepackage{subfig}
\usepackage{algcompatible}
\usepackage{newfloat}
\usepackage[usenames,dvipsnames]{xcolor}
\usepackage{footnote}
\usepackage{hyperref}
\usepackage[capitalise]{cleveref}

\DeclareFloatingEnvironment[
    fileext=loa,
    listname=List of Algorithms,
    name=ALGORITHM,
    placement=htbp,
]{algorithm}

\begin{document}

\title{A Dynamical Systems Perspective Reveals Coordination\\in Russian Twitter Operations}

\date{(Preprint - \today - Distribution Limited)}

\author{Sarah Rajtmajer}
\email[Sarah Rajtmajer: ]{srajtmajer@ist.psu.edu}
\author{Ashish Simhachalam}
\email[Ashish Simhachalam: ]{vus275@psu.edu}
\author{Thomas Zhao}
\email[Thomas Zhao: ]{tyz5069@psu.edu}
\affiliation{College of Information Science and Technology,
The Pennsylvania State University, University Park, PA 16802, USA}

\author{Brady Bickel}
\email[Brady Bickel: ]{brb162@arl.psu.edu}
\author{Christopher Griffin}
\email[Christopher Griffin: ]{griffinch@ieee.org}
\affiliation{Applied Research Laboratory,
The Pennsylvania State University, University Park, PA 16802, USA}


\begin{abstract}
We study Twitter data from a dynamical systems perspective. In particular, we focus on the large set of data released by Twitter Inc. and asserted to represent a Russian influence operation. We propose a mathematical model to describe the per-day tweet production that can be extracted using spectral analysis. We show that this mathematical model allows us to construct families (clusters) of users with common harmonics. We define a labeling scheme describing \textit{user strategy} in an information operation and show that the resulting strategies correspond to the behavioral clusters identified from their harmonics. We then compare these user clusters to the ones derived from text data using a graph-based topic analysis method. We show that spectral properties of the user clusters are related to the number of user-topic groups represented in a spectral cluster. Bulk data analysis also provides new insights into the data set in the context of prior work. 
\end{abstract}

\maketitle
\section{Introduction}
In this paper, we analyze a data set purported to represent Russian troll twitter messages, provided by Twitter Inc. \cite{S17,T17} using techniques from dynamical systems analysis. There have been several reports on data released by Twitter representing what is believed to be disinformation originating from Russian state-controlled accounts. Of particular focus has been possible targeted interference in the 2016 U.S. Presidential election. However, evidence continues to mount that foreign influence campaigns operationalized through social media remain a persistent threat. 

The data discussed in this paper and closely related data have been analyzed by several other authors. Manual analysis using data scraped from Twitter by Clemson University suggests that Russian twitter agents were pro-NRA \cite{M18}. Griffin and Bickel \cite{GB18,R18} initially reported on the use of unsupervised machine learning for extracting information from a subset of the data analyzed in this paper that was provided by NBC News \cite{P18}. Kellner et al. \cite{KRWR19} analyze this data set to investigate influence in the German Federal election. Im et al. \cite{ICSL19} build language-based classifiers for twitter bots using this data set. Cable and Hugh \cite{CH19} as well as Lim et al. \cite{LLZ19} study the problem of detecting twitter bots using general machine learning techniques. In addition, several reports have been written on the subject including the well known Mueller Report \cite{M19}.

The objective of this paper is to investigate the data provided by Twitter Inc. from a dynamical systems perspective. We propose a mathematical model to describe the per-day tweet rate (as the dynamic system) for a subset of the user (bot) population and show how this model leads to a coherent set of user clusters based on the dynamic behavior. In addition, we construct a novel strategy assignment for message data and show that clustered accounts yield consistent strategic behaviors. We then compare these clusters to constructed topic clusters (using the method in \cite{GB18}) and show that message content diversity is directly linked to diversity in per-day tweet rate. Subsequently, we treat some of the assertions made in \cite{GB18} as registered hypotheses and use the larger data set to validate or refute them. 

The aim of this work is substantively different from the literature on bot and fake news detection \cite{chen2017hunting, beskow2018bot, 7837909, Radziwill2016BotON,KT18,OA20}, as it is from the work on general modeling of tweet dynamics \cite{7363824, Gao}. To our knowledge, we are the first to adopt a dynamical systems approach to describe information operations, and in particular, the strategic behavior of coordinated actors in this space. As the detrimental effects of these operations are continually brought to our attention (e.g., \cite{stilgherrian}), the need for robust, formal approaches to modeling increasingly sophisticated strategic actors has become clear.

The remainder of this paper is organized as follows: In \cref{sec:Notation} we discuss notation and some pre-requisite information. Bulk data analysis is considered in \cref{sec:Bulk}. We then turn to specific dynamical systems inspired analysis of the data set in \cref{sec:Spectral}. The results from this analysis is compared and expanded using topic analysis in \cref{sec:Topic}. Conclusions and future directions are presented in \cref{sec:Conclusion}. Two appendices include specific information generated from the unsupervised clustering approaches discussed.

\section{Notation and Preliminaries}\label{sec:Notation}
In this section we establish notation and fundamental definitions needed for the remainder of this paper. All vectors are assumed to be column vectors unless stated otherwise and are denoted by boldface symbols in lower case; matrices are denoted by boldface symbols in upper case. An $m$ dimensional vector $\mathbf{x} = \langle{x_1,\dots,x_m}\rangle$ has components that are unbolded and indexed. If $\mathbf{X} \in \mathbb{R}^{m \times n}$ is a matrix, then $\mathbf{X}^T \in \mathbb{R}^{n \times m}$ is its transpose. If $\mathbf{x} = \{x_t\}_t$ is a time series, then its discrete Fourier transform is denoted $\hat{\mathbf{x}} = \{\hat{x}_\omega\}_{\omega} = \mathcal{F}(\mathbf{x})$. 

\subsection{Bag of Words Space} 
We use a bag-of-words (BOW) formalism in this paper because we are interested in a dynamical system analysis of data that is linked to text. This approach has been extensively used. The interested reader may consult \cite{MS99} as a reference. Assume there is a finite dictionary of words (or symbols) $\mathcal{W}$ with size $m$. Assume $\mathcal{W} = \{w^1,w^2,\dots,w^m\}$ is ordered, $i$ is the index of word $w^i$, and associate with word $w^i$ the Euclidean unit vector $\mathbf{e}_i$ via the map $g(w^i) = \mathbf{e}_i$. In a BOW model, the representation $g$ is extended to any sequence of words $(w_1\cdot w_2\cdots w_k)$ in the following way:
\begin{equation}
\mathbf{x} = g(w^1\cdot w^2\cdots w^k) = \sum_{i=1}^k g(w^i)
\end{equation}
The cosine metric between two vectors $\mathbf{x}$ and $\mathbf{y}$ in $\mathbb{R}^m$ is given by:
\begin{displaymath}
d(\mathbf{x},\mathbf{y}) = 1 - \frac{\langle{\mathbf{x},\mathbf{y}}\rangle}{\lVert\mathbf{x}\rVert\cdot\lVert\mathbf{y}\rVert},
\end{displaymath}
where $\lVert\cdot\rVert$ denotes the standard Euclidean metric and $\langle{\cdot,\cdot}\rangle$ is the standard Euclidean inner product. When $\mathbf{x},\mathbf{y} \geq 0$ (and non-zero), then $d(\mathbf{x},\mathbf{y}) \leq 1$ and $1 - d(\mathbf{x},\mathbf{y})$ is the cosine similarity.

\subsection{Twitter Dataset}
We assume a passing familiarity with the micro-blogging service Twitter. Each Twitter user generates short broadcast messages (tweets) that can refer to any other user by their username prepended with the \texttt{@} symbol. A Twitter user who follows another Twitter user will receive the broadcast tweets made by that user. A tweet is defined as a \textit{retweet} if it is a repost of another user's tweet. Users can be directionally connected through the act of retweeting forming a \textit{retweet network}. See \cite{TwitterHowTo}.

For an arbitrary sampling frequency (e.g., once per hour, once per day, etc.) the output of a Twitter user creates two discrete time series of interest. These can be thought of as observations from a dynamic system. Fix a user $i$. During sampling period $t$ a word sequence $M^{i}_t$ is observed which contains the concatenated tweets sent by user $i$ during period $t$. 
\begin{equation}
\mathbf{X} = \left\{g\left(M^{i}_t\right)\right\}_{t}
\end{equation}  
The second time series of interest is the tweet count series which simply counts the number of tweets sent by user $i$ during period $k$. We denote this $\left\{\nu^{i}_t\right\}_t$ or in functional form $\nu^{i}(t)$. In Section \ref{sec:Topic}, we use two timescales for topic analysis to compensate for the fact that tweets are short texts and therefore, short timescales are not instructive for understanding the behavior of the dynamical system in message space.

In this work, we are interested in the generation and representation of content by the IRA user group within Twitter, as identified by by Twitter Inc. For this reason, we categorize tweets in one of three ways:

\begin{description}
\item[Original ] A tweet by an IRA user is \textit{original} if it is \textit{not} a retweet of any other user's content.

\item[Spreading ] A tweet by an IRA user is \textit{spreading} if it is a retweet of another user's content and that user is \textit{within} the IRA user group.

\item[Amplifying ] A tweet by an IRA user is \textit{amplifying} if it is a retweet of another user's content and that user is \textit{outside} the IRA user group.

\end{description}
Assigning these descriptors to a tweet can be done automatically. The data set provides information on whether a tweet is a retweet and if so who is being retweeted. It is then straightforward to check whether the user who is retweeted is part of the IRA data set.

\subsection{User Strategy Space}
Define:
\begin{displaymath}
\Delta_n = \left\{\mathbf{x} \in \mathbb{R}^n : \sum_i x_i = 1,\,x_i \geq 0\right\}
\end{displaymath} 
This is the (n-1)-dimensional simplex embedded in $\mathbb{R}^n$ and used extensively in three-strategy evolutionary games (see e.g., \cite{GB17}). 

\begin{figure}[htbp]
\centering
\includegraphics[width=0.65\columnwidth]{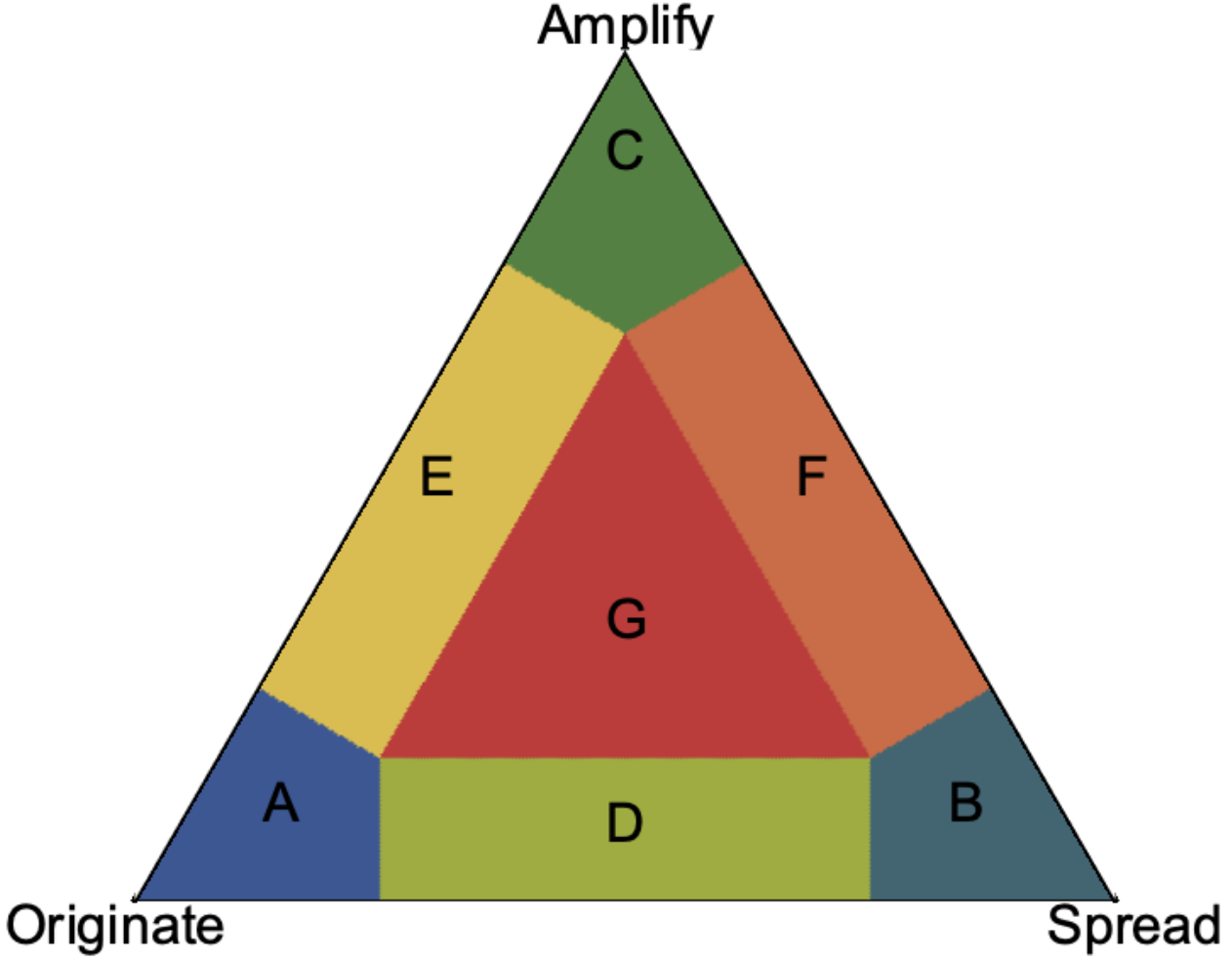}
\caption{A symbolization of $\Delta_3$ that will be used in analyzing user behavior.}
\label{fig:Symbol}
\end{figure}

We represent user $i$'s decision to post an original, spreading or amplifying tweet as strategic. Any time $\nu^i(t) > 0$, we can divide user $i$'s tweets into these three categories and calculate the three dimensional strategy vector:
\begin{equation}
\bm{\pi}^i(t) = \left\langle{\frac{\nu^i_1(t)}{\nu^i(t)},\frac{\nu^i_2(t)}{\nu^i(t)},\frac{\nu^i_3(t)}{\nu^i(t)}}\right\rangle,
\end{equation}
where $\nu^i_j(t)$ is the number of times strategy $j$ is selected by user $i$ during time period $t$. Each strategy vector $\bm{\pi}^i(t) \in \Delta_3$. Thus, $\{\bm{\pi}^i_t\}_t$ is a time series of mixed strategies in $\Delta_3$. To understand this dynamical system, we will adopt a symbolization of $\Delta_3$ allowing us to study (at a gross scale) the distribution of user strategies over longer time periods in the data. The symbolization is shown in \cref{fig:Symbol}.
Denote by $\left\{\sigma^i_t\right\} \subset \Delta_3$ the symbolized strategy sequence of User $i$, where $\sigma^i_t \in \mathcal{A} = \{A,B,C,D,E,F,G\}$.

\section{Bulk Data Analysis}\label{sec:Bulk}
Twitter Inc. maintains the Twitter Elections' Integrity Hub.\footnote{Twitter maintains a current repository of all public, non-deleted tweets and media from accounts believed to be connected to state-backed information operations here: \url{https://transparency.twitter.com/en/information-operations.html}.} This site contains tweets and accounts alleged to be part of a Russian active measures campaign against the West, led by the Internet Research Agency (IRA). In the datasets shared on the site, all user names for accounts with less than $5000$ followers are \textit{hashed}, that is, user names have been obscured through a string hashing function. Un-hashed versions of this data can be obtained upon request to Twitter Inc. Data analysis in this paper was performed with the un-hashed version. User account information will be protected\footnote{Usernames which were hashed in the publically released dataset are represeted in the paper as \texttt{hashed\_xx} and a mapping to the full hashed usernames is provided in Appendix D.} unless it is already clear that the user is a known figure and has appeared in a publicly available data set; e.g., the user \texttt{TEN\_GOP} is mentioned in the Mueller report \cite{M19}.

\begin{figure}[htbp]
\centering
\includegraphics[width=0.75\columnwidth]{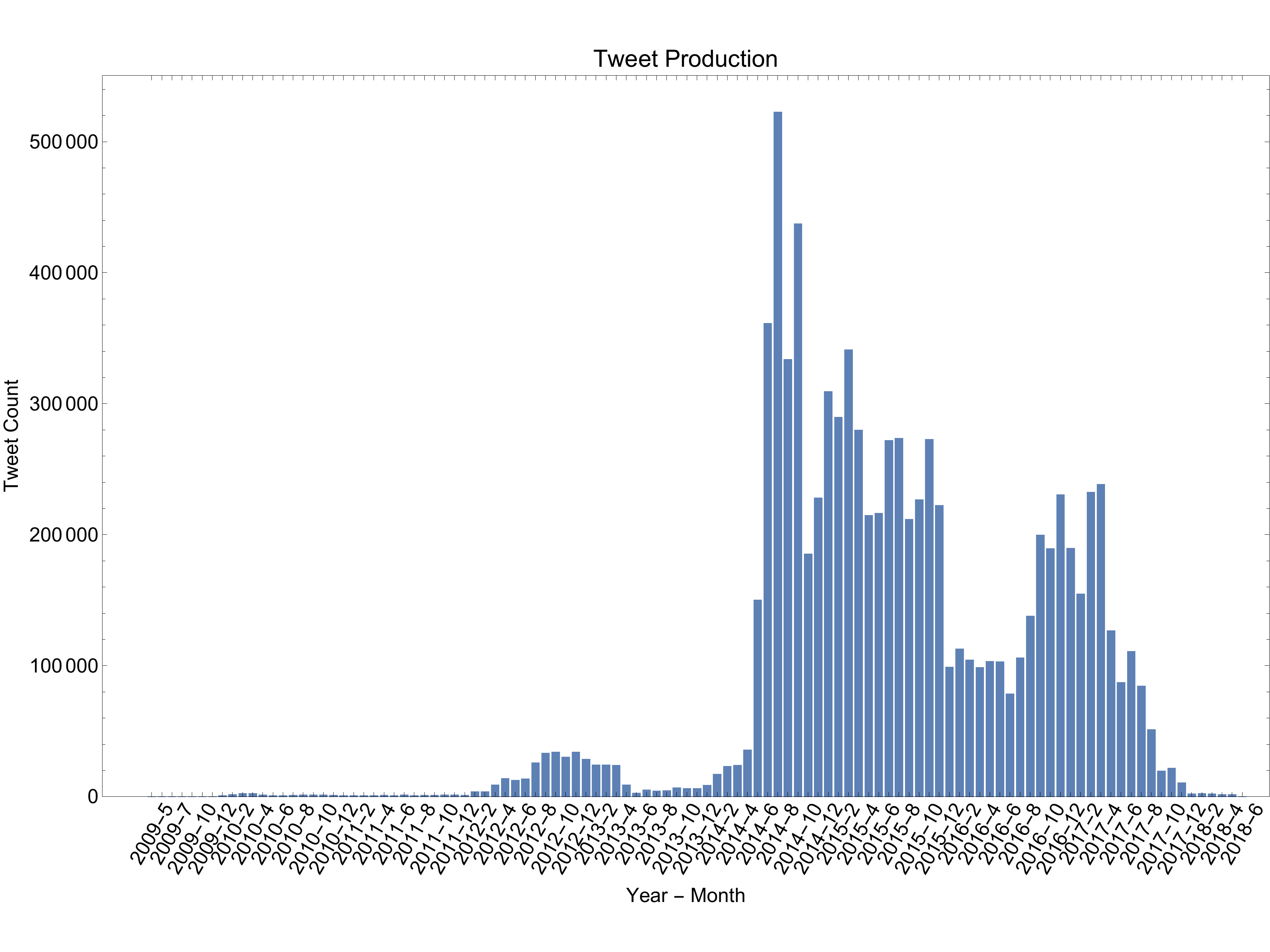}
\caption{A temporal histogram showing the total distribution of IRA tweets over a period from May-2009 to June 2018.}
\label{fig:TotalTweetCounts}
\end{figure}

The data set we use is an aggregate of the October 2018, January 2019, and June 2019 batch releases on the Elections' Integrity Hub. Our combined data set consists of $8,768,633$ individual tweets spanning over 9 years. There are $3,116$ users identified in the data. \cref{fig:TotalTweetCounts} shows the number of tweets per month over the time spanned by the data set. Others authors \cite{GB18,KRWR19,Lerman, Armchair1, Armchair2, Armchair3} have spent substantial time correlating twitter activity to world events and there is non-trivial correlation to these events. 

\begin{figure}[htbp]
\centering
\includegraphics[width=0.95\columnwidth]{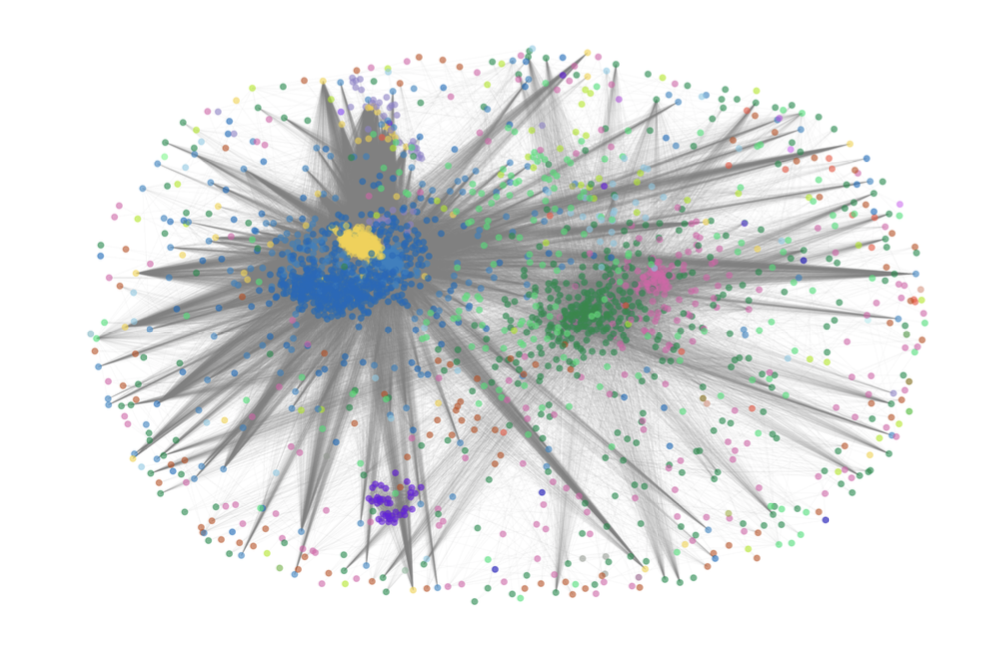}
\caption{The retweet network amongst IRA accounts over the complete dataset. Using modularity-based community detection, we identify 28 communities and attain modularity of 0.395.}
\label{fig:RetweetNetwork}
\end{figure}

The internal retweet network of IRA users (see \cref{fig:RetweetNetwork}) suggests a somewhat fragmented network with the top retweeted accounts tweeting primarily in Russian. We identify 28 communities in the retweet network using modularity-based community detection.
Fragmentation within the retweet network will be mirrored in the dynamical systems analysis in the sequel, suggesting that a consistent picture emerges at multiple scales of analysis. 

\subsection{English Language Tweets}
\begin{figure}[htbp]
\centering
\includegraphics[width=0.75\columnwidth]{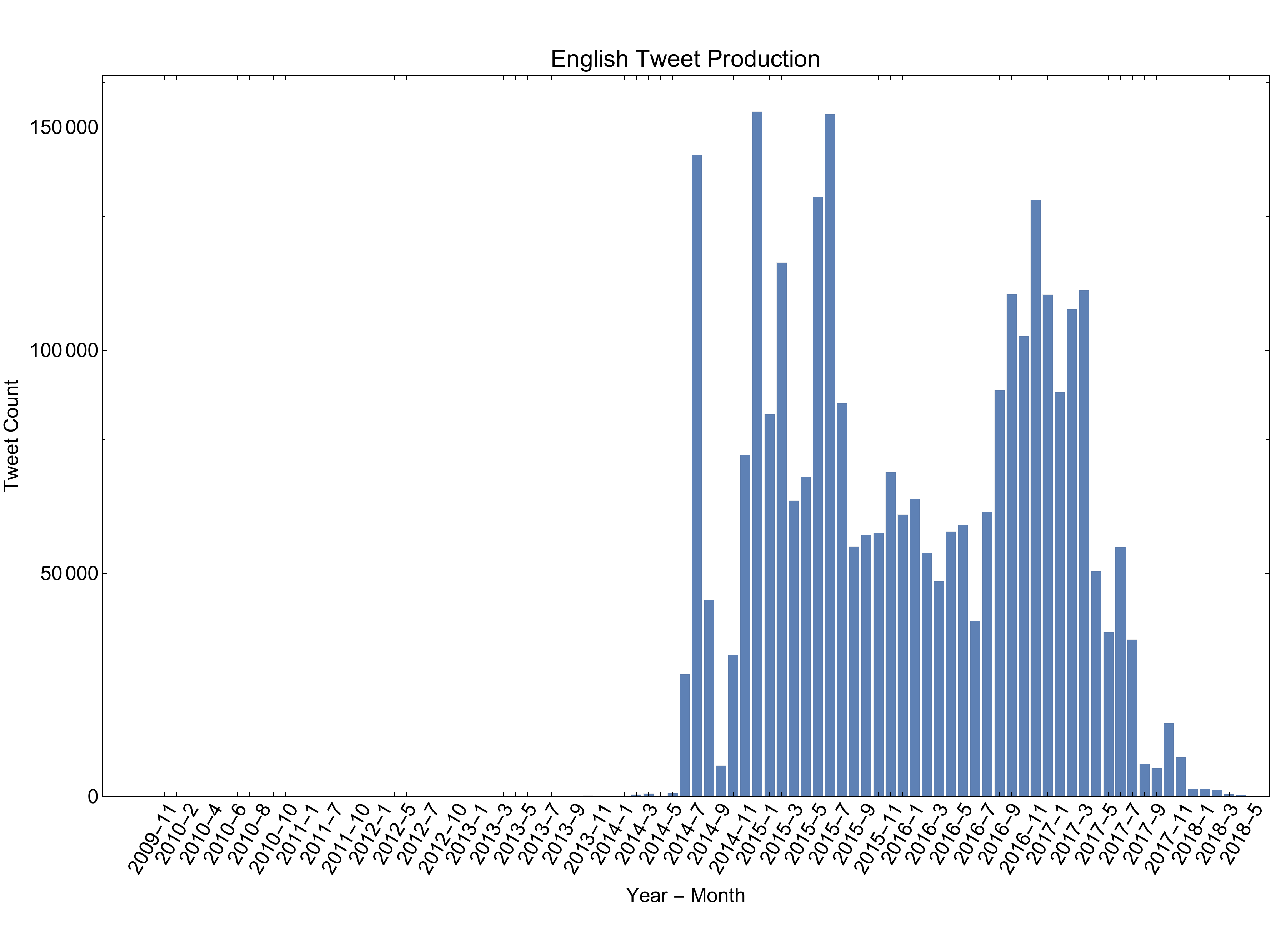}
\caption{A temporal histogram showing the total distribution of English language IRA tweets over the data set.}
\label{fig:EnglishTweets}
\end{figure}

We now focus on English language tweets because this will allow us to test hypotheses posed in \cite{GB18} using the full data set provided by Twitter Inc. In \cite{GB18} the Mathematica language detector was used to classify tweet language. In our present dataset, Twitter Inc. has provided a language for each tweet in the data set under investigation. We note that the accuracy of Twitter's language detector is unknown. A temporal histogram of English language tweets is shown in \cref{fig:EnglishTweets}.
For simplicity, we focus our study in English on the time period between January 1, 2015 and December 31, 2017. This includes the 2016 US presidential election as well as the majority (88.86\%) of English tweets in the data set. In the sequel, we will focus even more specifically on users who tweet sufficiently often to ensure that derived time series are not subject to small sample problems. 

\subsubsection{Model of tweet frequency}

\begin{figure}[htbp]
\centering
\includegraphics[width=0.75\columnwidth]{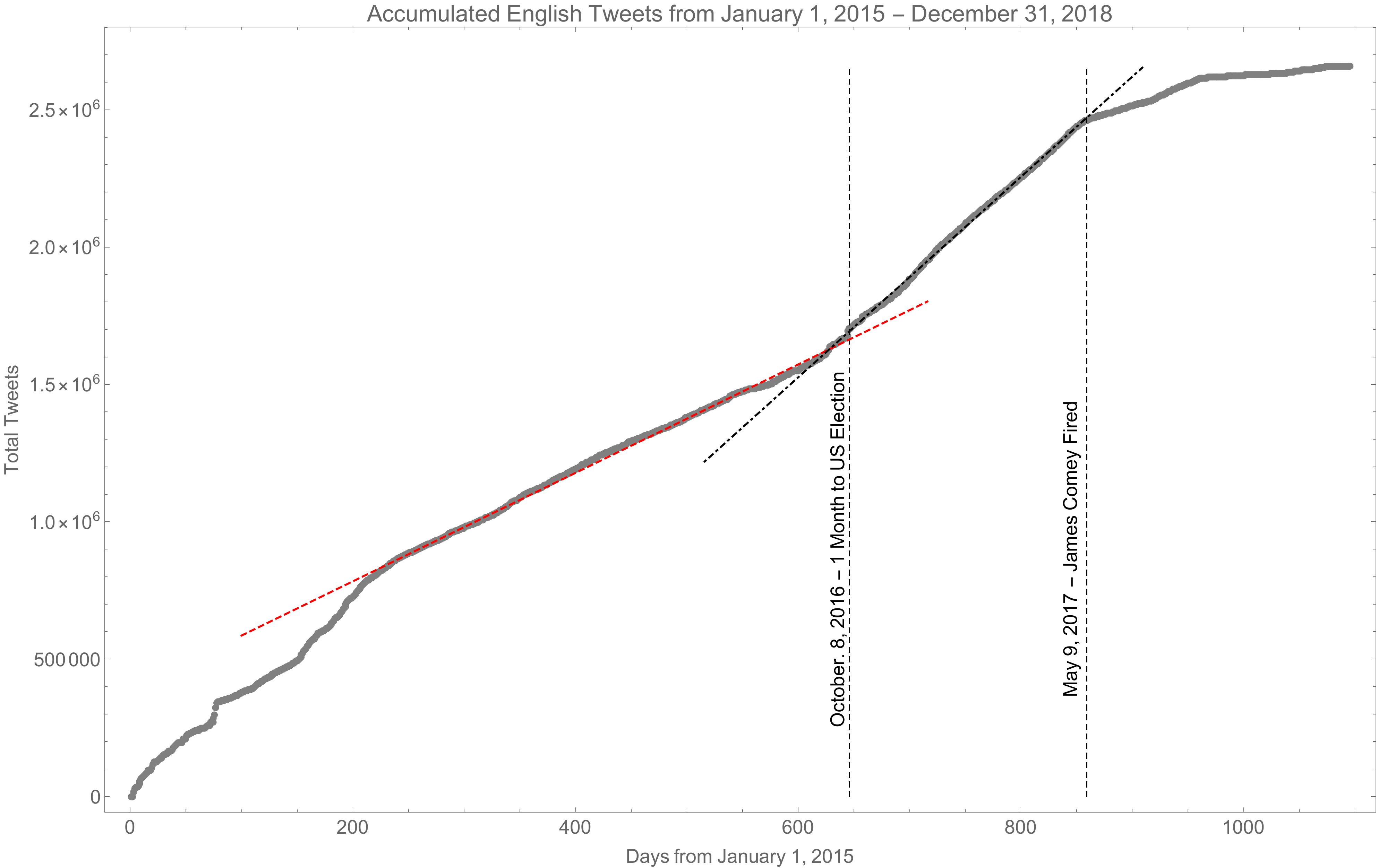}
\caption{The English Tweet accumulation function from January 1, 2015 and December 31, 2017 appears piecewise linear.}
\label{fig:EnglishAccumulation}
\end{figure}

Let $\{N(t)\}_t$ denote the total number of tweets posted in English each day between January 1, 2015 and December 31, 2017 and let the accumulation function be given by:
\begin{equation}
S(t) = \sum_{s\leq t} N(s)
\end{equation} 
This function is shown in \cref{fig:EnglishAccumulation}.
Visual inspection suggests that the function is piecewise linear. Griffin and Bickel \cite{GB18} propose the hypothesis that there is a change point in the behavioral (tweet frequency) dynamics on or about the start of the Republican National Convention (July 18, 2016). The complete data picture suggests that this hypothesis is not correct. However, we can support the hypothesis that there is a change point in the total messaging velocity $N(t)$ at some point near the 2016 presidential election. To see this, we fit two linear models of the form:
\begin{equation}
\hat{S}^j(t) = \beta_0^j + \beta_1^j(t - t_0^j)
\label{eqn:PiecewiseFunction}
\end{equation}
The first model uses data from $S(t)$ starting on July 20, 2015 and ending on September 8, 2016. The second model uses data from $S(t)$ starting on September 9, 2016 and ending on May 9, 2017. The fits for these models are both shown in \cref{fig:EnglishAccumulation}. Both models have $r^2-\text{Adj} > 0.99$ (i.e., more than 99\% of the variance in the data is explained by the linear models). Moreover statistically we can show that a change point occurs at some point in the data set near September 8 by observing the model parameters: 
\begin{description}
\footnotesize
\item[Model 1 Parameters]
\begin{displaymath}
\begin{array}{l|lll}
 \text{} & \text{Estimate} & \text{Std. Error} & \text{Confidence Interval} \\
\hline
1 & 783392. & 1250.53 & (777648.,789136.) \\
t-200 & 1973.81 & 5.20356 & (1949.91,1997.71) \\
\end{array}\end{displaymath}
\item[Model 2 Parameters]
\begin{displaymath}
\begin{array}{l|lll}
 \text{} & \text{Estimate} & \text{Std. Error} & \text{Confidence Interval} \\
\hline
1 & 1.58438\times 10^6 & 950.379 & (1.57996\times 10^6,1.5888\times 10^6) \\
t-616 & 3647.54 & 6.76713 & (3616.05,3679.03) \\
\end{array}
\end{displaymath}
\normalsize
\end{description}
The $5\sigma$ confidence intervals on $\beta_1^1$ and $\beta_1^2$ are non-intersecting, thus suggesting that these two linear models are different. Thus, if $S(t)$ is modeled by a piecewise function of the form in \cref{eqn:PiecewiseFunction}, then with high confidence we can assert there is a change point in this function near $t_0^2 = 616$ (September 9, 2016).

\subsubsection{Model of tweet strategy}

\begin{figure}[htbp]
\centering
\subfloat[Before September 9, 2016]{\includegraphics[width=0.45\columnwidth]{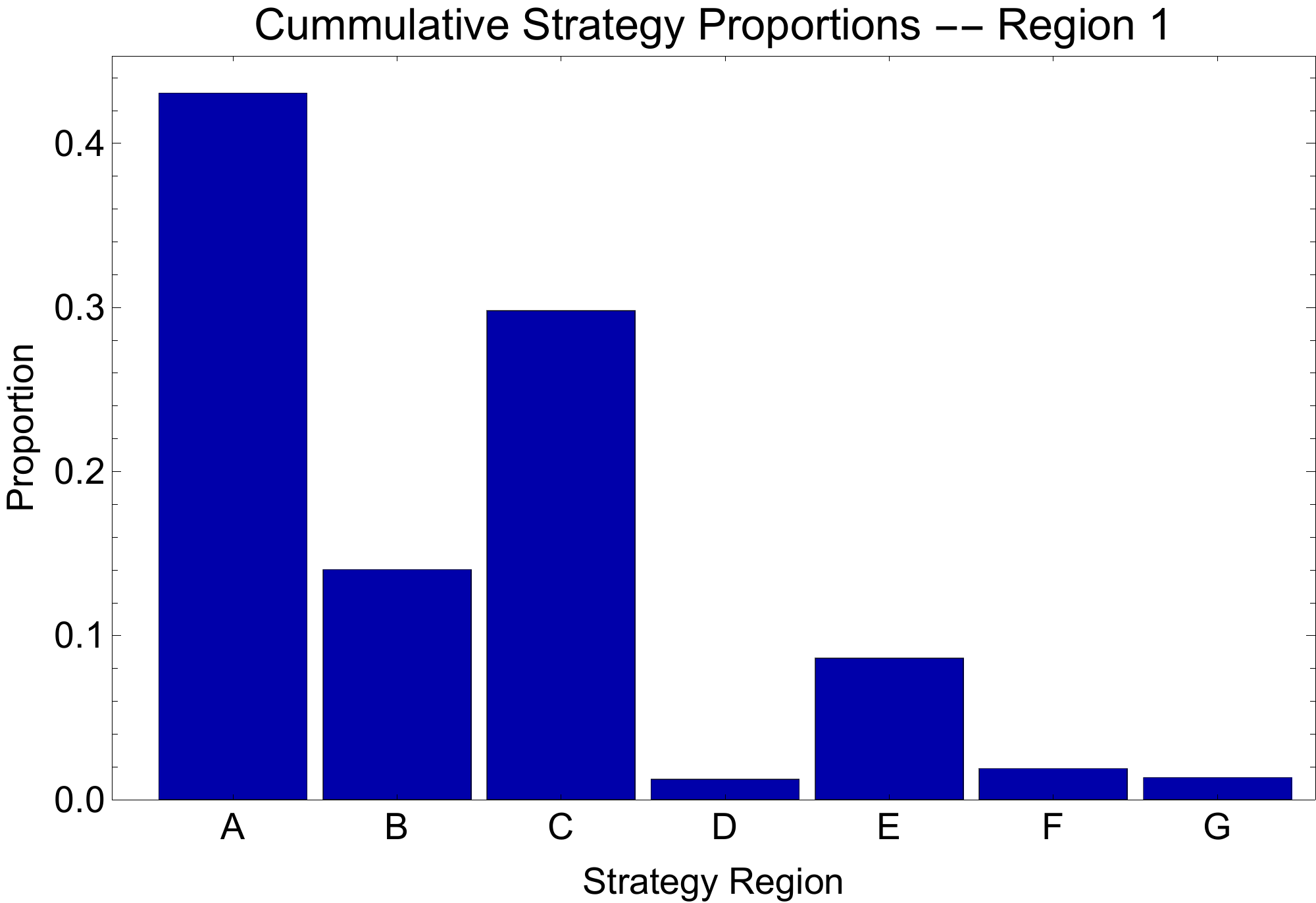}}\quad
\subfloat[After September 9, 2016]{\includegraphics[width=0.45\columnwidth]{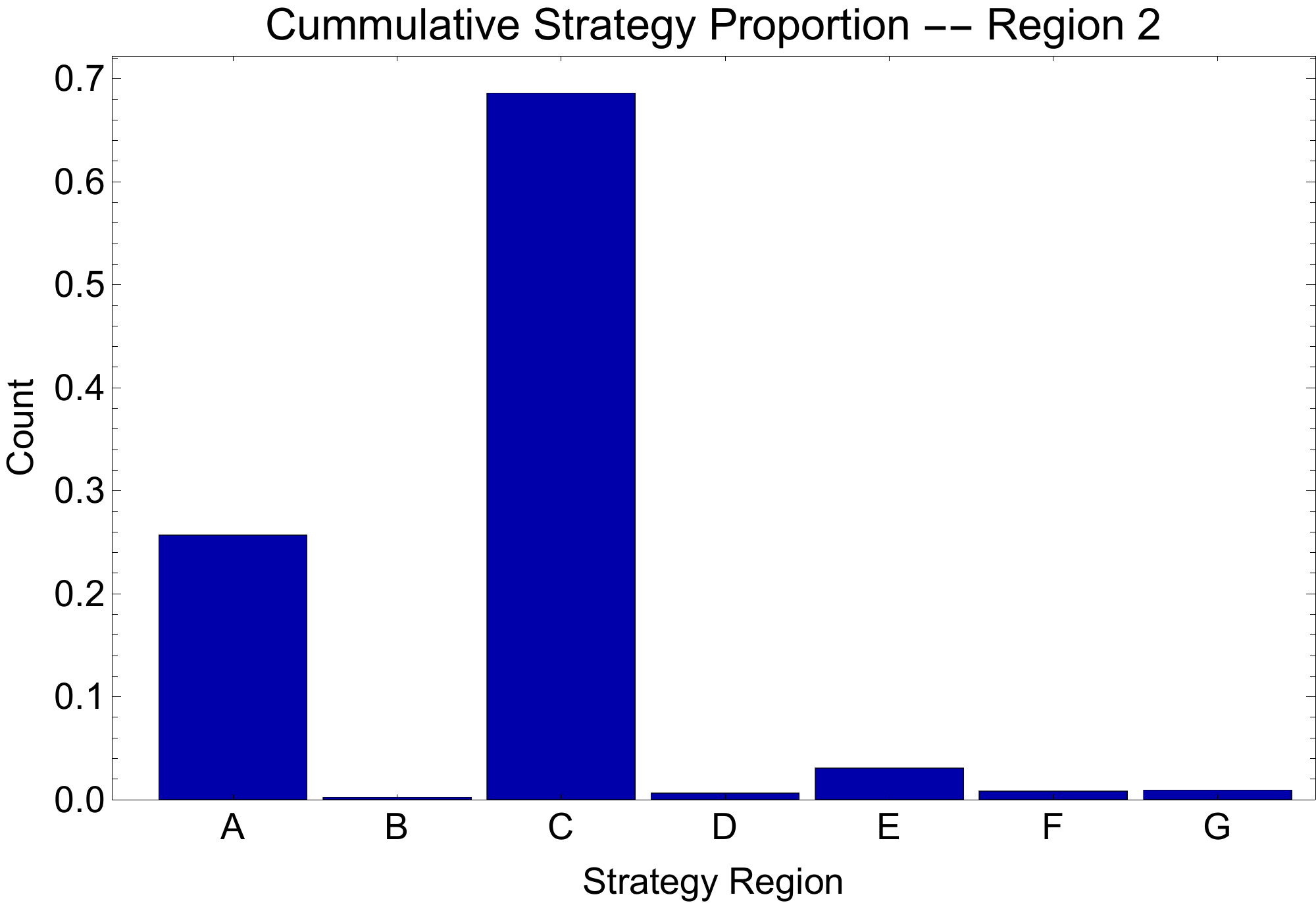}}
\caption{The discrete distributions of visiting patterns of the symbolic dynamical systems $\{\sigma^i_t\}_t$ in two temporal regions identified.}
\label{fig:SymbolDist}
\end{figure}

We illustrate that in addition to a change in messaging rate, there is also a change point in the strategy used by the IRA users. Here we consider the symbolic dynamic systems $\{\sigma^i_t\}_t$ for all users $i$ sampled daily. We compare the distribution of visited partition elements in $\mathcal{A}$ before and after September 9, 2016, the change point used in the previous analysis. Define the indicator:
\begin{displaymath}
I_\alpha(\sigma^i_t) = \begin{cases}
1 & \text{$\sigma^i_t = \alpha$}\\
0 & \text{otherwise}
\end{cases},
\end{displaymath}
and define the discrete distributions:
\begin{align*}
p^1(\alpha) &= \frac{\sum_i\sum_{t \in [t_0^1,t_0^2)} I_\alpha(\sigma^i_t)}{\sum_\alpha\sum_i\sum_{t \in [t_0^1,t_0^2)} I_\alpha(\sigma^i_t)}\\
p^2(\alpha) &= \frac{\sum_i\sum_{t \in [t_0^2,t_f)} I_\alpha(\sigma^i_t)}{\sum_\alpha\sum_i\sum_{t \in [t_0^2,t_f)} I_\alpha(\sigma^i_t)}
\end{align*}
where $t_f$ is May 10, 2017. All data \textit{up to} $t_f$ are considered. The two distributions are shown in \cref{fig:SymbolDist}.
It is clear (visually) that these two distributions are distinct. However, we compute $\chi^2 = 30,827$ using $p^1(\alpha)$ as the reference distribution. Thus, with extremely high certainty we can assert that a strategy change occurred at some point between 2016 and 2017. In particular, IRA users shifted their most common strategy from posting original tweets to posting amplifying tweets. This pattern is investigated further in the sequel when we focus on the subset of IRA users who post the highest volume of tweets. 

\section{Spectral Analysis of the Dynamical System}\label{sec:Spectral}
\begin{figure}[htbp]
\centering
\includegraphics[width=0.65\columnwidth]{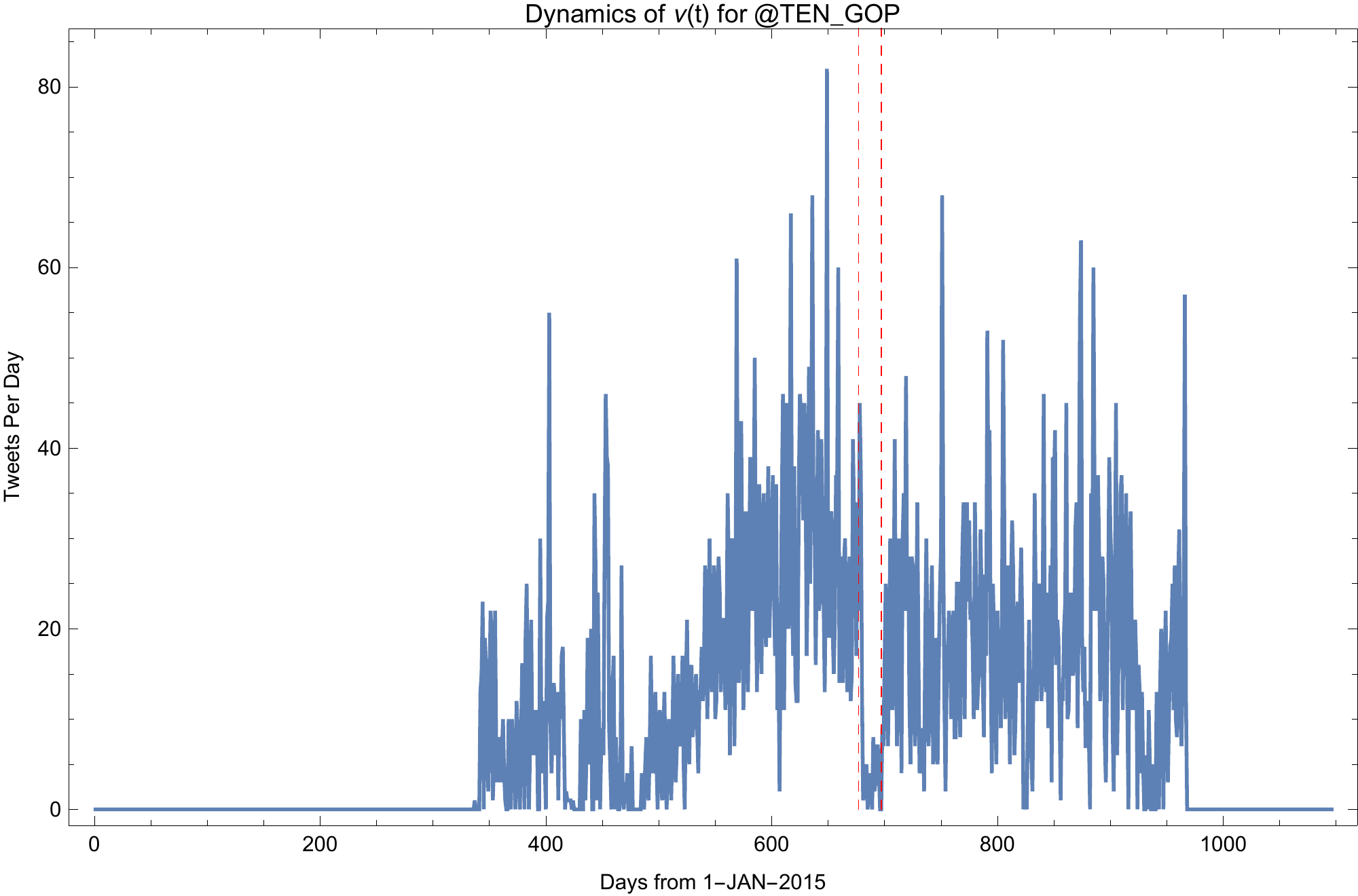}
\caption{The non-stationary dynamics of \texttt{TEN\_GOP} serve as an example of the non-stationary behavior of most users. The drop in activity after the election is shown with the dashed lines.}
\label{fig:Example}
\end{figure}

Observing samples from $\nu^i(t)$ for (almost) any arbitrary user illustrates that the underlying dynamics of users is non-stationary. This is illustrated in \cref{fig:Example} using \texttt{TEN\_GOP}, however any user would suffice.  
To analyze the data set as a dynamical system, we now restrict our attention to two shorter time periods and a smaller set of users. We select users in the $90^\text{th}$ percentile for total volume of tweets from January 1, 2015 through December 31, 2017. In particular, these users posted at least $1,093$ tweets during this period. Restricting to this set of users ensures there is sufficient text information available to make analysis of message content meaningful. We refer to this set of 312 users by the set $\mathcal{U}^*$. Moreover, we restrict our investigation to the time periods from March 9, 2016 to November 8, 2016 and November 29, 2016 to July 31, 2017. Both of these time periods are 244 days long. This ensures we have a sufficient time span for a meaningful time series analysis. Note we omit the period immediately after the 2016 US presidential election because many users had a dramatic decrease in their output. This is illustrated in \cref{fig:Example}. Finally, in order to ensure we have sufficient observations within each of the 244-day time periods, we consider only those users who tweet at least once per day over 60\% of the days surveyed. As a result, we analyze 24 user dynamics during the March 9, 2016 - November 8, 2016 time period and 117 user dynamics during the November 29, 2016 - July 31, 2017 time period. The usernames considered within each of the two periods are provided in \cref{sec:UserId's}. These two sets of users will be denoted $\mathcal{U}^*_1$ and $\mathcal{U}^*_2$. 

The objective of the remainder of this section is to illustrate that the behavior of some users in $\mathcal{U}^*_1$ and $\mathcal{U}^*_2$ can be described by noisy quasi-periodic functions and can be grouped by common operational frequencies. We further show how these groups produce consistent topical content and use consistent strategies. 

\subsection{Families of Quasi-Periodic Behavior}
Using a small subset of the data used in this paper, \cite{GB18} observed that some Twitter users exhibit periodic behavior. In this section we test this observation for multiple users in $\mathcal{U}^*$, including \texttt{TEN\_GOP}, the user discussed in \cite{GB18}. We show that in this larger data set the following model describes a subset ($88.8 \%$) of per-day user tweet volume in both the pre- and post-election time periods we consider:
\begin{equation}
\nu^i(t)= \mu^i(u) + \sum_j A^i_j(u)\cos\left(\omega^{[i]}_jt +\varphi^i_j\right) + \epsilon^i(t).
\label{eqn:Model}
\end{equation}
Here $\mu^i(t)$ is a drift term affected by a (hidden) control signal $u$, $A^i(u)$ is an amplitude affected by the same $u$, $\varphi^i$ is a phase specific to $i$ and $\omega^{[i]}$ is a frequency common to members of a group $[i]$ to which user $i$ belongs. The term $\epsilon^i(t) \sim (0,\sigma^i)$ is a mean-zero noise term. More importantly, the parameter $\omega^{[i]}$ exhibits some consistency over the two time periods, though some users do switch their fundamental frequencies. Furthermore, the data suggest that the  number of terms in the Fourier expansion (indexed by $j$) is generally small ($6$ or fewer); therefore we assume the sum in \cref{eqn:Model} is finite. We note that \textit{not all users} can be modeled this way, with some users exhibiting extremely noisy or bursty behavior. 

\begin{figure}[htbp]
\centering
\includegraphics[width=0.65\columnwidth]{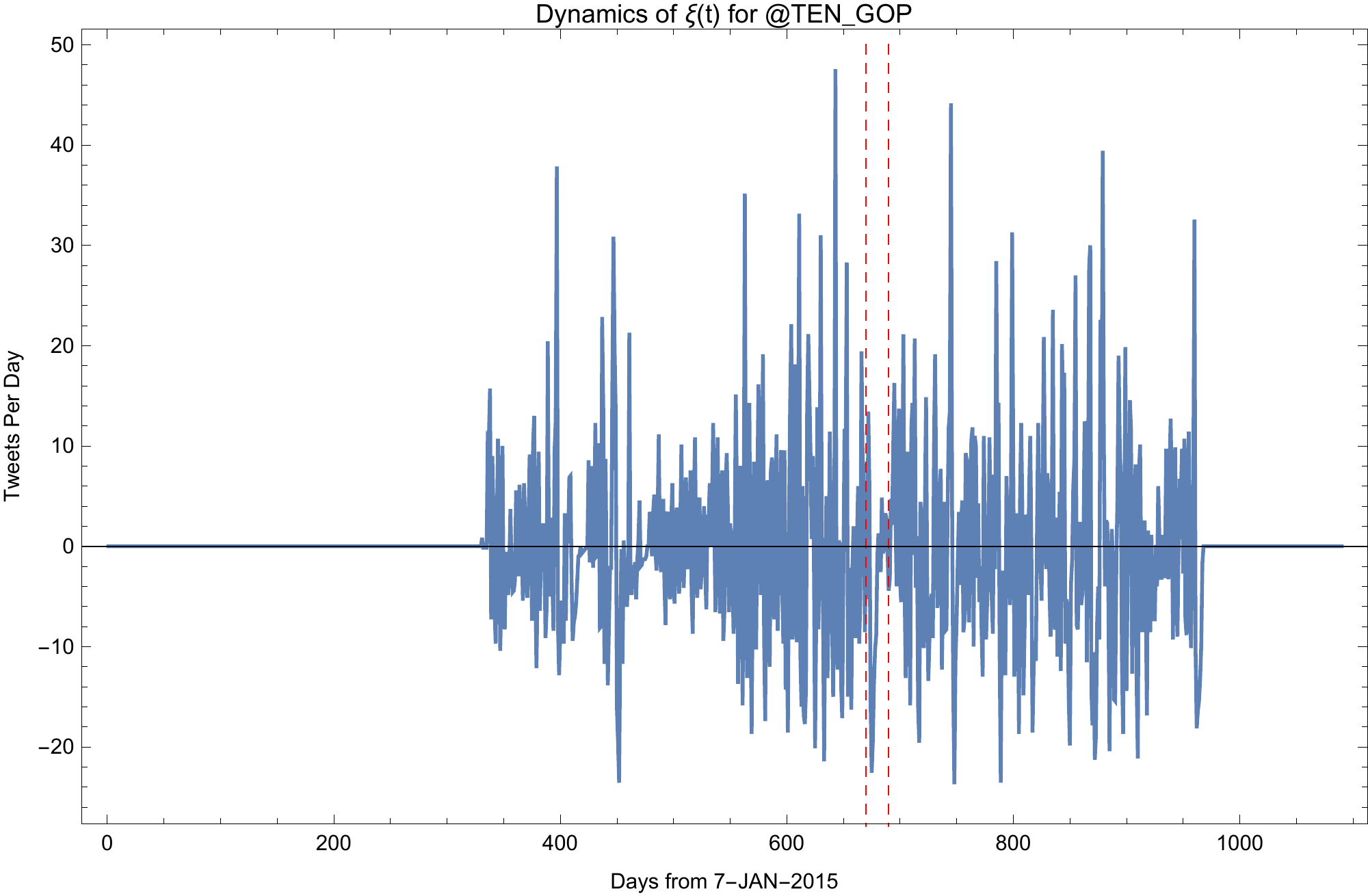}
\caption{The detrended dynamics of \texttt{TEN\_GOP} removes some of the linear growth shown around the run up the 2016 US presidential election.}
\label{fig:DeTrend}
\end{figure}

\begin{figure}[htbp]
\centering
\subfloat[Pre-Election Spectra]{\includegraphics[width=0.45\columnwidth]{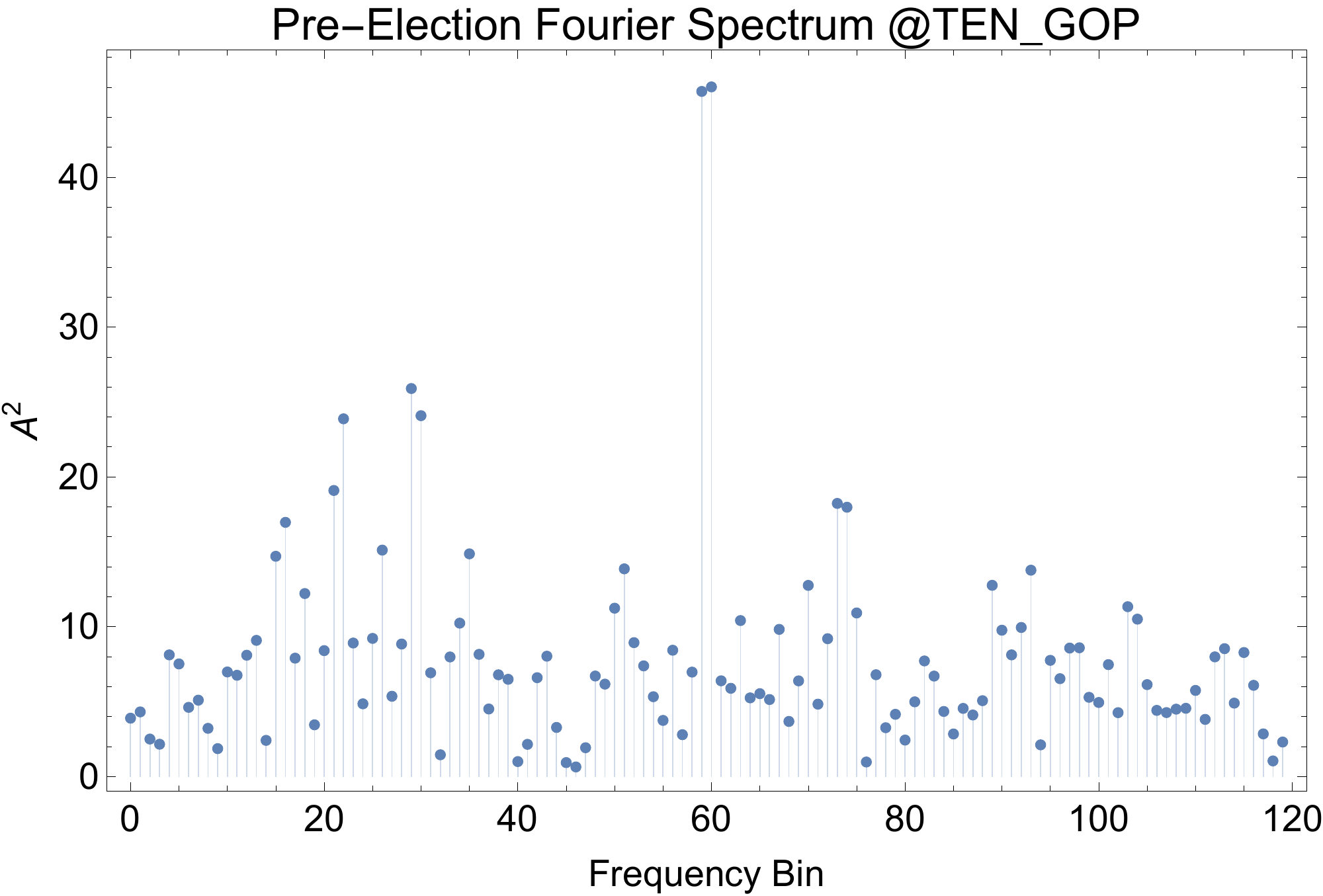}}\quad
\subfloat[Post-Election Spectra]{\includegraphics[width=0.45\columnwidth]{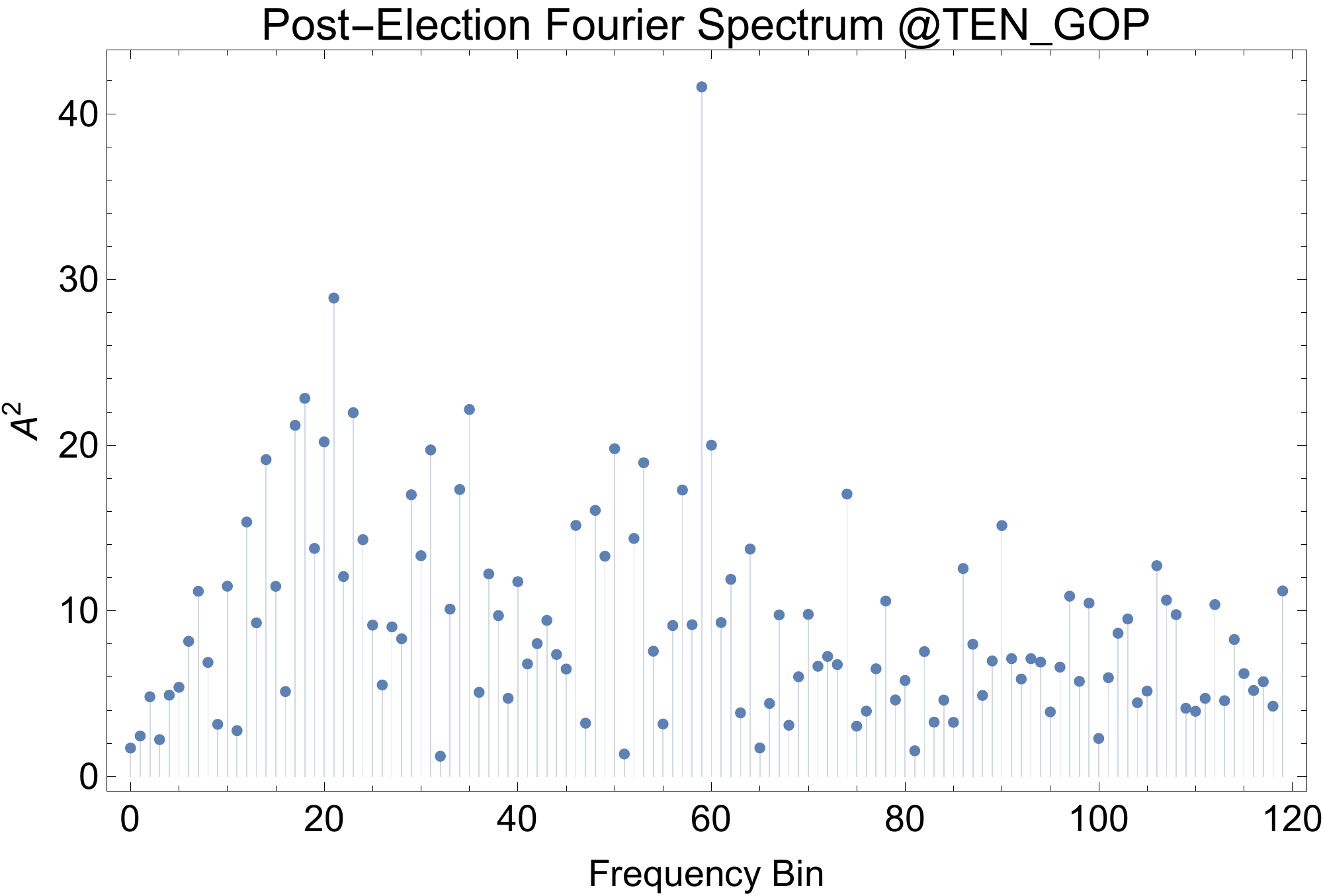}}\\
\subfloat[Pre-Election Fit]{\includegraphics[width=0.45\columnwidth]{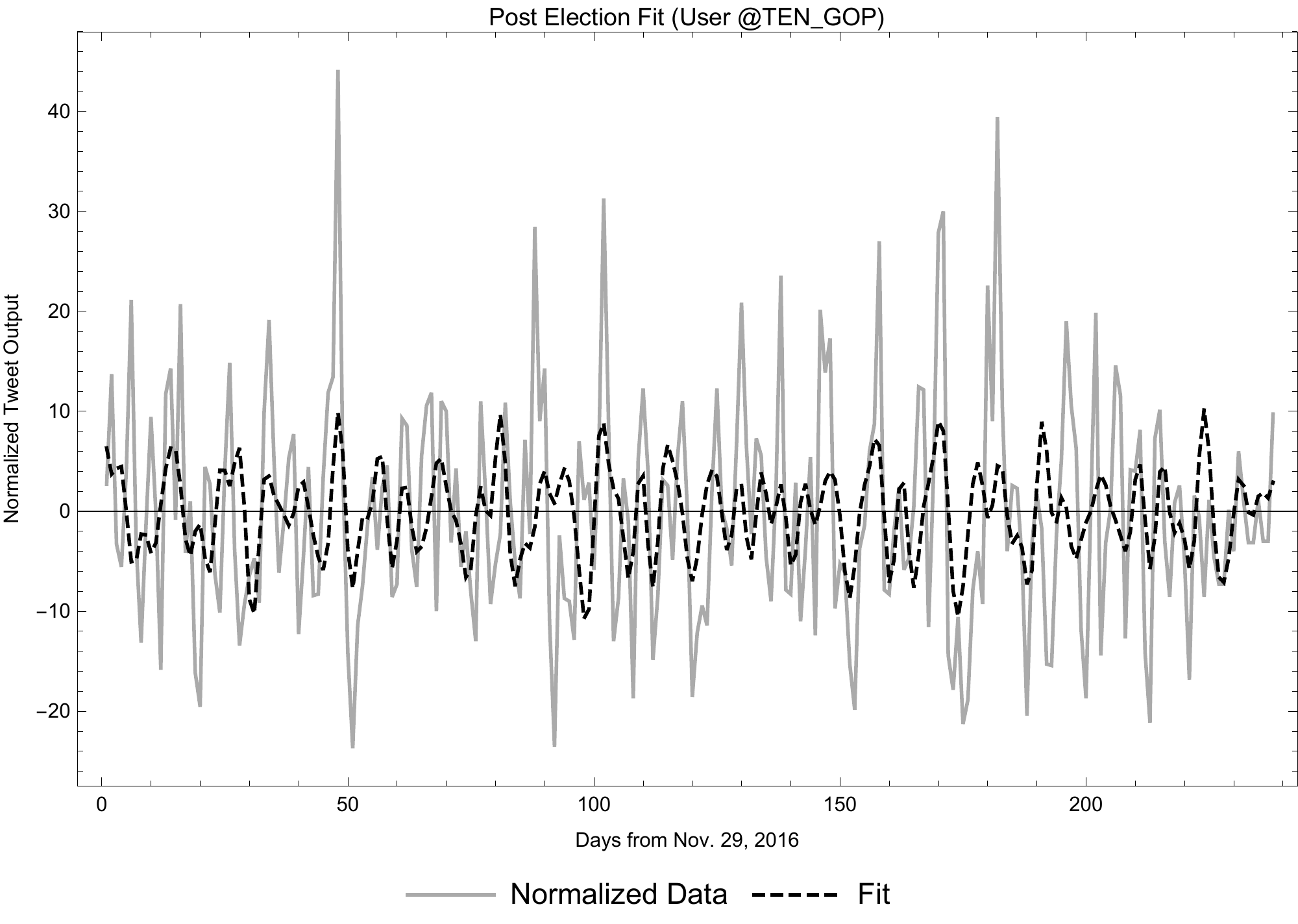}}\quad
\subfloat[Post-Election Fit]{\includegraphics[width=0.45\columnwidth]{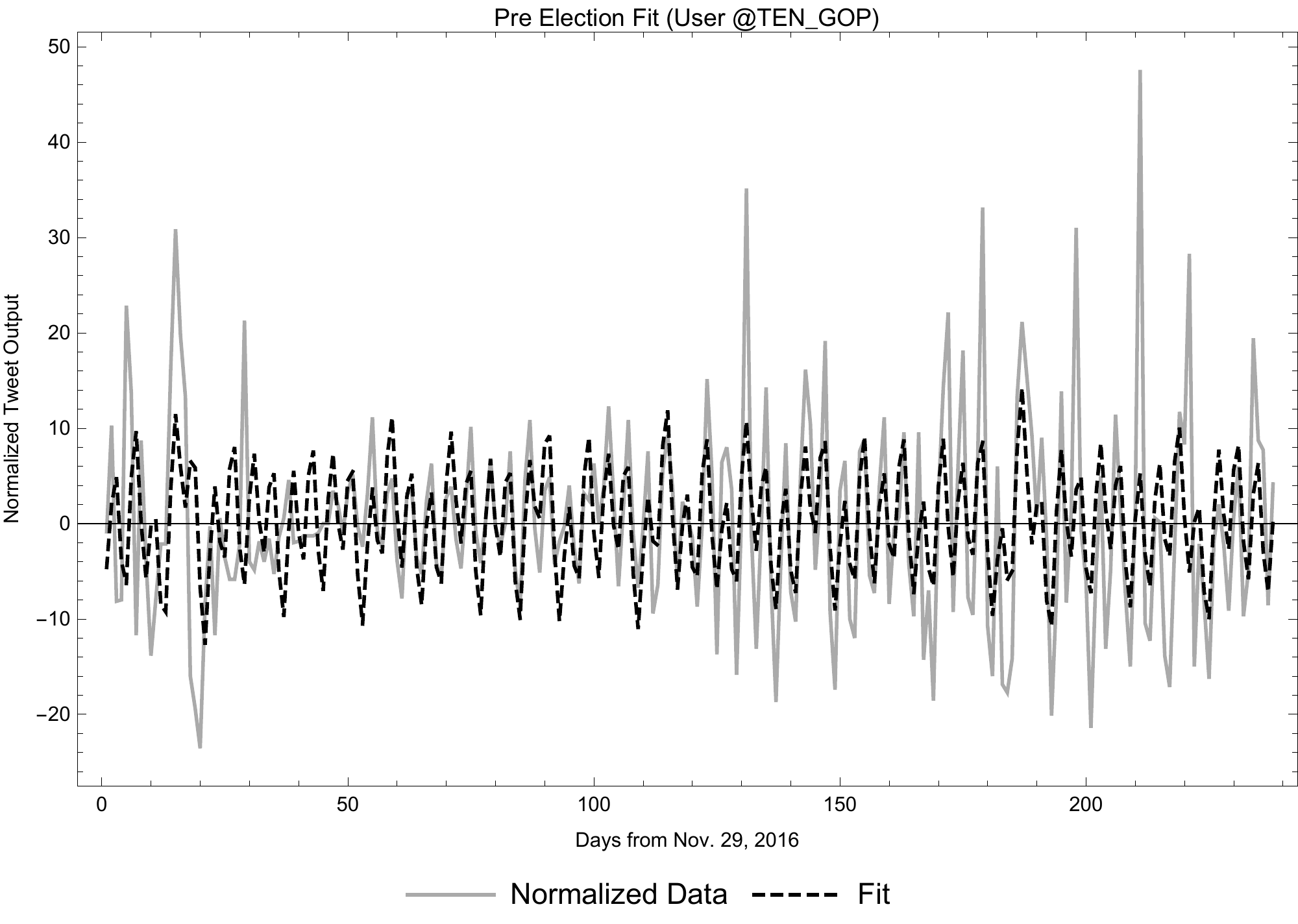}}
\caption{(Top) Fourier spectra for \texttt{TEN\_GOP} before and after the 2016 US Presidential Election. (Bottom) Corresponding fit using a 6 term Fourier sequence.}
\label{fig:TENGOP-Spectra}
\end{figure}

Before discussing the full set of data from $\mathcal{U}^*$, we illustrate the phenomena we discuss using \texttt{TEN\_GOP}. This allows us to confirm the hypothesis set out in \cite{GB18} that this user has a periodicity of 4 days in his/her tweet pattern. Note first from \cref{fig:Example} that there is a trend in the underlying data. To compensate for the trend, we use a detrended price oscillator approach \cite{D13} replacing $\nu^i(t)$ (the daily tweet volume) with:
\begin{displaymath}
\xi^i(t) = \nu^i(t) - \bar{\nu}^{i}_{\text{SM},7}(t).
\end{displaymath}
Here $\bar{\nu}^{i}_{\text{SM},7}$ is the simple 7 day moving average prior to $t$. The result of detrending is shown in \cref{fig:DeTrend}.
Notice the de-trended data do not show a linear increase in the run up to the 2016 US presidential election. Removing these trends also allows dominant frequencies in the spectrum to appear more clearly. Thus we can replace \cref{eqn:Model} with the more convenient expression:
\begin{equation}
\xi^i(t)= \sum_j A^i_j(u)\cos\left(\omega^{[i]}_jt +\varphi^i_j\right) + \epsilon^i(t).
\label{eqn:Model2}
\end{equation}
The Fourier transform and a fit using 6 terms for $\xi(t)$ for \texttt{TEN\_GOP} is shown in \cref{fig:TENGOP-Spectra} for the period before and after the 2016 US elections.
We note that in both spectra there is a high amplitude frequency at bin 60 indicating a cyclic behavior recurring (approximately) every 4 days. This is consistent with the results found in \cite{GB18}. We also note that there are a reasonably small number of high amplitude frequencies, particularly before the election. For completeness we illustrate the spectra and fit of \texttt{KansasDailyNews} in \cref{sec:Kansas}. This user has much more periodic behavior and a cleaner spectrum. 

\subsection{Behavior Spectrum Clustering}
Before proceeding, we layout the general analytic procedure to be followed in determining families of behavior describing $\xi^{i}(t)$. 
\begin{enumerate}
\item Cluster users based on the spectra of $\xi^i(t)$, thereby constructing the family $[i]$.  
\item Compare the derived clusters between the two time periods.
\item Examine the trajectories $\bm{\pi}^{i}(t)$ for those individuals in $[i]$ to determine whether there is a correlation between tweet strategy and tweet volume. 
\end{enumerate}
In Section \ref{sec:Topic}, we discuss clustering users in word space and relate these clusters to the behavioral clusters. 

Even after detrending the data, the spectra in \cref{fig:TENGOP-Spectra} are noisy. This is common to all users and is described by the noise term $\epsilon^i(t)$. To address noise in the spectra, we apply a de-noising procedure and following \cite{FGNP+03}, we use a principal components analysis to project the spectra onto a lower-dimensional space. 

\begin{figure}[htbp]
\centering
\subfloat[Pre-Election]{\includegraphics[width=0.45\columnwidth]{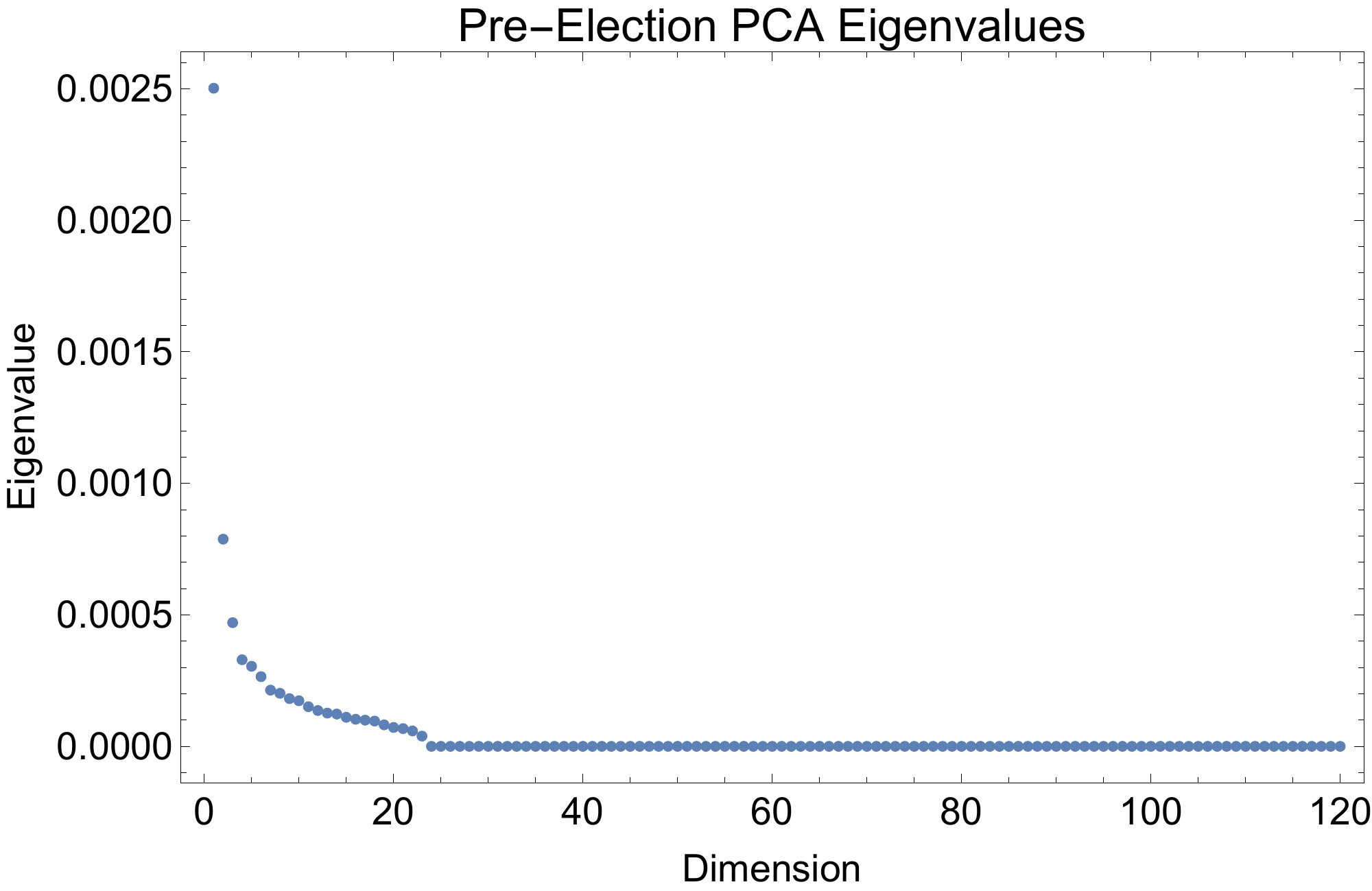}}\quad
\subfloat[Post-Election]{\includegraphics[width=0.45\columnwidth]{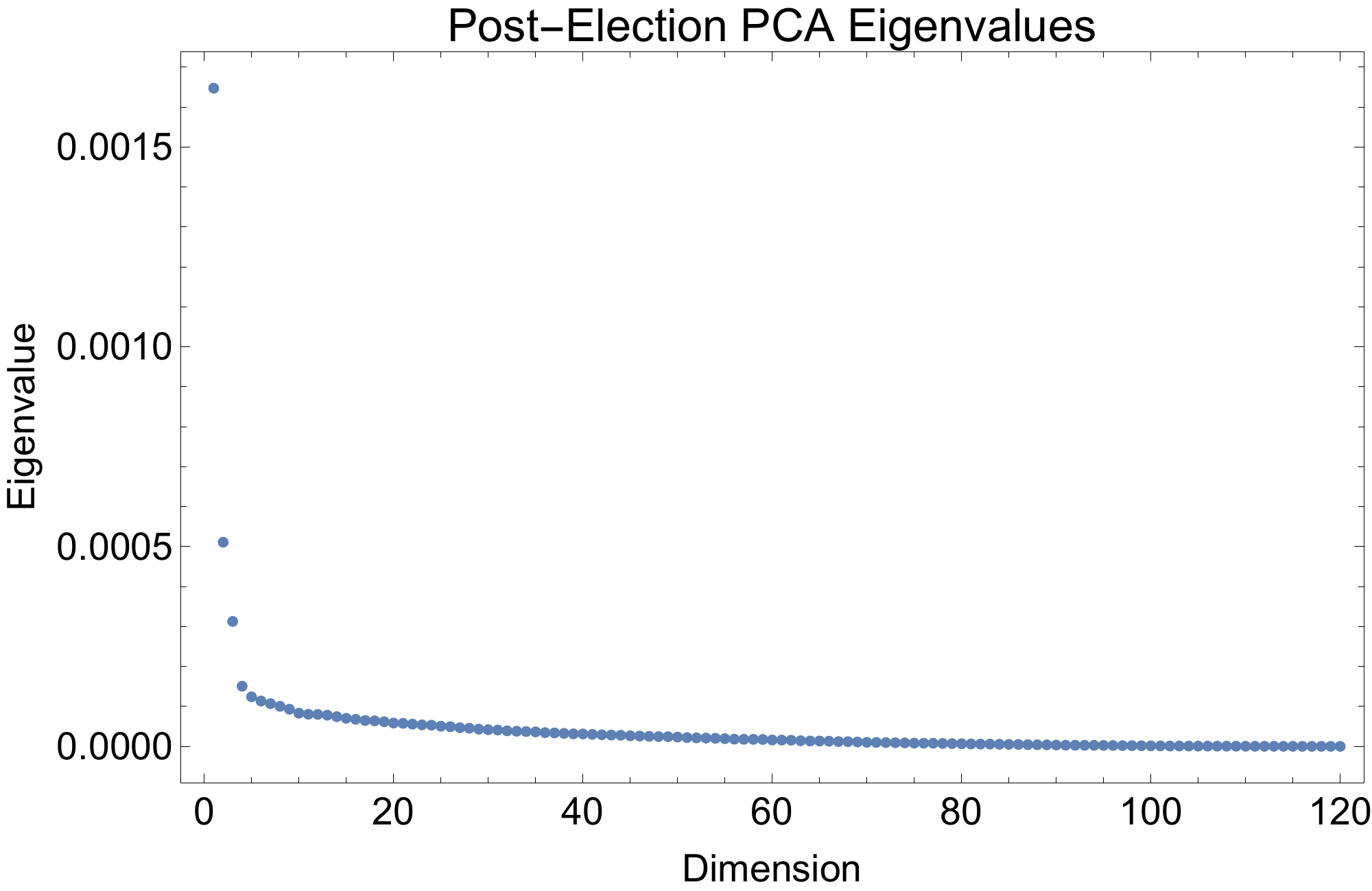}}
\caption{The eigenvalues of the covariance matrices obtained during principal components analysis of the de-noised Fourier transforms of the de-trended per-day volume data suggest three components adequately describe the data.}
\label{fig:Eigenvalues}
\end{figure}

\begin{figure}[htbp]
\centering
\subfloat[Pre-Election]{\includegraphics[width=0.45\columnwidth]{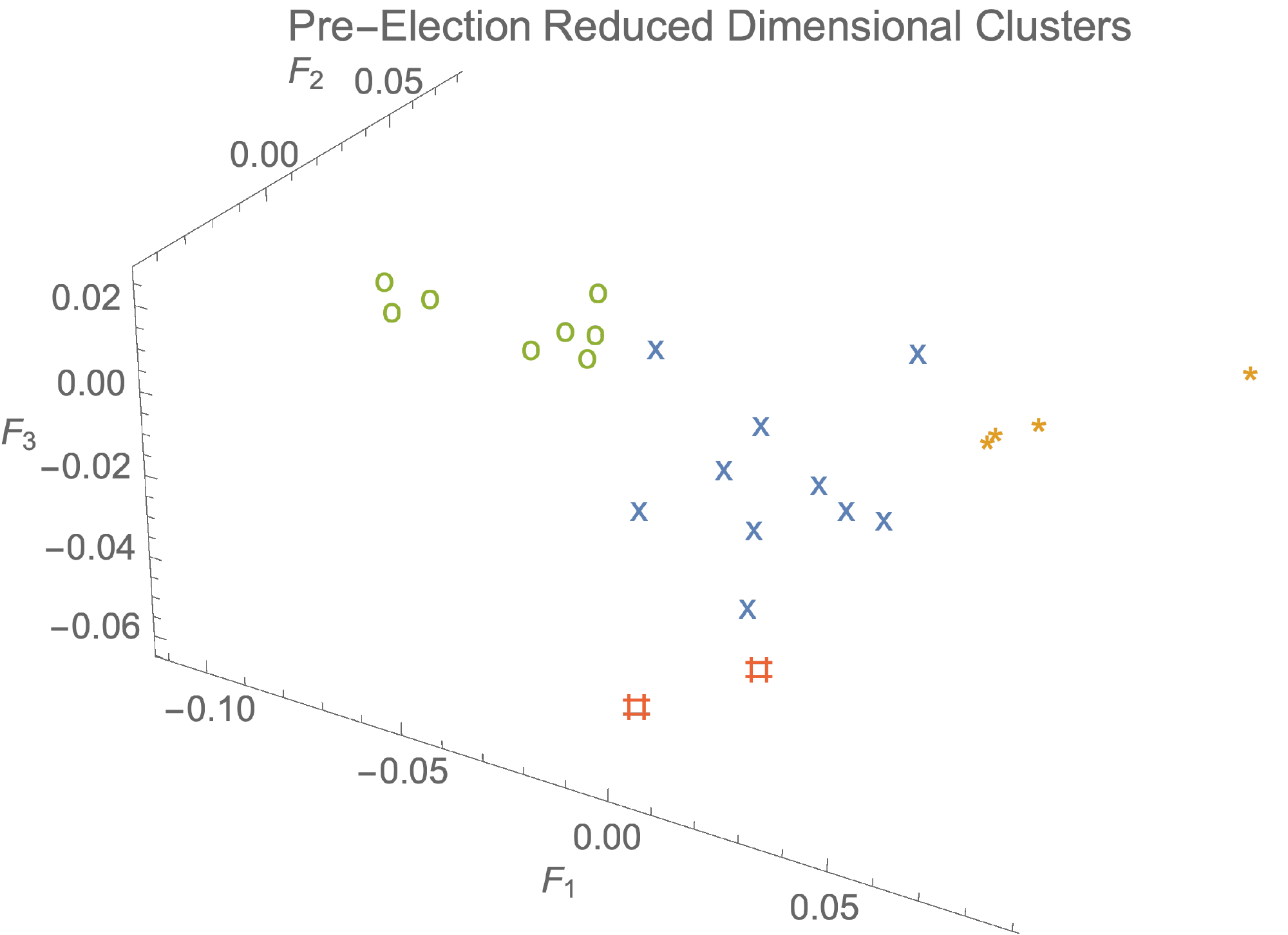}}\quad
\subfloat[Post-Election]{\includegraphics[width=0.45\columnwidth]{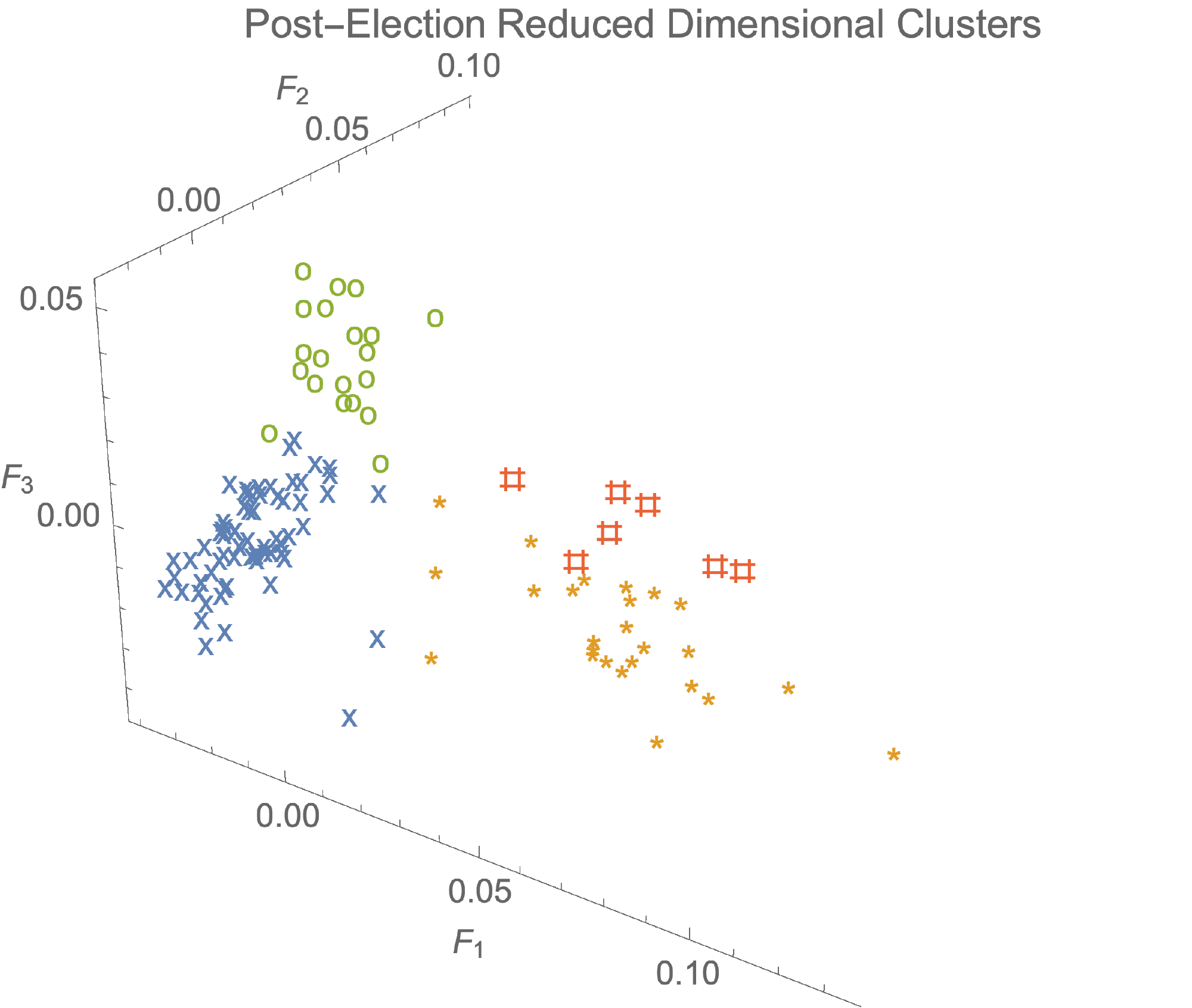}}
\caption{The reduced dimension clusters of spectral data show reasonable separation in three-dimensional space for both pre- and post-election data.}
\label{fig:Clusters3D}
\end{figure}

The spectral de-noising procedure is simple: we zero out frequency bins whose square amplitude is in the bottom $q$-percentile. For this data, we used $q = 0.33$. Let $D(\hat{\mathbf{x}},q)$ denote the de-noising procedure on a Fourier transform $\hat{\mathbf{x}}$ using quantile $q$ as a threshold. For each user $p_i \in \mathcal{U}_p^*$ ($p \in \{1,2\}$) this leaves a matrix:
\begin{displaymath}
\bm{\Xi}_p = 
\begin{bmatrix}
\hat{\bm{\xi}}_{p_1}^T\\
\vdots\\
\hat{\bm{\xi}}_{p_m}^T
\end{bmatrix},
\end{displaymath}
where:
\begin{displaymath}
\hat{\bm{\xi}}^{p_i} = D\left[\mathcal{F}\left(\bm{\xi}^{p_i}\right),q\right]
\end{displaymath}
is the de-noised Fourier transform of the time series $\bm{\xi}^{p_i} = \left\{\xi^{p_i}_t\right\}_t$. We then apply principal components analysis \cite{J11} to the data in $\bm{\Xi}_p$. The result is a reduced dimensional data set that exhibits good separation. The eigenvalues of the covariance matrix computed during the principal components analysis for both the pre- and post-election data sets are shown in \cref{fig:Eigenvalues}.
We note in both cases that there is a knee in the curve of the eigenvalues after the third eigenvalue. Therefore, we reconstruct the data in three dimensional space. Visual inspection of the reduced dimensional data suggests that the data may divide into four clusters. We use the $k$-medoid method \cite{KRD87} to cluster the data. The results are shown in \cref{fig:Clusters3D}.  

\begin{figure*}[p]
\centering
\subfloat[Pre-Election Cluster 1]{\includegraphics[width=0.23\textwidth]{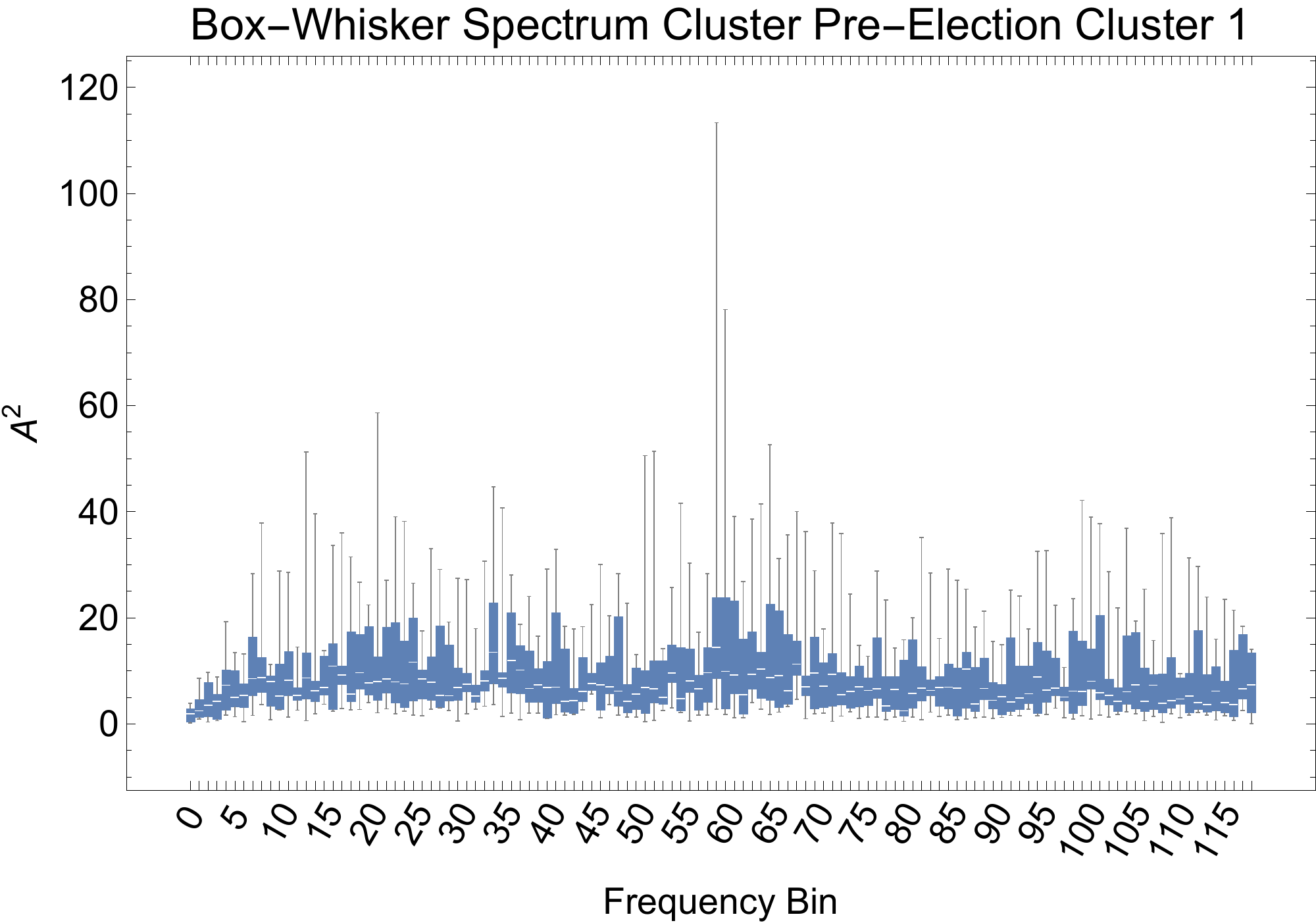}}
\subfloat[Pre-Election Cluster 2]{\includegraphics[width=0.23\textwidth]{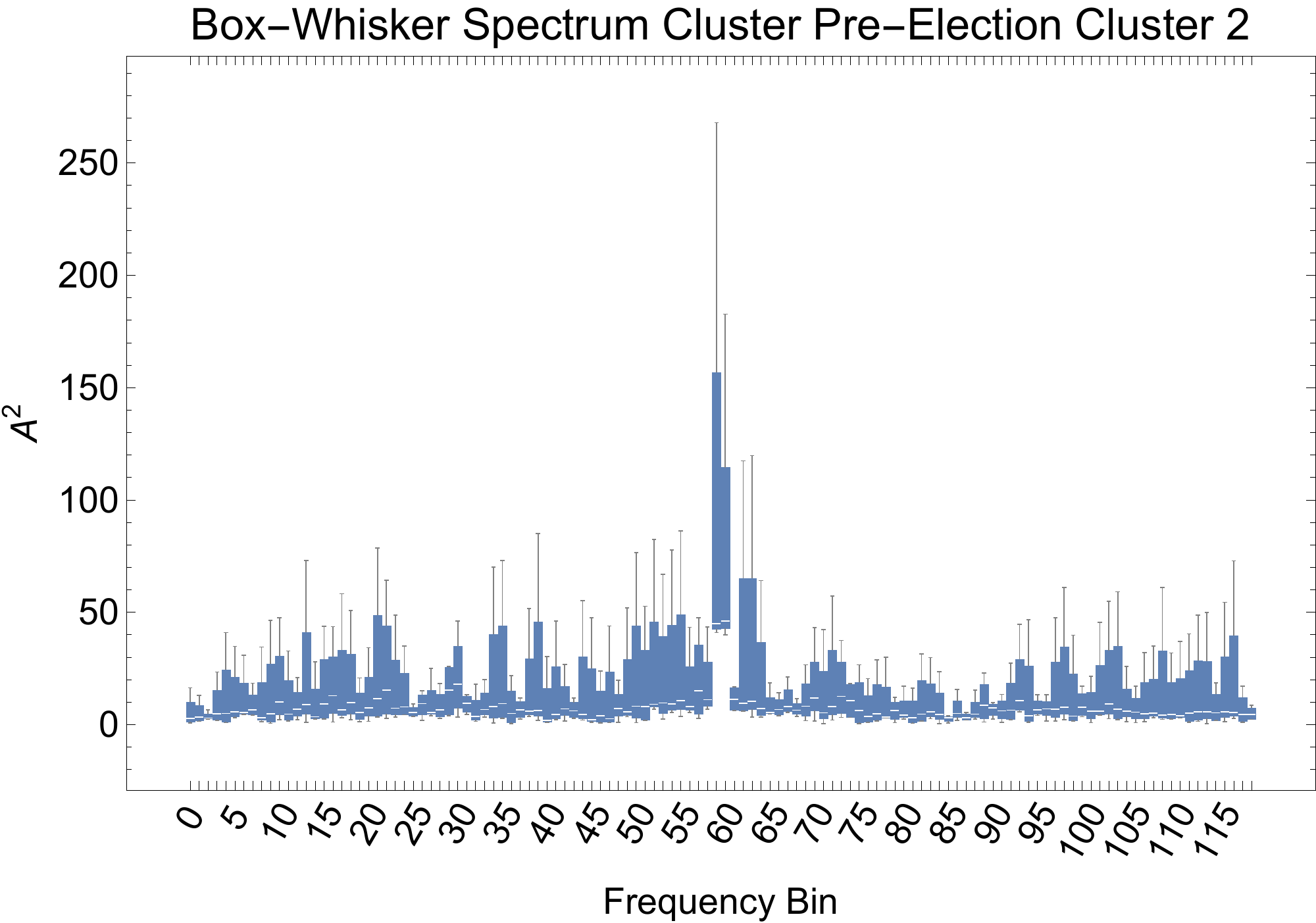}}
\subfloat[Pre-Election Cluster 3]{\includegraphics[width=0.23\textwidth]{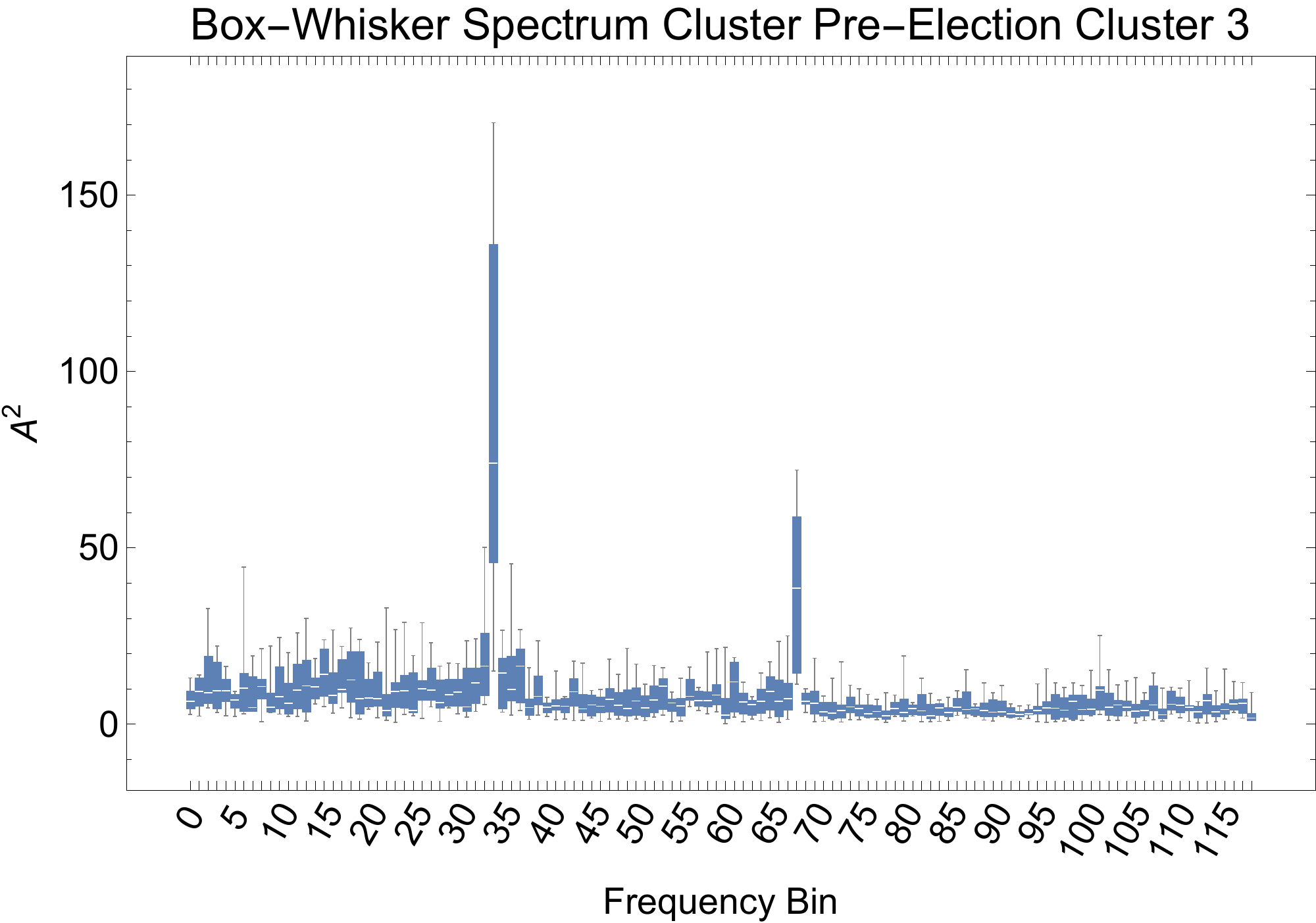}}
\subfloat[Pre-Election Cluster 4]{\includegraphics[width=0.23\textwidth]{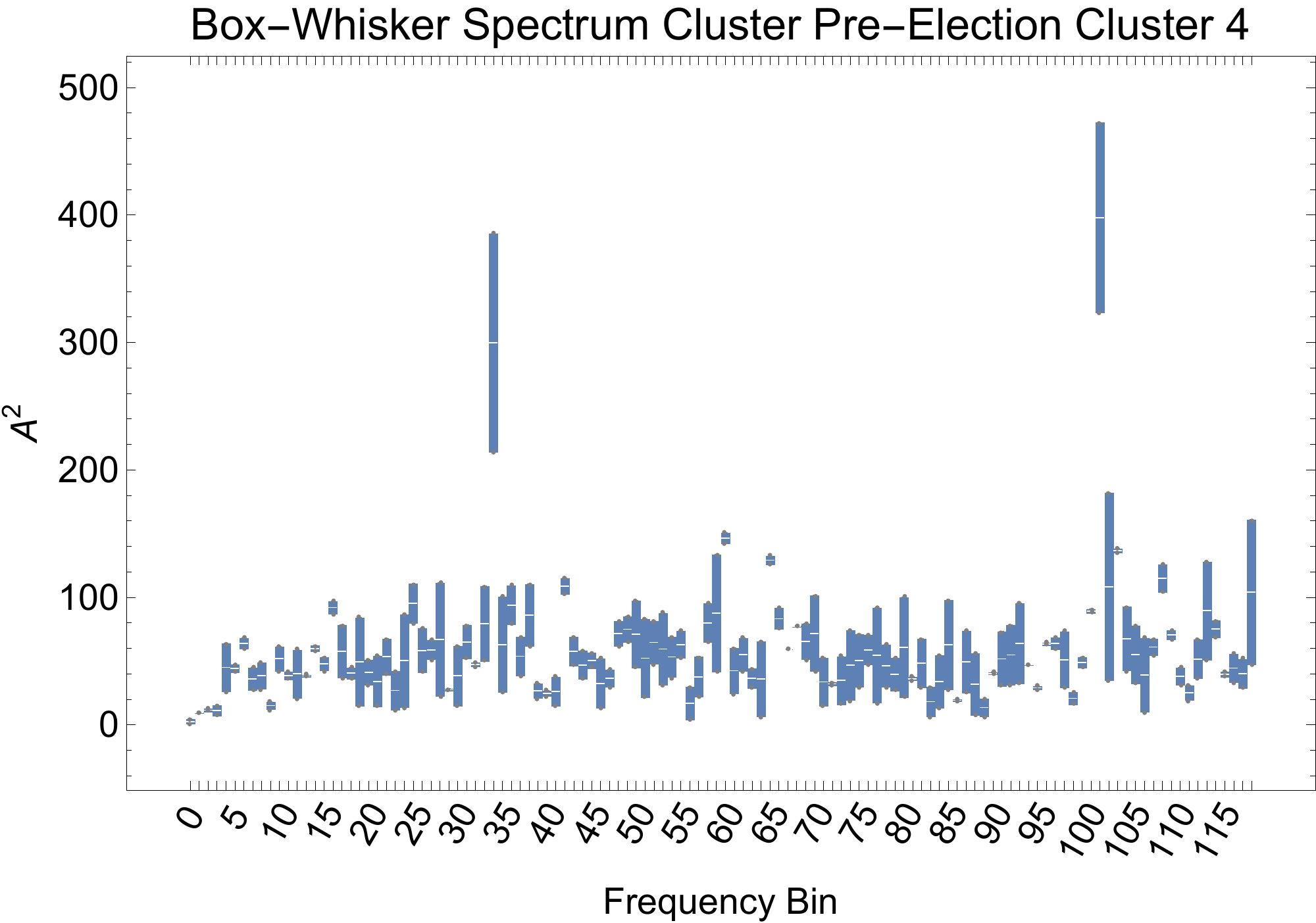}}\\
\subfloat[Pre-Election Strategy 1]{\includegraphics[width=0.23\textwidth]{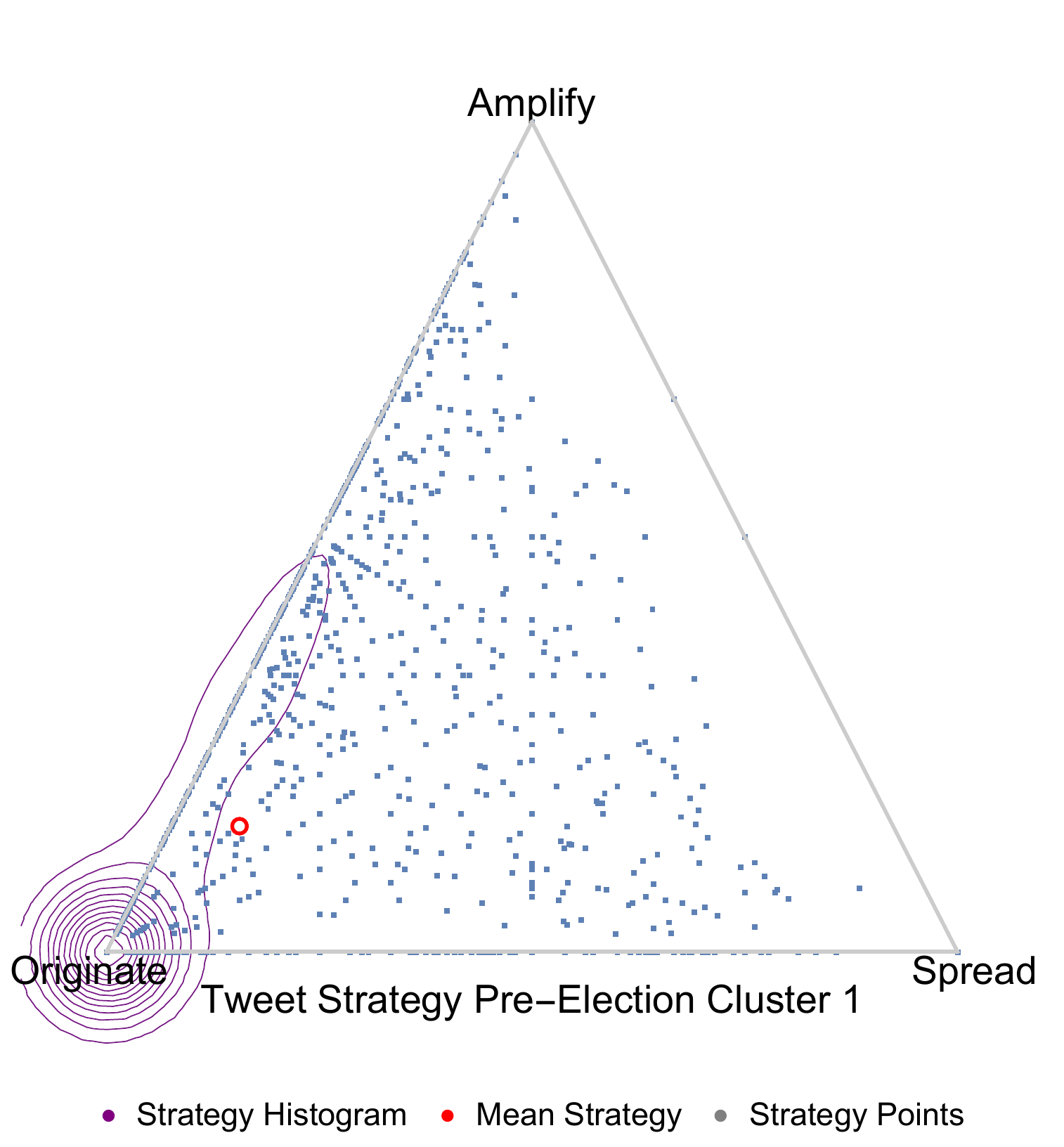}}
\subfloat[Pre-Election Strategy 2]{\includegraphics[width=0.23\textwidth]{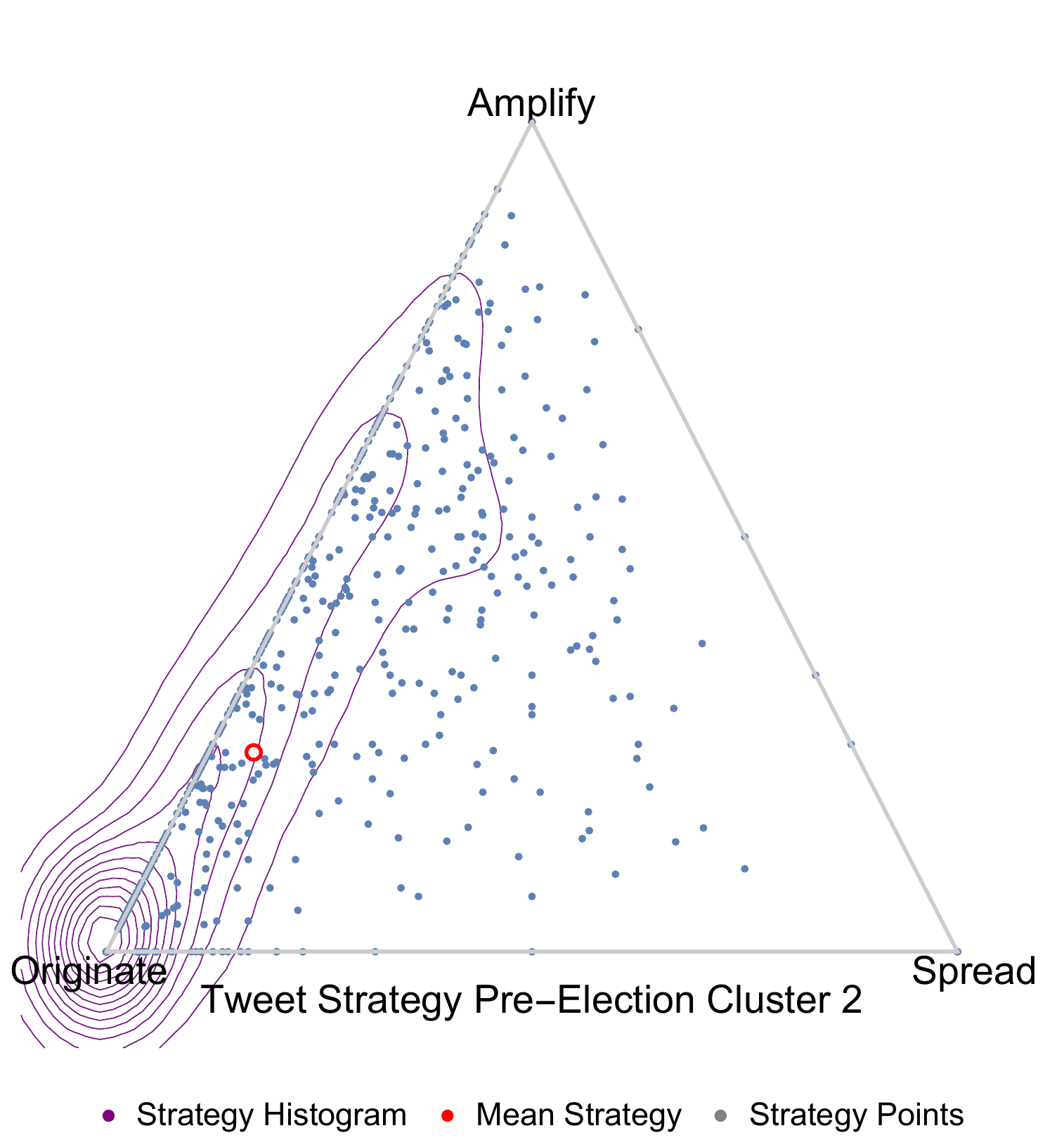}}
\subfloat[Pre-Election Strategy 3]{\includegraphics[width=0.23\textwidth]{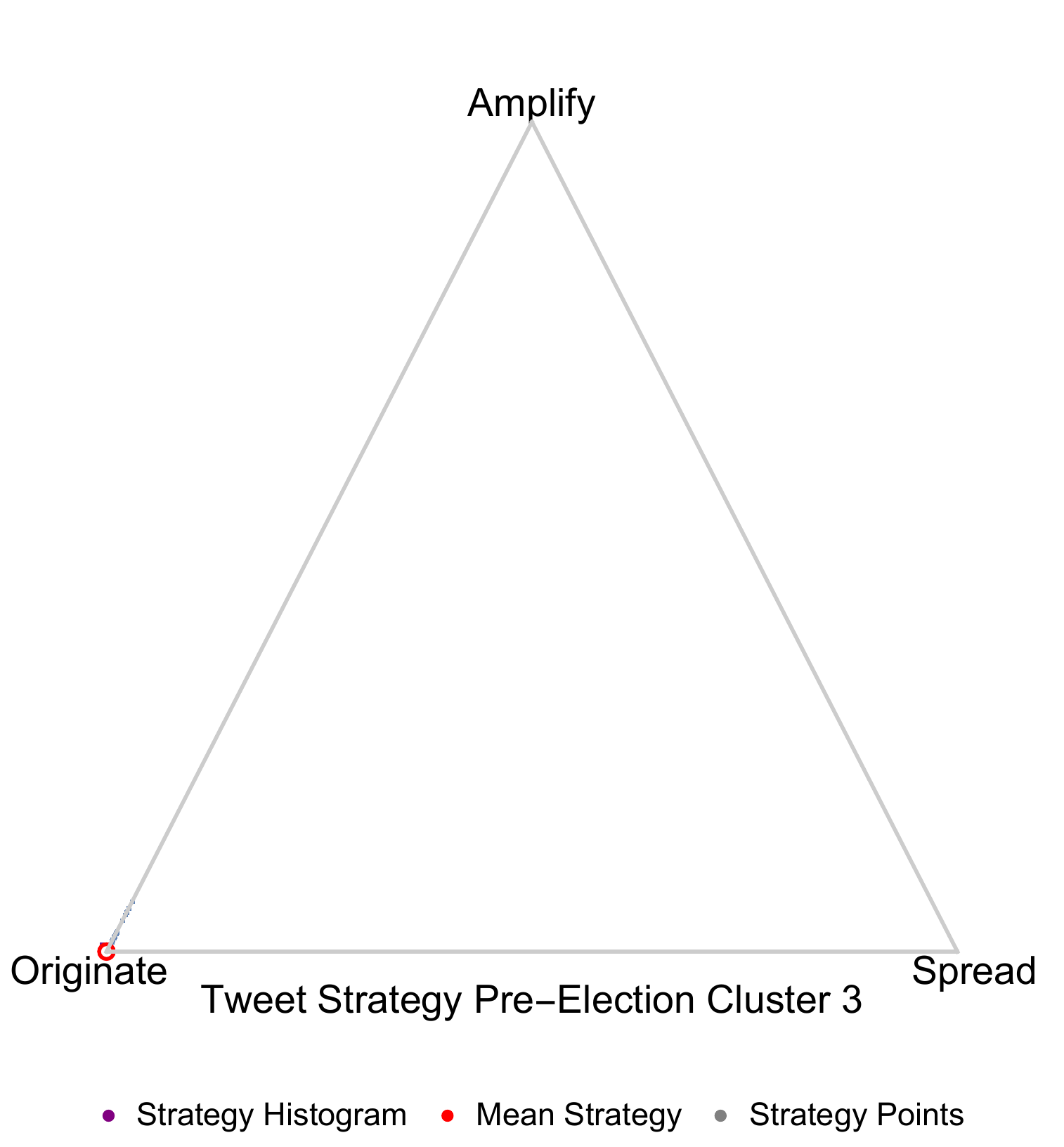}}
\subfloat[Pre-Election Strategy 4]{\includegraphics[width=0.23\textwidth]{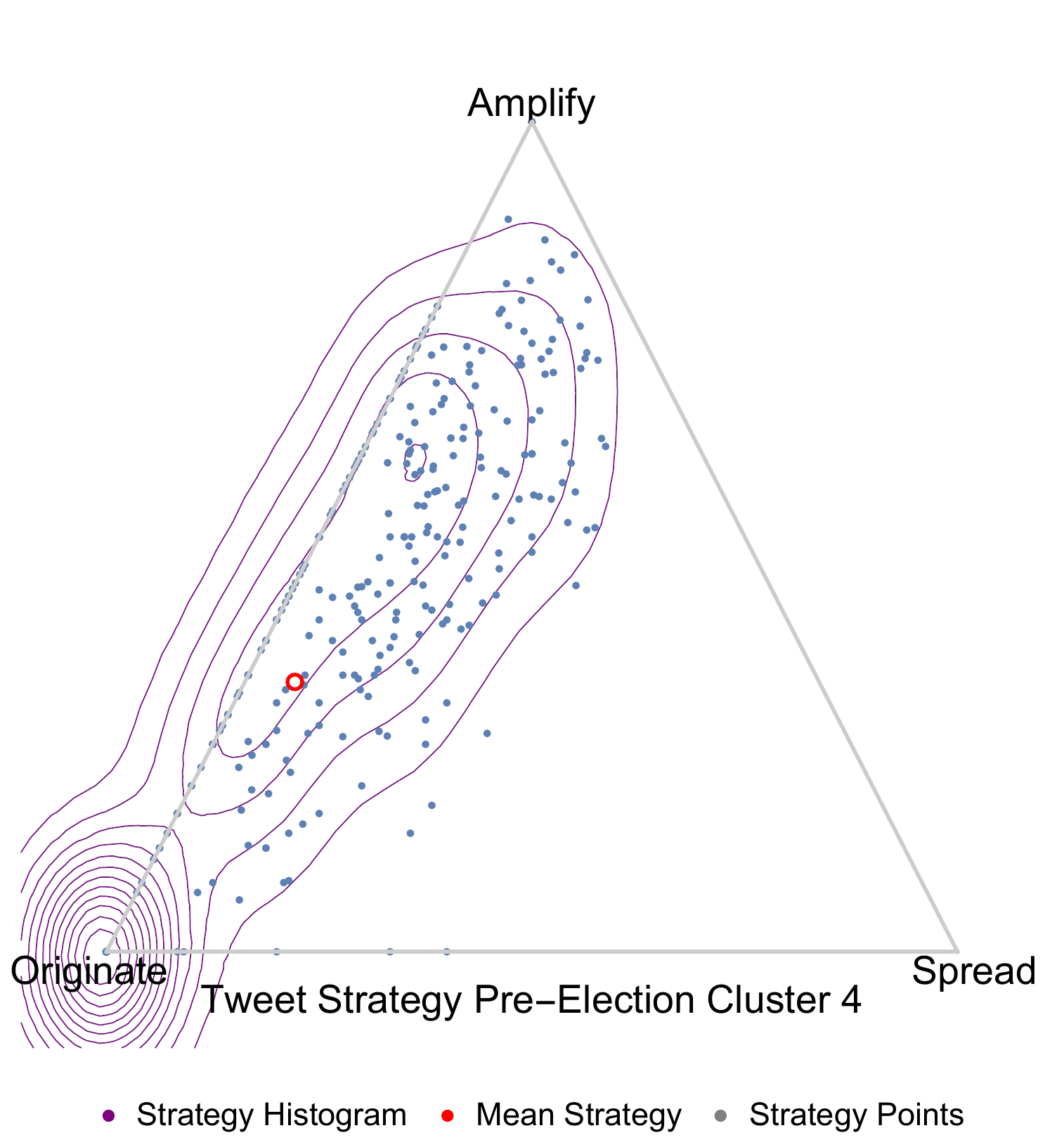}}\\
\subfloat[Post-Election Cluster 1]{\includegraphics[width=0.23\textwidth]{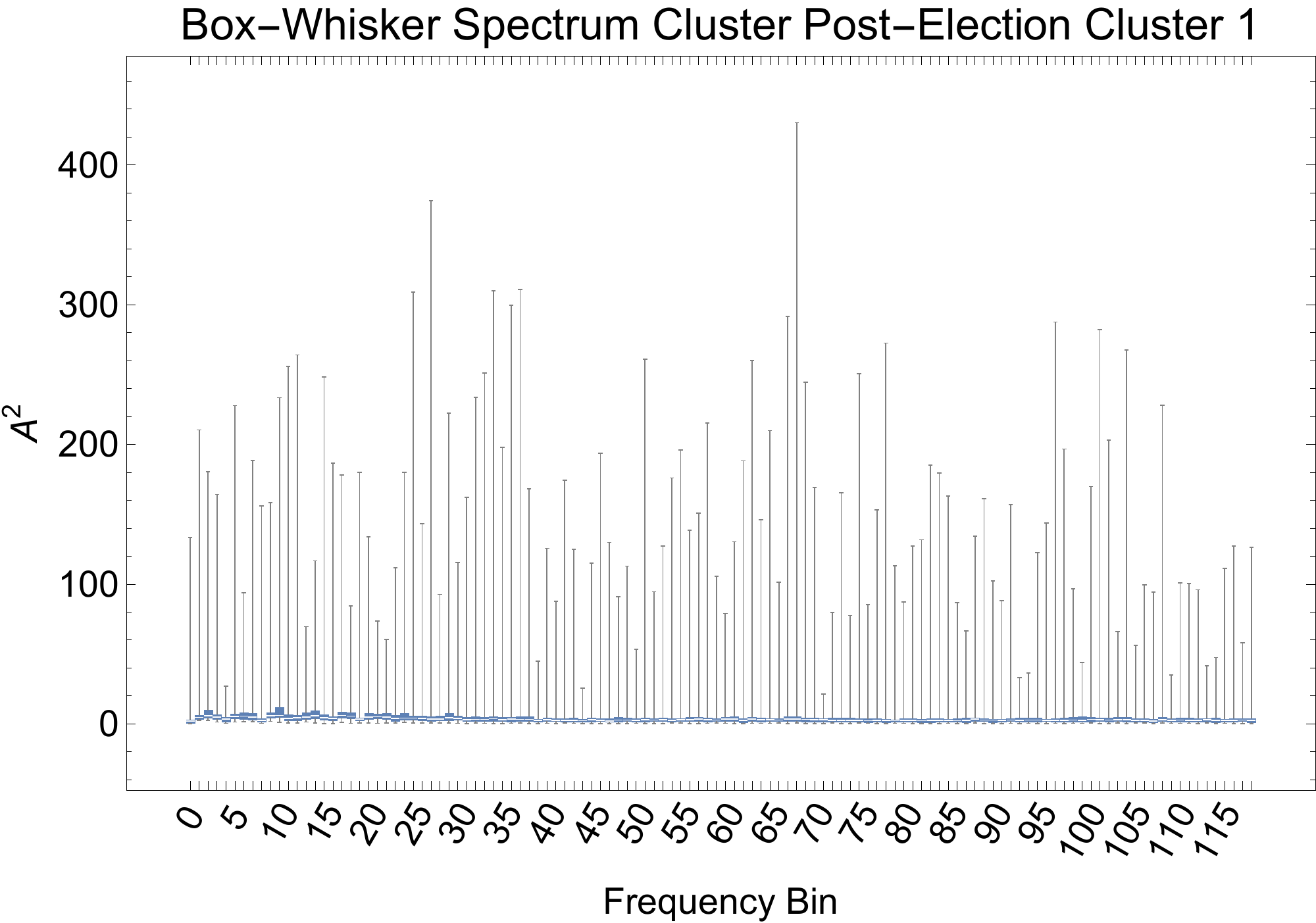}}
\subfloat[Post-Election Cluster 2]{\includegraphics[width=0.23\textwidth]{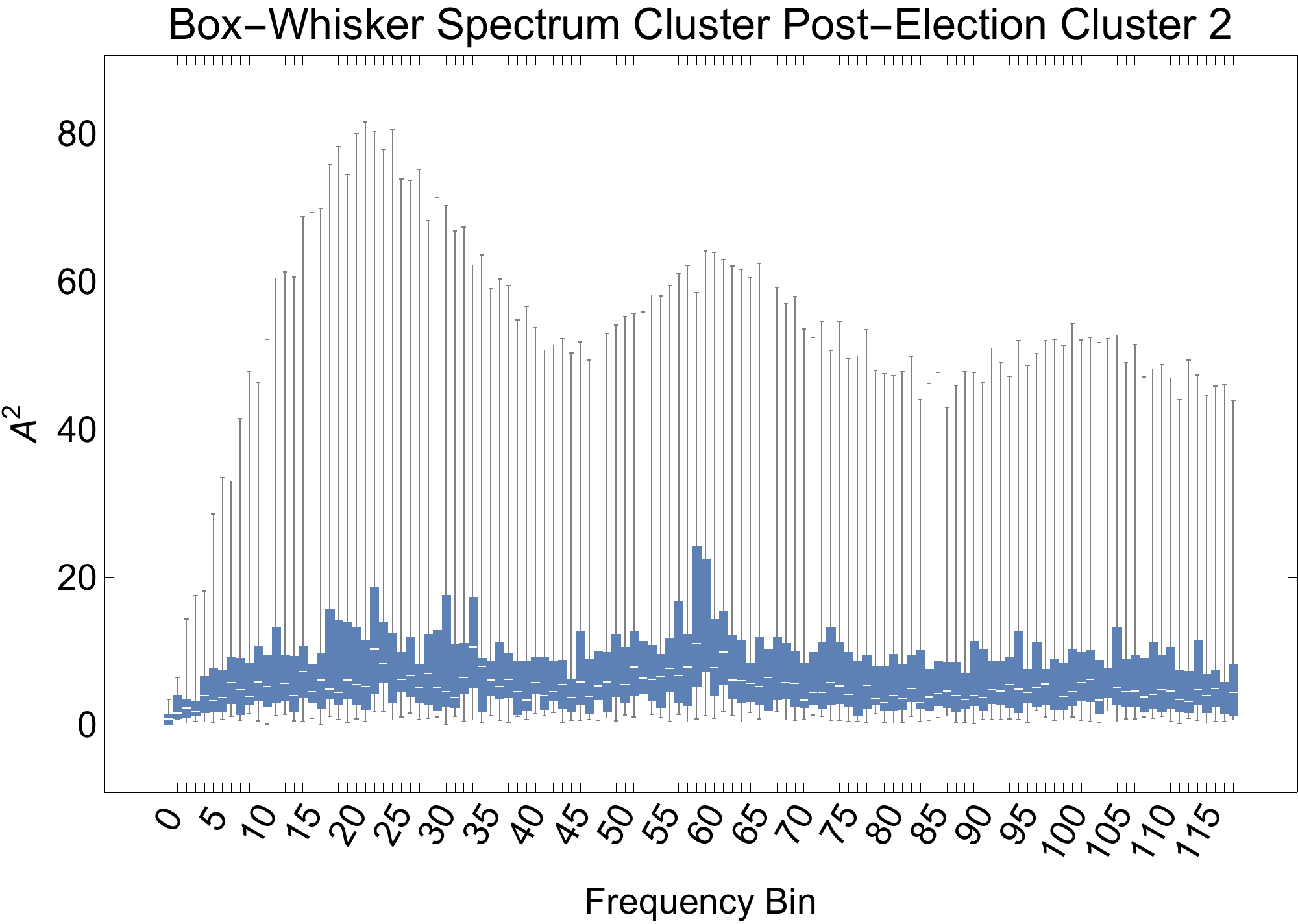}}
\subfloat[Post-Election Cluster 3]{\includegraphics[width=0.23\textwidth]{Figures/PreElectionBoxWhiskers3.pdf}}
\subfloat[Post-Election Cluster 4]{\includegraphics[width=0.23\textwidth]{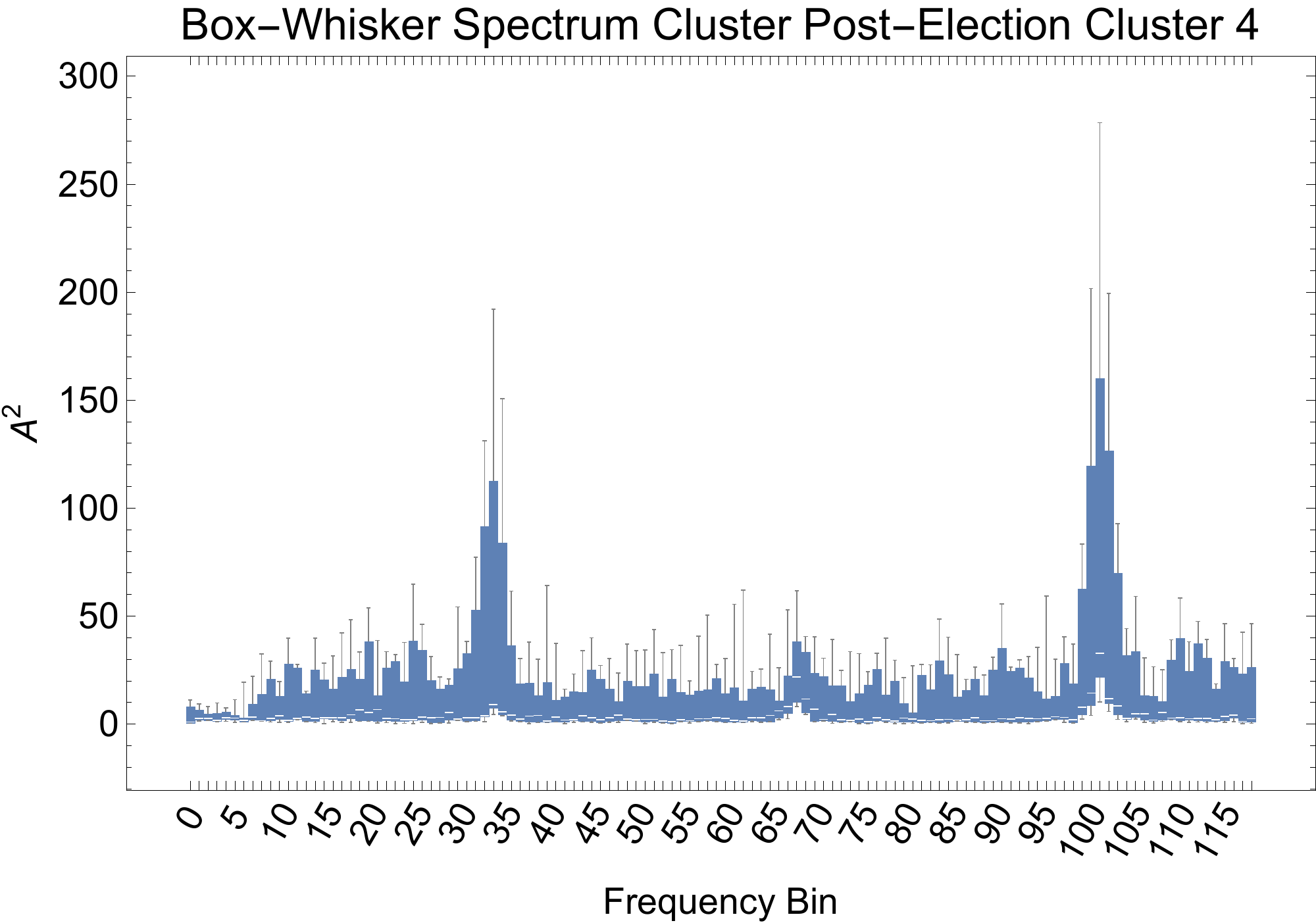}}\\
\subfloat[Post-Election Strategy 1]{\includegraphics[width=0.23\textwidth]{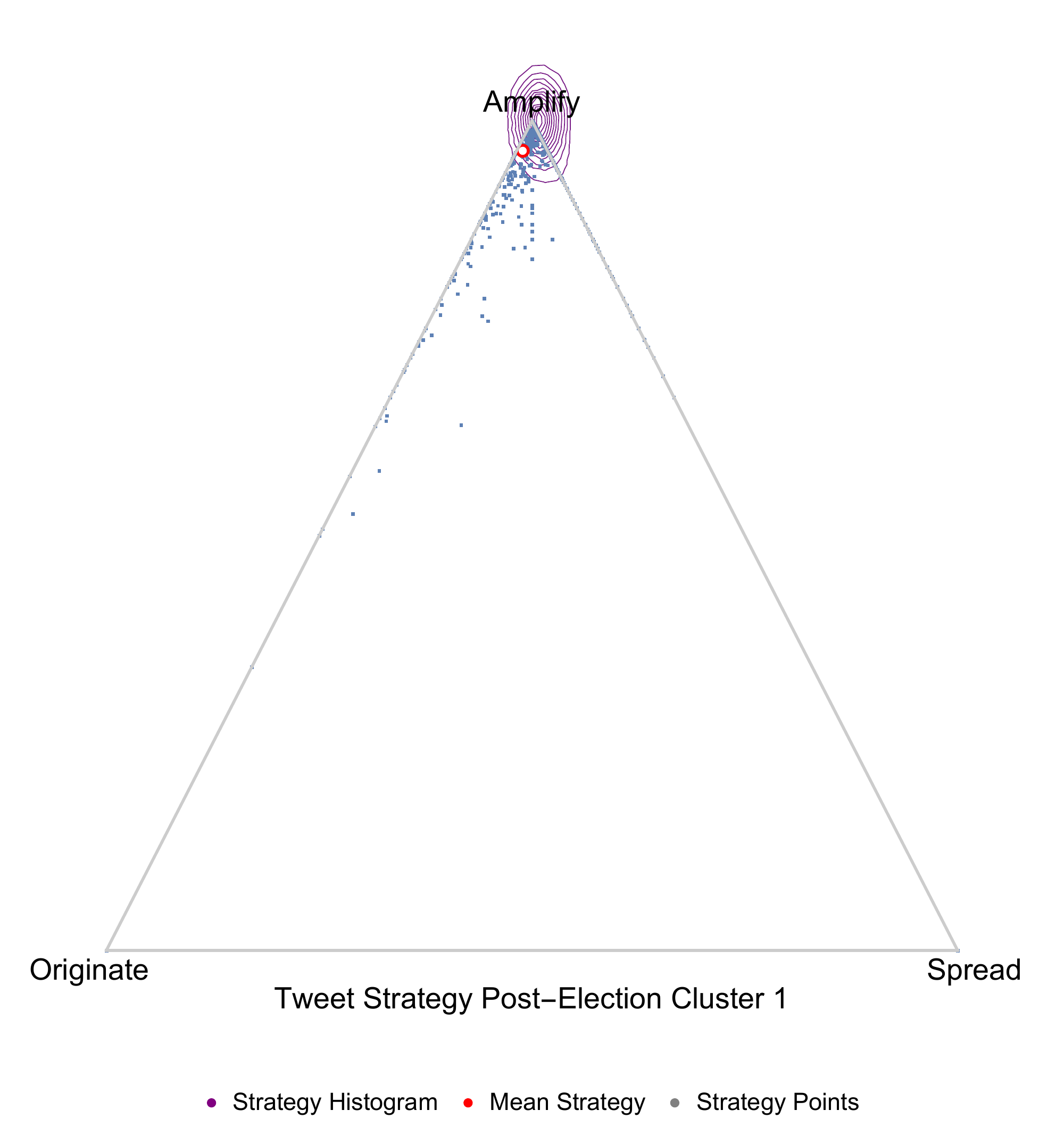}}
\subfloat[Post-Election Strategy 2]{\includegraphics[width=0.23\textwidth]{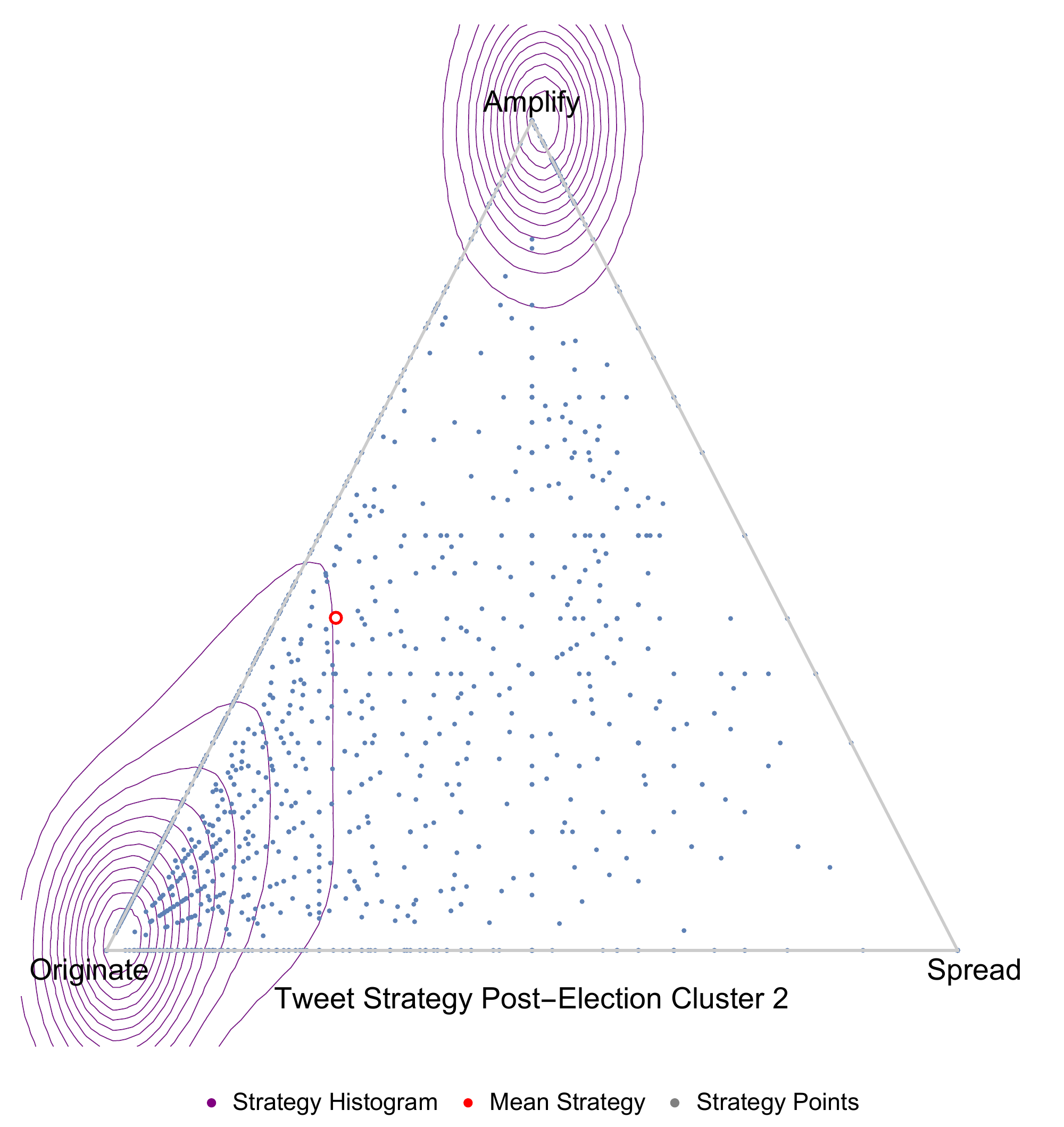}}
\subfloat[Post-Election Strategy 3]{\includegraphics[width=0.23\textwidth]{Figures/PreElectionDensity3.pdf}}
\subfloat[Post-Election Strategy 4]{\includegraphics[width=0.23\textwidth]{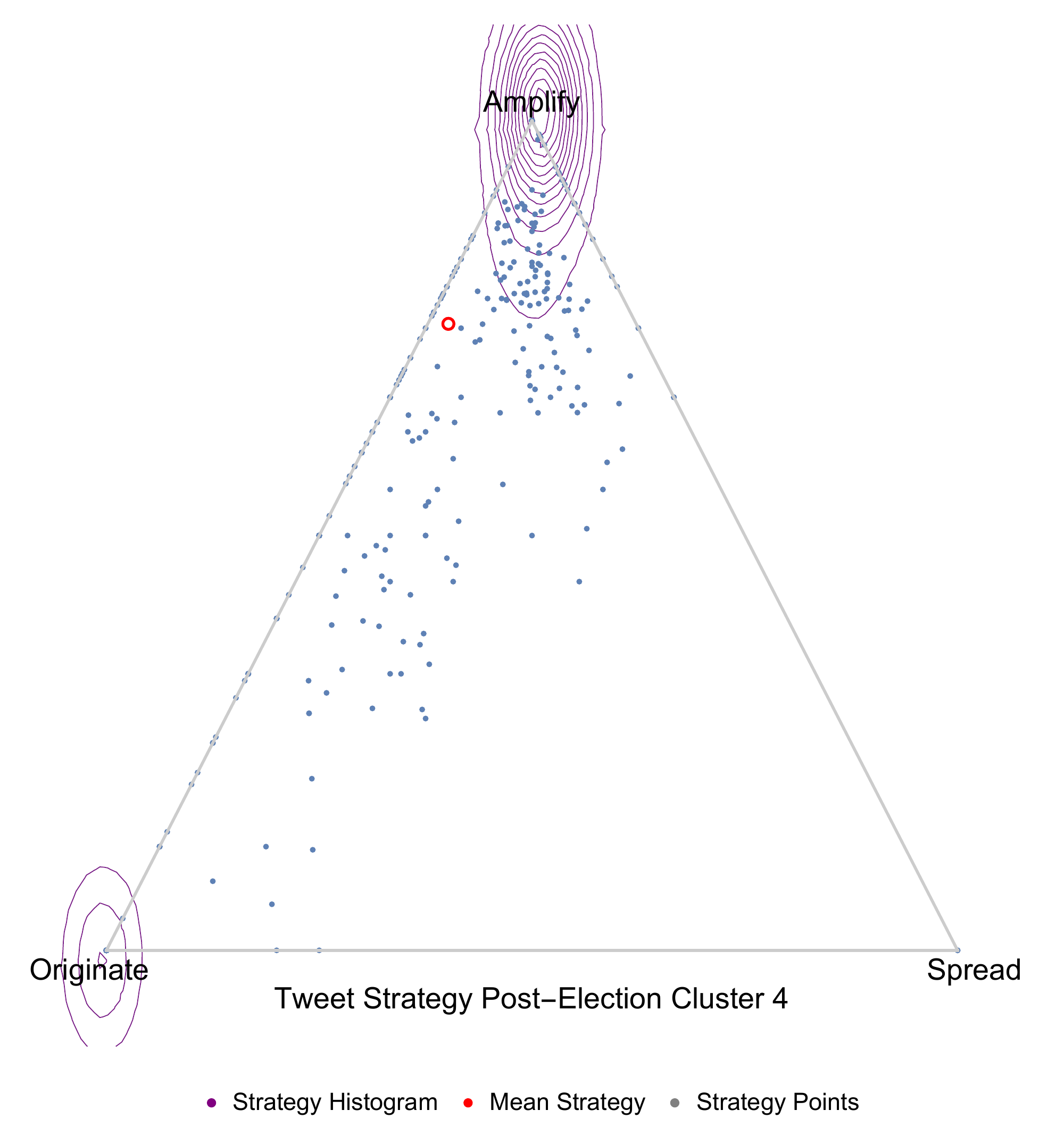}}
\caption{Both pre- and post-election spectra and the corresponding strategies are illustrated. The four clusters in the pre-election period can be matched the four clusters in the post-election period showing consistent dynamics across time periods exit. Strategies show a clear transition from originating to amplifying in some clusters but not in others.}
\label{fig:ReallyBigSpectra}
\end{figure*}

The resulting clusters each have specific spectral properties that define them. These are shown in \cref{fig:ReallyBigSpectra}, where we use box-and-whisker plots to show the ranges of each frequency bin over all elements of the cluster. Cluster 1 is composed of users whose behavior is not modeled by \cref{eqn:Model2}, except to say that their per-day tweet volume is composed of noise; i.e., there is no dominant frequency in their spectra. Cluster 2 is composed of users like \texttt{TEN\_GOP} whose behavior is harmonic with period (approximately) 4 days. This cluster could be sub-divided in the post-election period since we can see a substantial amount of variation in the lower-frequency amplitudes. The 4 day periodicity is clearer in the pre-election period. Clusters 3 and 4 both have extremely sharp defining principal frequencies between bins 30 and 35 (corresponding to a 7 day periodicity). Cluster 3 has a second frequency spike between bin 60 and 65, implying a periodicity of 4 days like Cluster 2. Cluster 4 has a second  frequency spike between bins 100 and 105, corresponding to a faster 2-3 day cycle, which is difficult to explain from a practical perspective. These results support \cref{eqn:Model2} as modeling the tweet rate of the users in this data set and also support the idea that these accounts are controlled Twitter bots \cite{MA16,CH19,KRWR19}.

\begin{figure}[htbp]
\centering
\subfloat[\texttt{DetroitDailyNew}]{\includegraphics[width=0.45\columnwidth]{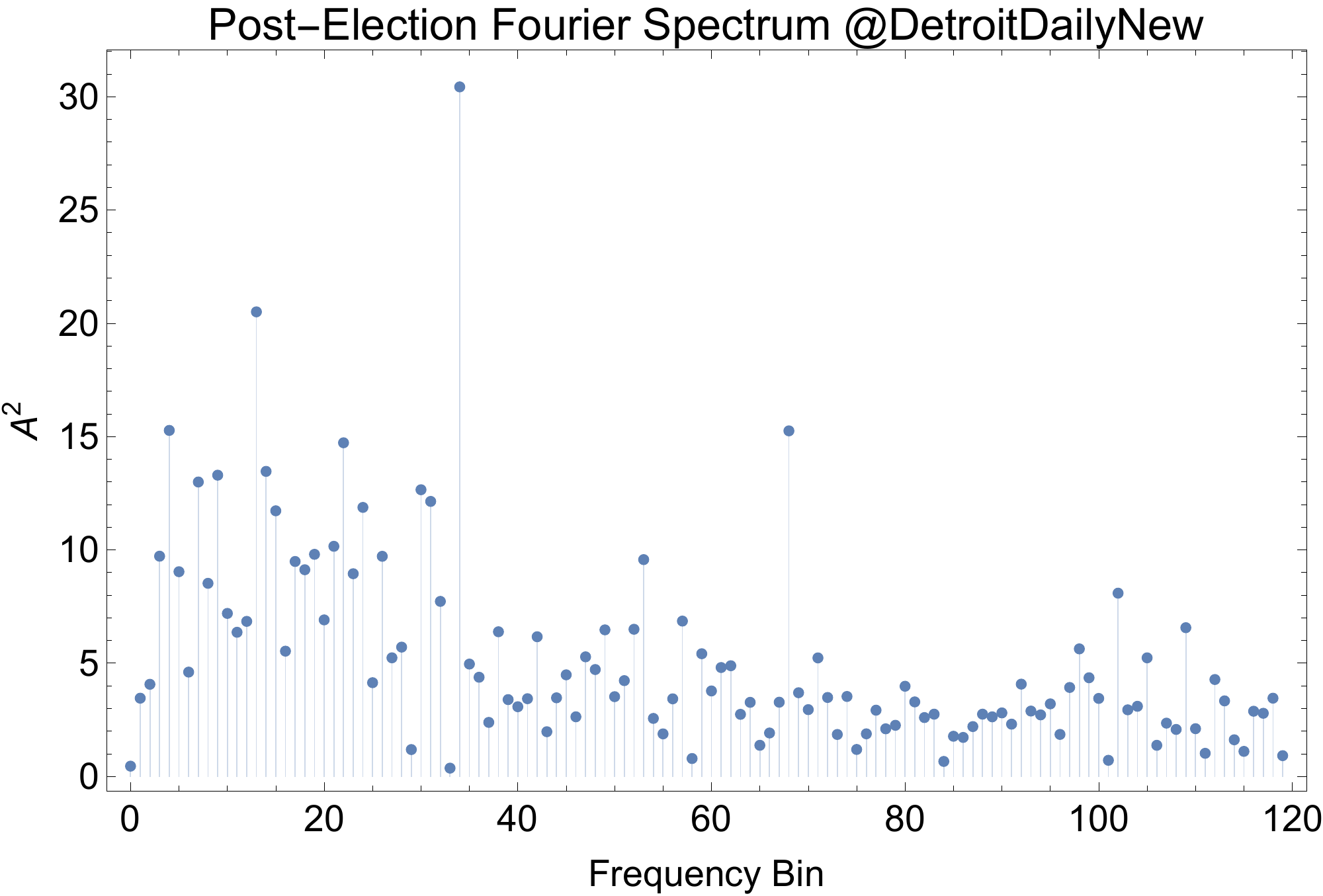}}\quad
\subfloat[\texttt{DailySanJose}]{\includegraphics[width=0.45\columnwidth]{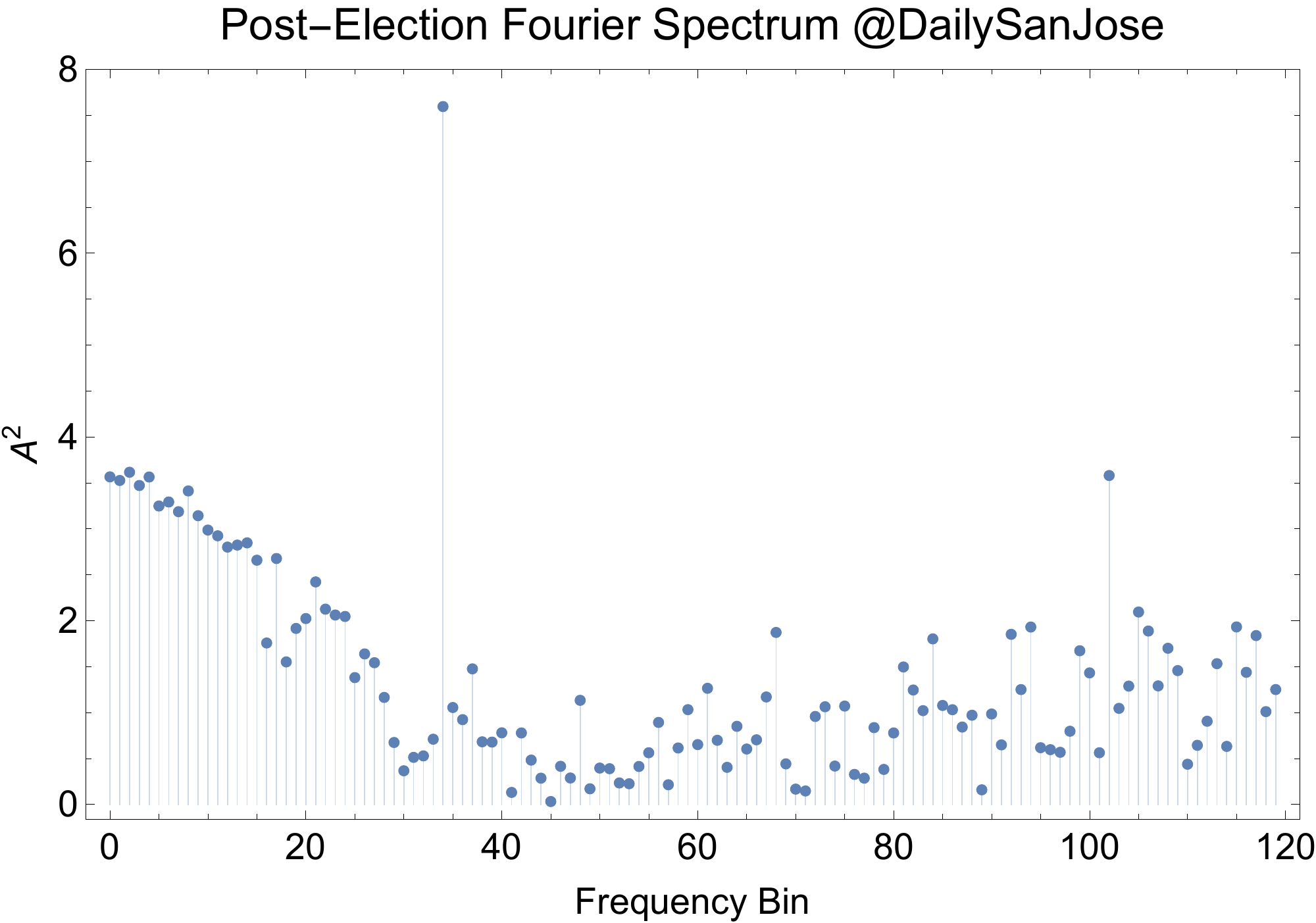}}
\caption{Two anomalies appear in the automatic behavior clustering. From a user name perspective, \texttt{DetroitDailyNew} and \texttt{DailySanJose} should be in Cluster 3.}
\label{fig:Anomalies}
\end{figure}

The user names in the themselves have interesting semantic properties. In both the pre-election period and the post-election period, Cluster 3 consists of user names that appear to be regionally selected news sources. For example, \texttt{OnlineMemphis}, \texttt{KansasDailyNews} and \texttt{ChicagoDailyNew} are all in Cluster 3. However, there are some interesting anomalies to this rule. \texttt{DetroitDailyNew} and \texttt{DailySanJose} both appear in Cluster 1 in the post-election time period. This can be explained by anomalies in their spectra. (See \cref{fig:Anomalies}.)
While both users have a spike between bins 30 and 35, \texttt{DetroitDailyNew} has a second frequency spike between bin 60 and 65 like \texttt{TEN\_GOP}, while \texttt{DailySanJose} has some low frequency noise and a frequency spike between bins 100 and 105, consistent with Cluster 4. The confusion results in these two users being placed in Cluster 1. 

We analyze the strategic behavior present in the clusters in \cref{fig:ReallyBigSpectra}. The three strategies are shown along with the sample points and a smoothed density histogram of the strategies. The weighted mean point of the strategies is shown as an open circle (in red). Consistent with the results illustrated in \cref{fig:SymbolDist}, there is a transition from the \textit{originate} strategy to the \textit{amplify} strategy for users in Cluster 1, 2 and 4. We note, however, that Cluster 3, which contains users with news-related names, continues to follow the \textit{originate} strategy both before and after the US presidential election. It is also interesting to note that Cluster 2 (which contains \texttt{TEN\_GOP}) maintains a balance between the \textit{originate} and \textit{amplify} strategies in contrast to users in Clusters 1 and 4.

\section{Topical Extraction Behavior Correlation}\label{sec:Topic}
In this section, we construct a second set of clusters during the pre- and post-election periods based on the topics mentioned in posted tweets. We investigate the word-space dynamical system over two long time periods, and then compare the resulting clusters of behaviors (and extracted topics) with the clusters derived in the previous sections. We show evidence to support the hypothesis that the better-modeled a cluster is by \cref{eqn:Model2}, the fewer topical-user clusters are represented, that is, the greater the topical cohesion observed. 

\subsection{Algorithm for User Clustering by Text}\label{sec:TopicClustering}
We discuss the algorithm (\cref{alg:LangCluster}) used to generate clusters of users by their topical signatures. \cref{alg:LangCluster} is a compilation of several pre-existing methods in natural language processing \cite{MS99} and manifold learning \cite{TSL00,BN02} with a novel variant that makes use of graph analytics developed in the physics community \cite{N06,N16,N13}. For this reason, we present the algorithm in detail.

We assume some minimal knowledge from statistical language processing. In particular, a \textit{stopword} is a word with little value to text meaning; examples in English include ``the'' and ``a''. Word stemming removes common endings to ground the text into a more systematic vocabulary. For example, ``ending'' is replaced with ``end''.

We assume the input is a sequence of tweets and users. In Line 1, the tweets from User $i$ are concatenated to form a synthetic document. For the remainder of this algorithm, if $w$ is a word, let $\#(w,i)$ be the number of times User $i$ uses word $w$. In Line 2, the words in the document are passed through a stemmer. Since we did not know that multiple languages were in use \textit{a priori}, we used the standard English (Porter) stemmer available in Mathematica 11.3. However, multiple stemmers and language detectors could be used at this step. In Line 3, we use a custom stopword list consisting of the (stems of) most common words in a language. Again, this was specialized to English, however it is possible to generalize this step. The remaining non-stopwords were used to create term count vectors for each user in Line 4. In Line 5, a binary term-user matrix is constructed. The entries are indicators determining whether a term is used by a user. A \textit{dynamic stopword set} is initialized in Step 7. In Lines 8 - 10, we iterate through the term-user matrix and determine whether a term is used by more than $p \times 100\%$ of the users, with $p \in (0,1)$. For our study we set $p = 0.5$. The rows corresponding to the words that were used by more than $p \times 100\%$ are added to $T$.
\begin{algorithm}
\begin{flushleft}
    \textbf{Input: }Raw tweets with users\\
    \textbf{Result: }Clusters of Users
\end{flushleft}
    \begin{algorithmic}[1]
    \State Concatenate tweets per user to make documents
    \State Stem all words in documents
    \State Remove stopwords by their stems
    \State Generate word counts per user
    \State Create binary term-user matrix:
    \begin{displaymath}
      \mathbf{X}_{ij} \leftarrow \begin{cases}
        1 & \text{If User $j$ uses term $i$} \\
        0 & \text{otherwise}
    \end{cases}
    \end{displaymath}
    \State Define $\{n_r,n_c\}\leftarrow \mathrm{Dimensions}[\mathbf{X}]$
    \State Define $T \leftarrow \emptyset$\;
    \FOR{$i \in \{1,\dots,n_r\}$}
     \IF{$\sum_j\mathbf{X}_{ij} > p \cdot n_c$}
     \State $T\leftarrow T \cup \{i\}$
    \ENDIF
    \ENDFOR
    \State Delete all terms corresponding to indexes in $T$
    \State Set $T \leftarrow \emptyset$.
    \FOR{$i \in \{1,\dots,n_c\}$}
    \State Compute mean word count $\bar{w}$ and word count variance $S_w^2$ for User $i$.
    \State Fit a Gamma distribution $\Gamma(k,\theta)$ for User $i$'s word use with:
    \begin{displaymath}
        k = \frac{\bar{w}^2}{S_w^2} \qquad \theta = \frac{S^2}{\bar{w}}
    \end{displaymath}
    \FOR{Each word used by User $i$}
      \IF{$\#(w,i)$ is in the $q\times 100\%$ percentile of $\Gamma(k,\theta)$}
      \State $T \leftarrow T \cup \{w\}$.
      \ENDIF
    \ENDFOR
    \ENDFOR
    \State Delete all terms not in $T$.
    \State Create new term-user matrix: $\tilde{\mathbf{X}}_{ij}$ is the number of times User $j$ uses term $i$.
    \State Replace each column of $\tilde{\mathbf{X}}_i$ with 
    $\tilde{\mathbf{X}}_i/\lVert{\tilde{\mathbf{X}}_i}\rVert$.
    \State Compute modified adjacency matrix: $\mathbf{A} \leftarrow \tilde{\mathbf{X}}^T\cdot\tilde{\mathbf{X}}$
    \State Compute $\tilde{\mathbf{A}}$ by zeroing the diagonal of $\mathbf{A}$
    \State Set $B_i \leftarrow k^\text{th}\text{ largest value in $\mathbf{A}_{i\cdot}$}$
    \FOR{$i \in \{1,\dots,n_c\}$}
      \FOR{$j \in \{1,\dots,n_c\}$}
        \IF{$\tilde{\mathbf{A}}_{ij} < \min\{B_i,B_j\}$}
         \State $\tilde{\mathbf{A}}_{ij} \leftarrow 0$
        \ENDIF
      \ENDFOR
    \ENDFOR
    \State Create weighted graph $G$ with weight matrix $\tilde{\mathbf{A}}$
    \State Cluster vertices using maximum modularity clustering.
\end{algorithmic}
\caption{Construct User Clusters from Linguistic Features}
\label{alg:LangCluster}
\end{algorithm}

After deleting the dynamic stopwords, we identify keywords across all documents in Lines 13 - 18 using frequency. While it is the case that many documents follow the Zipf-Mandelbrot law and their word counts follow a Zipf-Mandelbrot distribution \cite{Z32,M65}, we have found it useful to smooth these distributions and use a percentile cutoff. To do this, we fit a Gamma distribution to the word counts and then remove words that do not fall in the $q \times 100\%$ percentile of the distribution. For this study, we set $q = 0.9$. We chose the Gamma distribution because it has non-negative support, and being a two parameter distribution is widely adaptable. We note, there are alternate way of identifying keywords \cite{MS99}.

After removing non-keywords, the term and user set is fixed and a new term-user matrix $\tilde{\mathbf{X}}$ can be computed where each column is the (reduced term) term counts for the corresponding user. In Line 26, we then replace each column of $\tilde{\mathbf{X}}$, denoted $\tilde{\mathbf{X}}_i$ with its unit vector form. That is, column $i$ of  $\tilde{\mathbf{X}}$ becomes $\tilde{\mathbf{X}}_i/\lVert\tilde{\mathbf{X}}\rVert$.

To understand the remainder of the algorithm, we think of $\mathbf{X}$ as a sub-matrix of the weighted adjacency matrix $\mathbf{B}$ of a bipartite graph between terms and documents where:
\begin{displaymath}
    \tilde{\mathbf{B}} = \begin{bmatrix}
       \mathbf{0} & \tilde{\mathbf{X}}\\
       \tilde{\mathbf{X}}^T & \mathbf{0}
    \end{bmatrix}
\end{displaymath}
Suppose that $\mathbf{B}$ has the same structure as $\tilde{\mathbf{B}}$, but with all positive values replaced by $1$. It is a classic result in graph theory that $\mathbf{B}^2$ has in its $(i,j)$ position the number of walks of length $2$ from vertex $i$ to vertex $j$. Thus, if $\mathbf{X}$ is again the binarized form of $\tilde{\mathbf{X}}$, then:
\begin{displaymath}
  \mathbf{B}^2 = \begin{bmatrix}
  \mathbf{X} \cdot \mathbf{X}^T & \mathbf{0}\\
  \mathbf{0} & \mathbf{X}^T \cdot \mathbf{X}
\end{bmatrix}
\end{displaymath}
Here $\mathbf{X}^T\cdot\mathbf{X}$ is a square matrix whose $(i,j)$ position is the number of terms shared between User $i$ and User $j$ when $i \neq j$. By normalizing the columns of $\tilde{\mathbf{X}}$ in Line 26, we ensure that the $(i,j)$ element of $\tilde{\mathbf{X}}^T\cdot \tilde{\mathbf{X}}$ is just the cosine similarity measure of the term vectors used by User $i$ and User $j$ (when $i \neq j$).

In Lines 27-29, we create the adjacency matrix $\tilde{\mathbf{A}}$ by using the process described above and zero the diagonal. We then compute bounds (define neighbors) for each user, by finding the $k^\text{th}$ largest value. This is the graph theoretic equivalent of using $k$-nearest neighbors \cite{HSD00}. In our work we set $k = 10$. In Lines 30 - 36, we zero an edge weight (i.e., delete edges) at index $(i,j)$ if it is less than the bound computed for user $i$ or user $j$. In essence, this creates a version of a preferential attachment graph where (i) preference is based on shared topics, (ii) each user prefers to attach to $k$ other users but (iii) user $i$ may attach to more than $k$ users if there are more than $k$ other users who have ranked user $i$ as one of their top $k$ connections. If modeled as a directed graph, this would be an $(n_c, k)$ directed Barabasi-Albert network \cite{AB00} with a hidden preference based on topic.

Finally, in Lines 37 - 38, we form the graph and use maximum modularity clustering defined by Newman \cite{N06} to build user communities. This is a variation on the Isomap \cite{TSL00} and Multidimensional Scaling \cite{BG97} approaches to manifold learning.

\subsection{User-Topic Clustering Results}
\begin{figure*}[p]
\centering
\subfloat[Pre-Election User-Topic Graph]{\includegraphics[width=0.45\textwidth]{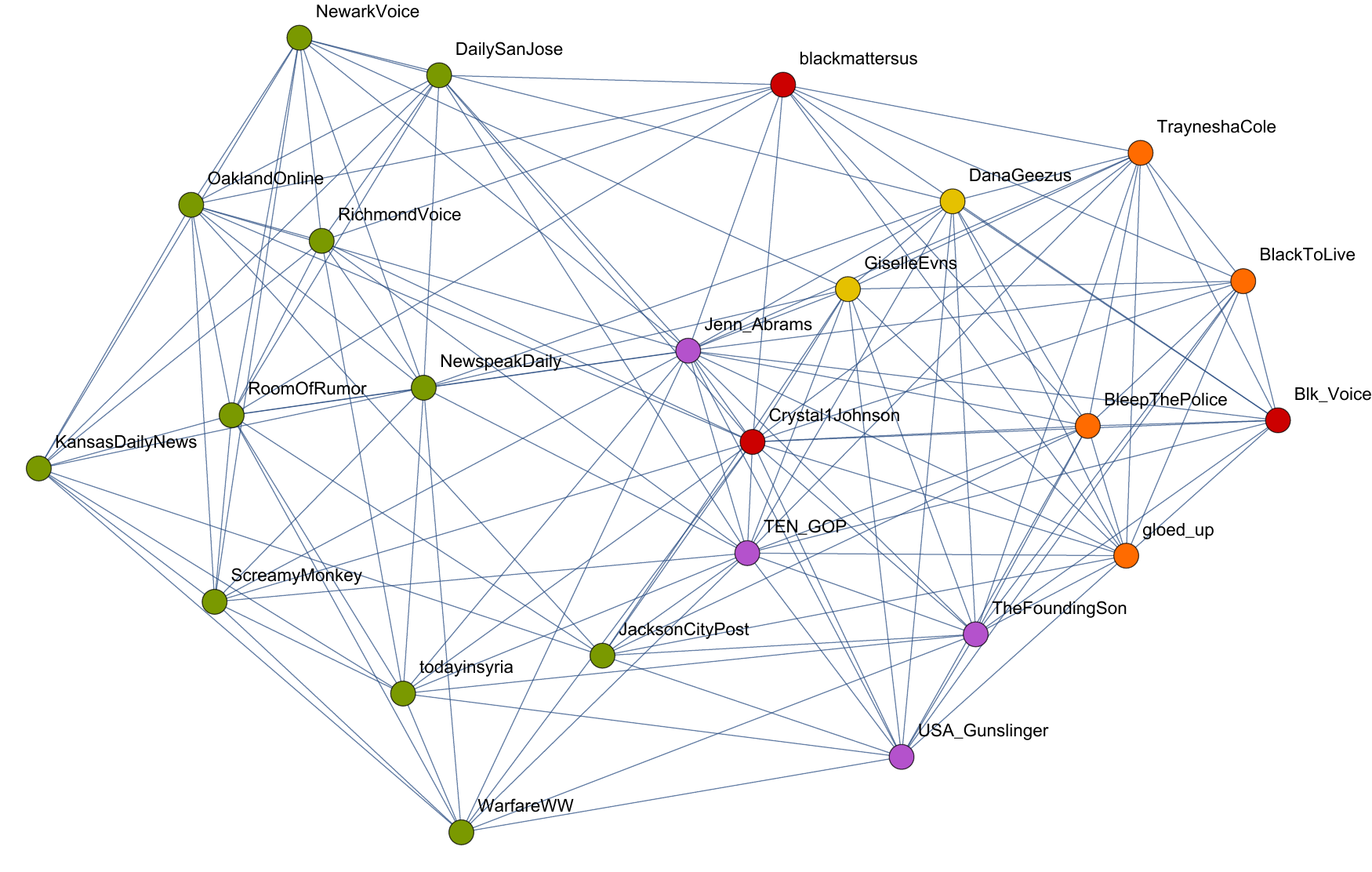}} \\
\subfloat[Post-Election User-Topic Graph]{\includegraphics[width=0.65\textwidth]{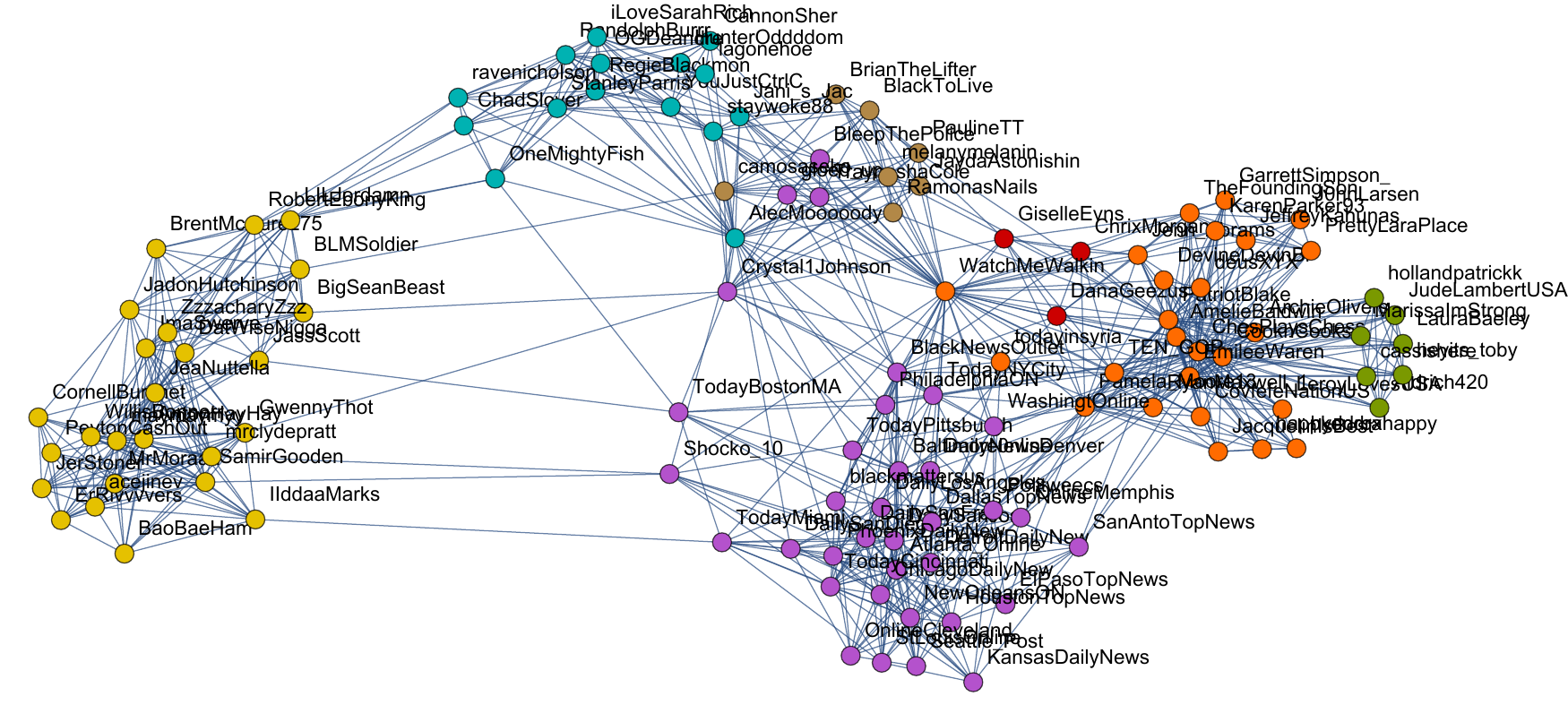}} \\
\subfloat[Pre-Election Word Clouds]{\includegraphics[width=0.85\textwidth]{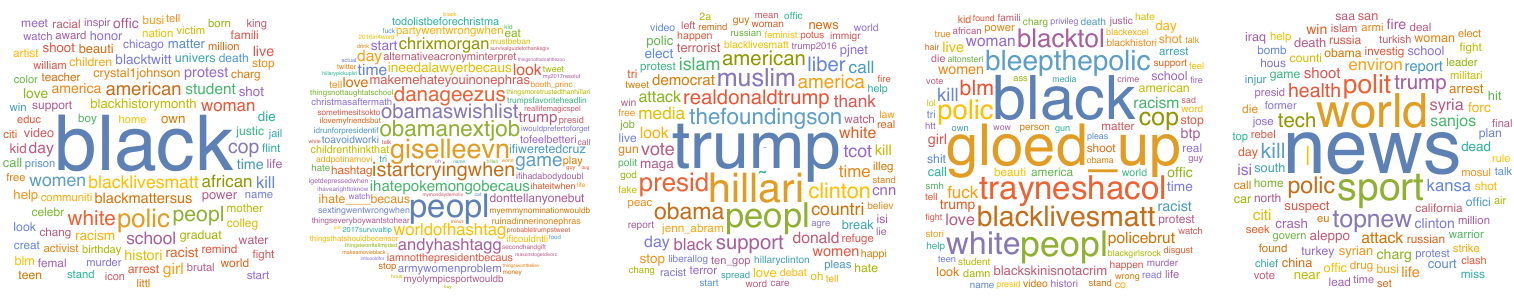}}\\
\subfloat[Post-Election Word Clouds]{\includegraphics[width=0.85\textwidth]{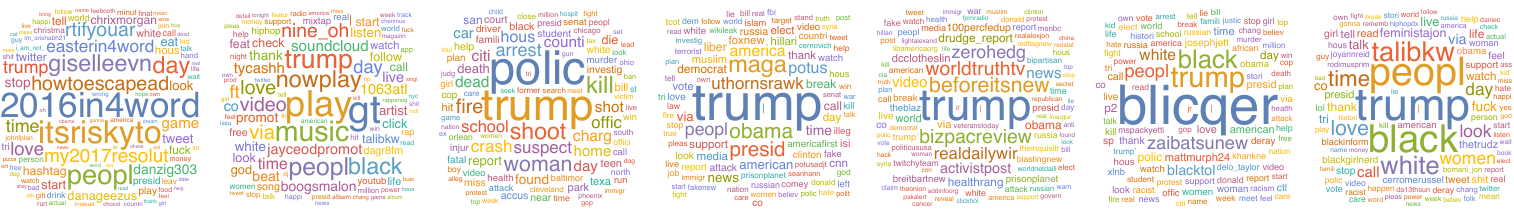}}
\caption{Topical user networks and clusters are generated by \cref{alg:LangCluster} for pre- and post-election data (a,b). Word clouds represent frequent topics in each cluster (c,d).}
\label{fig:WordClusters}
\end{figure*}

Applying \cref{alg:LangCluster} yields 5 language-based user clusters in the pre-election period and 7 language-based clusters in the post-election period. The resulting user graphs generated in Line 37 of \cref{alg:LangCluster} are shown in \cref{fig:WordClusters} along with the resulting maximum modularity clusters \cite{N06} generated in Line 38. The user clusters are listed in \cref{sec:TopicCluster}. 

We can generate word clouds from the user clusters by merging the stemmed, non-stop words used by the individuals in these clusters. This provides general information about the topical information that caused the users to be clustered. For the word clouds shown in \cref{fig:WordClusters}, we do not remove common words. 

Topics emergent in the pre-election time period are largely consistent with what is already known about this twitter data set \cite{M19,CH19,ICSL19,LLZ19}. There are two groups whose comments may be designed to stoke racial tension (Word Clouds 1 and 4), one group that focuses on news (Word Cloud 5), one group devoted to politics (Word Cloud 3) and in particular Donald Trump, and a final group that seems to re-tweet information from a variety of twitter handles (Word Cloud 2). 

Topics in the post-election time period are similar, but with some apparent changes in composition. In particular, one cluster (Word Cloud 1) seems to be the result of posts about the holiday(s) that occurred during the post-election period. There is a user group that focuses on news (Word Cloud 3). Interestingly, this group contains the majority of the news-themed user names (e.g., \texttt{KansasDailyNews}). There are at least two groups that post on race relations (Word Clouds 6 and 7). The remaining groups seem to focus on music (Word Cloud 2) and general topics related to Donald Trump (Word Clouds 5 and 6) with various subtle language differences separating these groups. As is to be expected based on the Mueller report, \texttt{TEN\_GOP} is a member of post election user group 5 and pre-election user group 3. 

\subsection{Comparison with Previous Clusters}
\begin{figure}[tb]
\centering
\subfloat[Pre-Election $C_1$]{\includegraphics[width=0.45\columnwidth]{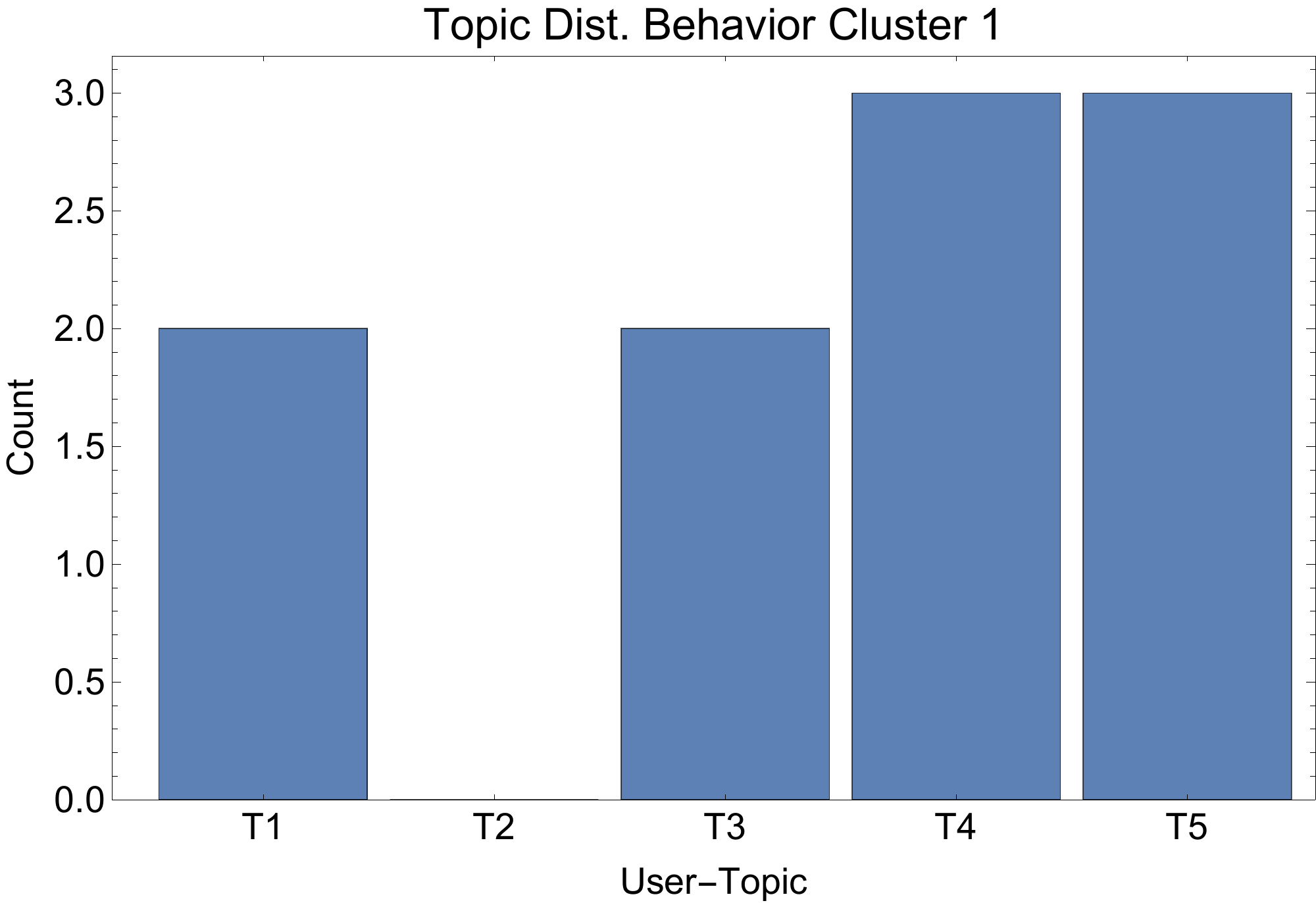}}\quad
\subfloat[Pre-Election $C_2$]{\includegraphics[width=0.45\columnwidth]{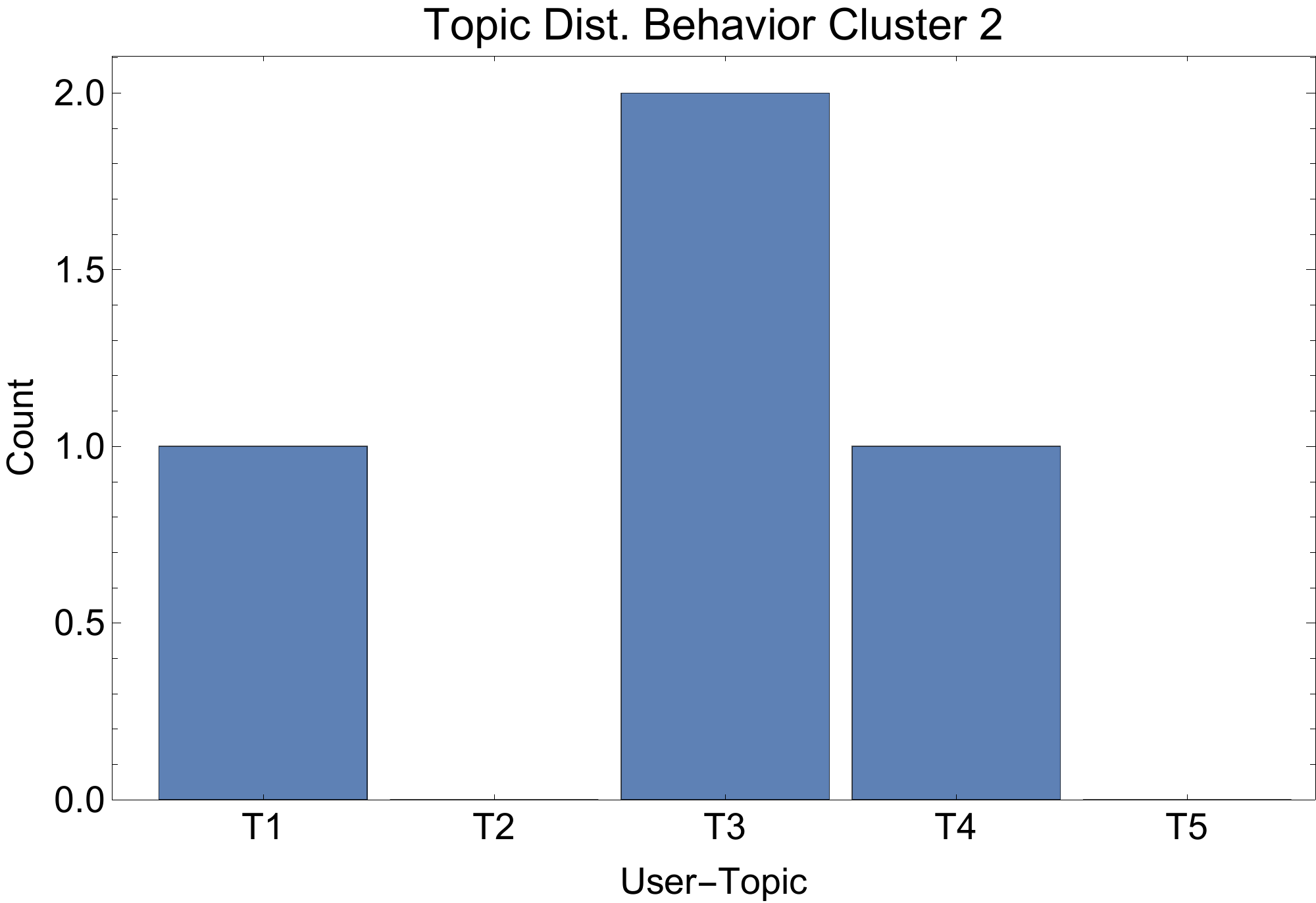}}\\
\subfloat[Pre-Election $C_3$]{\includegraphics[width=0.45\columnwidth]{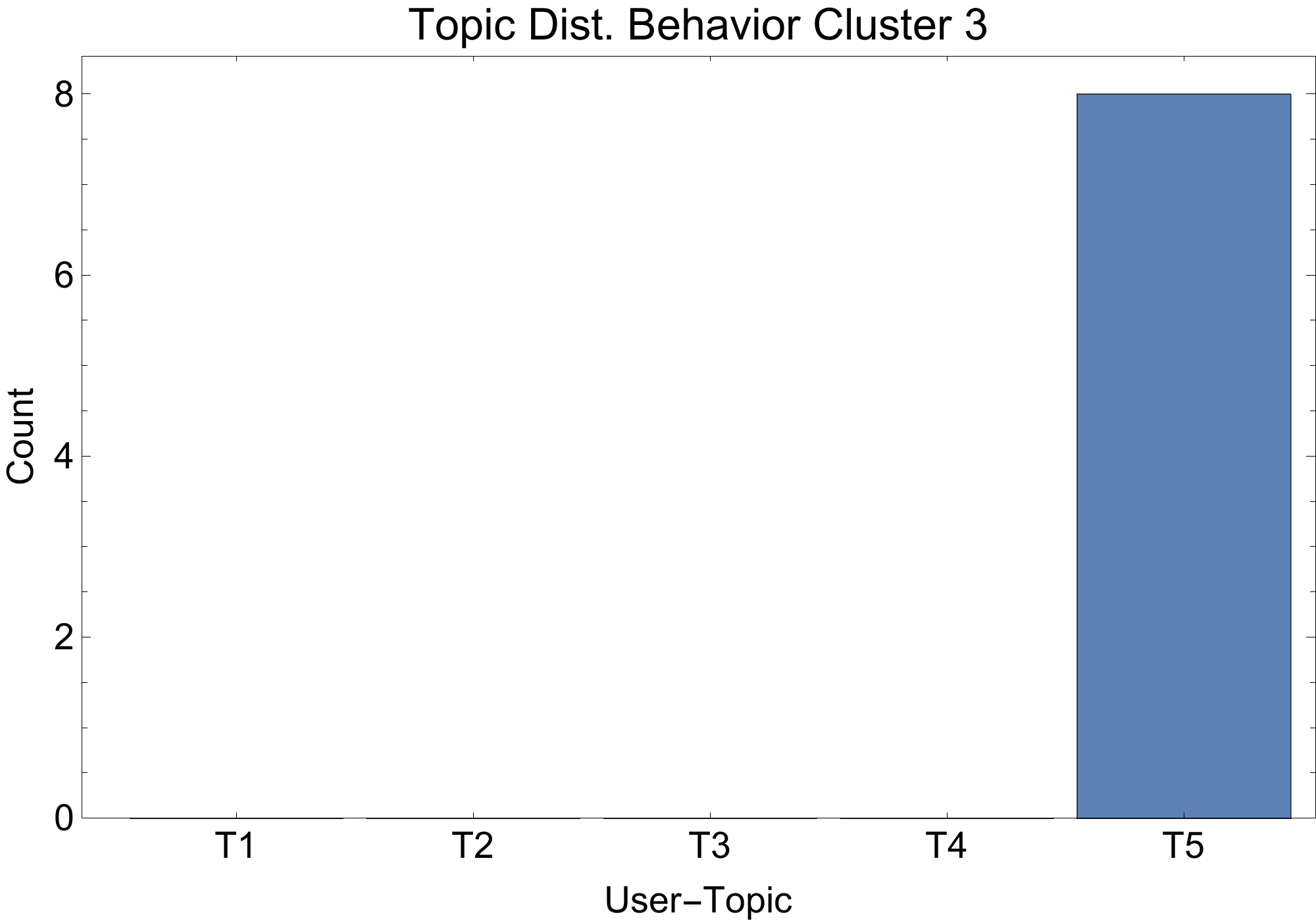}}\quad
\subfloat[Pre-Election $C_4$]{\includegraphics[width=0.45\columnwidth]{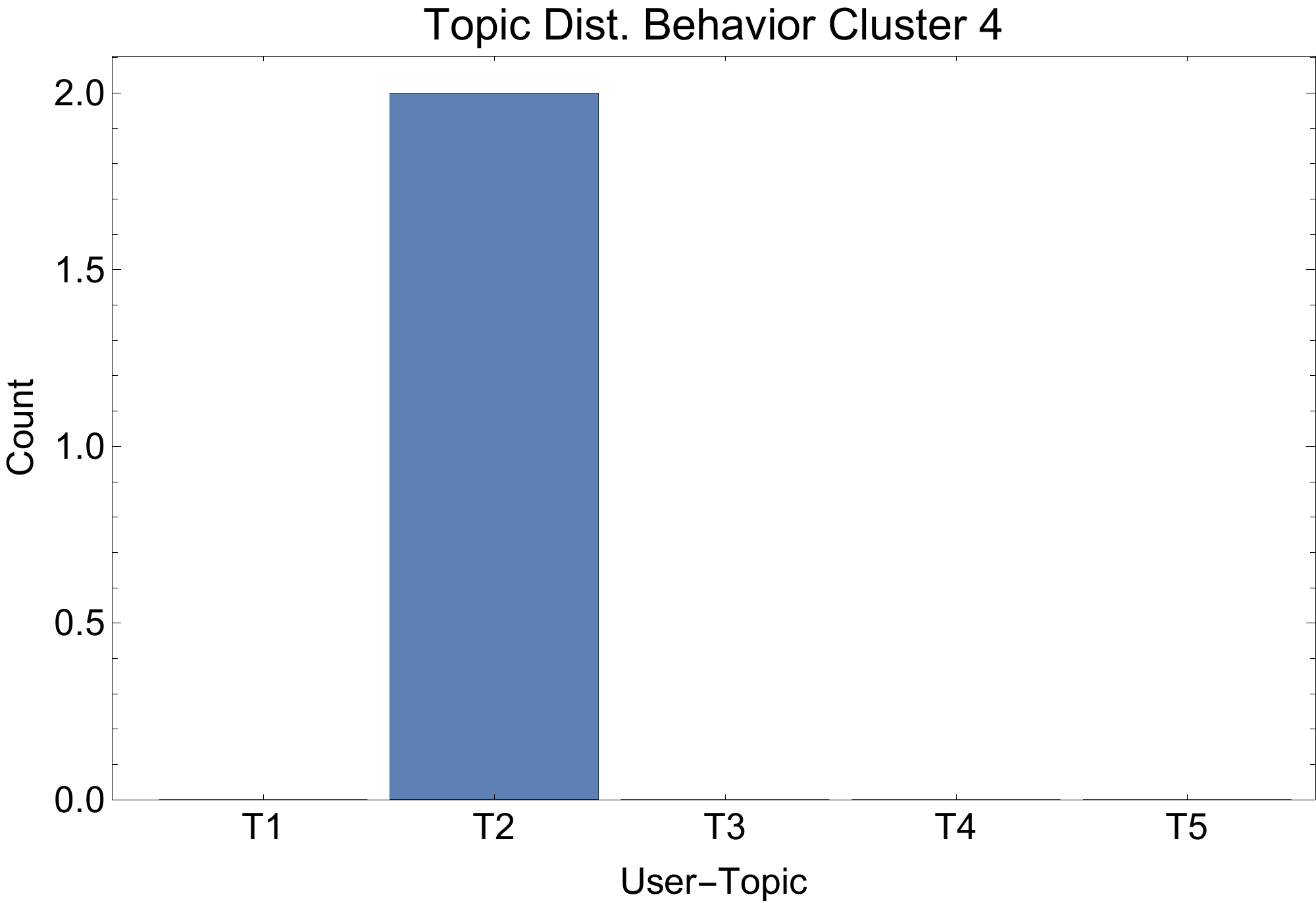}}\\
\subfloat[Post-Election $C_1$]{\includegraphics[width=0.45\columnwidth]{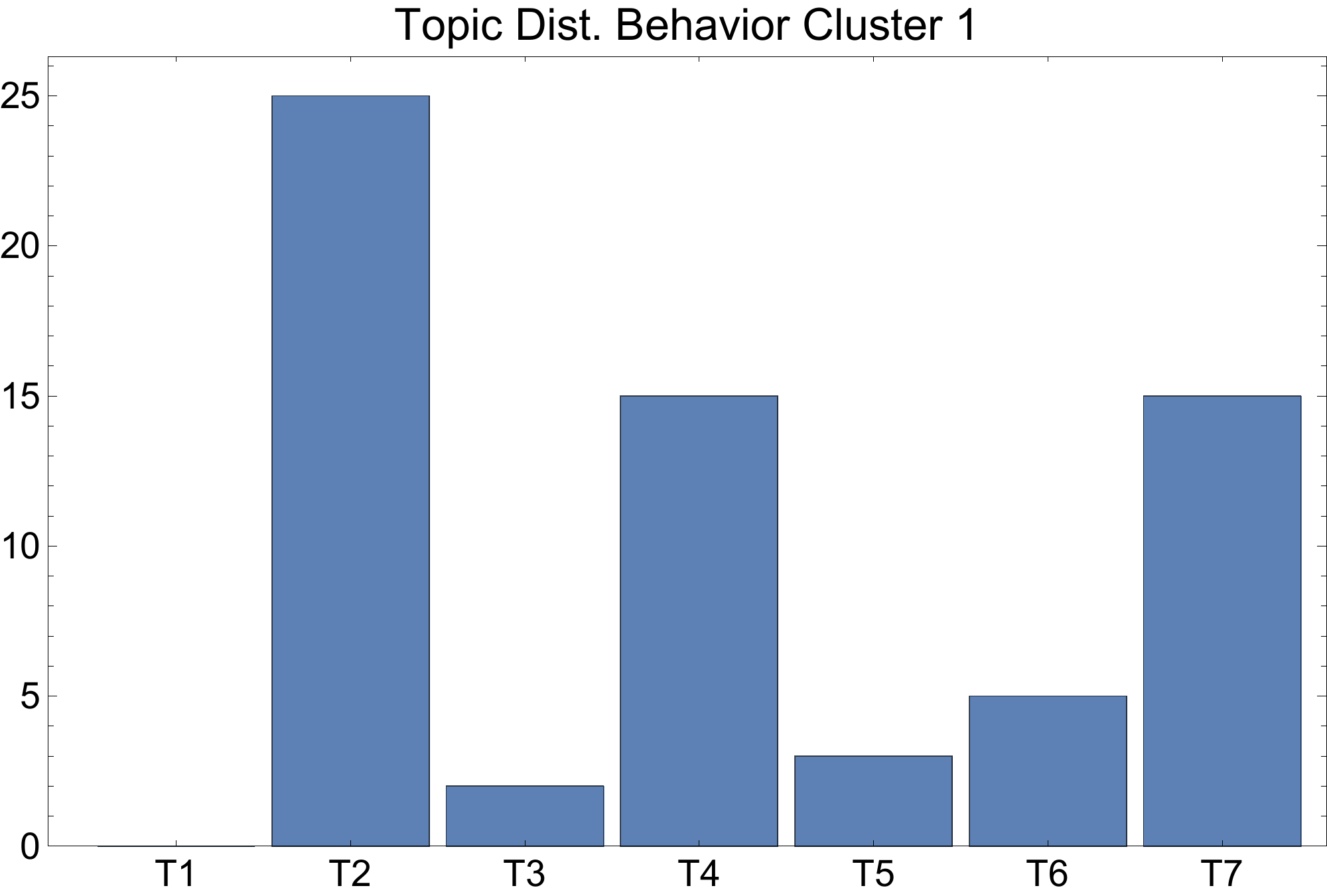}}\quad
\subfloat[Post-Election $C_2$]{\includegraphics[width=0.45\columnwidth]{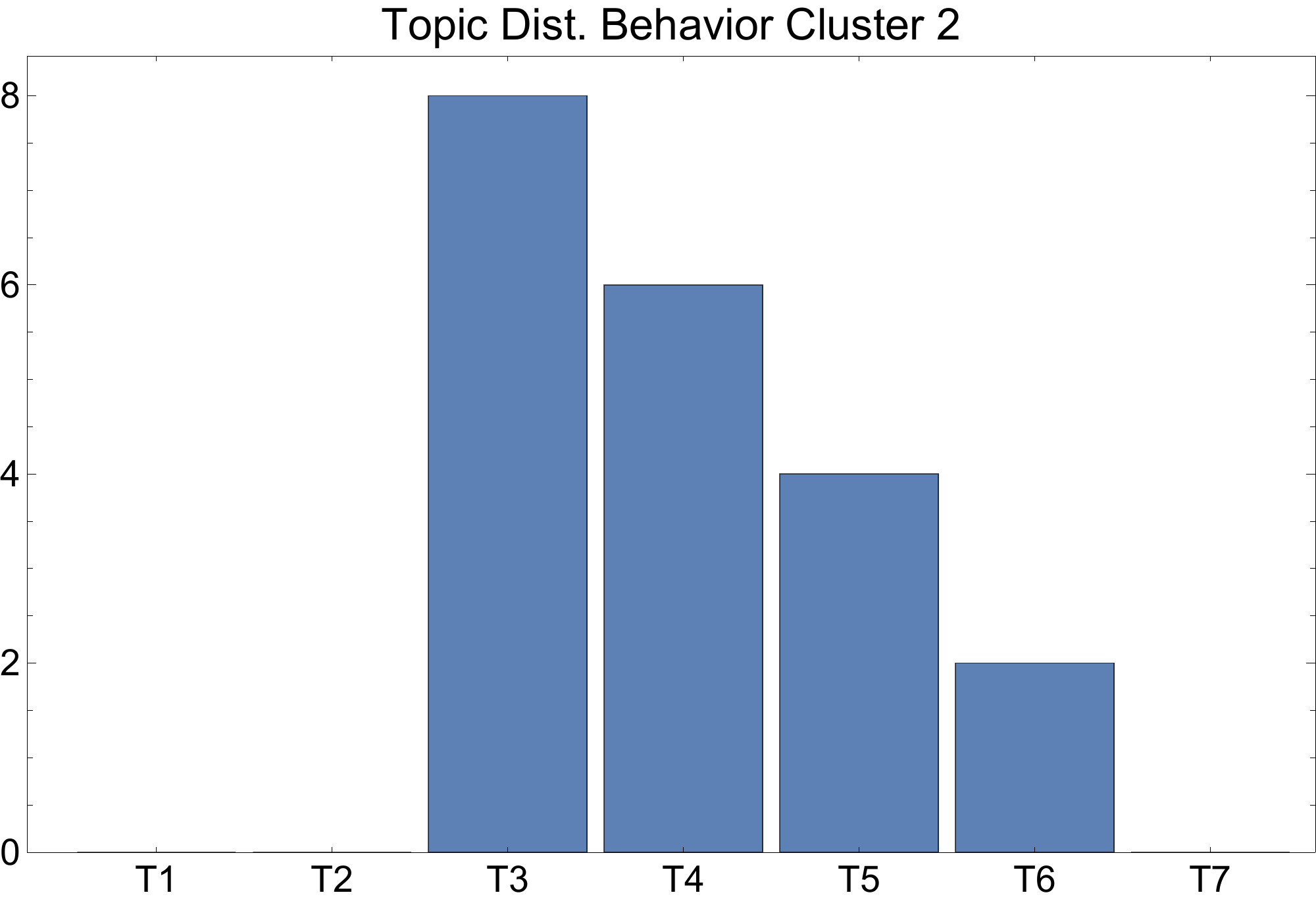}}\\
\subfloat[Post-Election $C_3$]{\includegraphics[width=0.45\columnwidth]{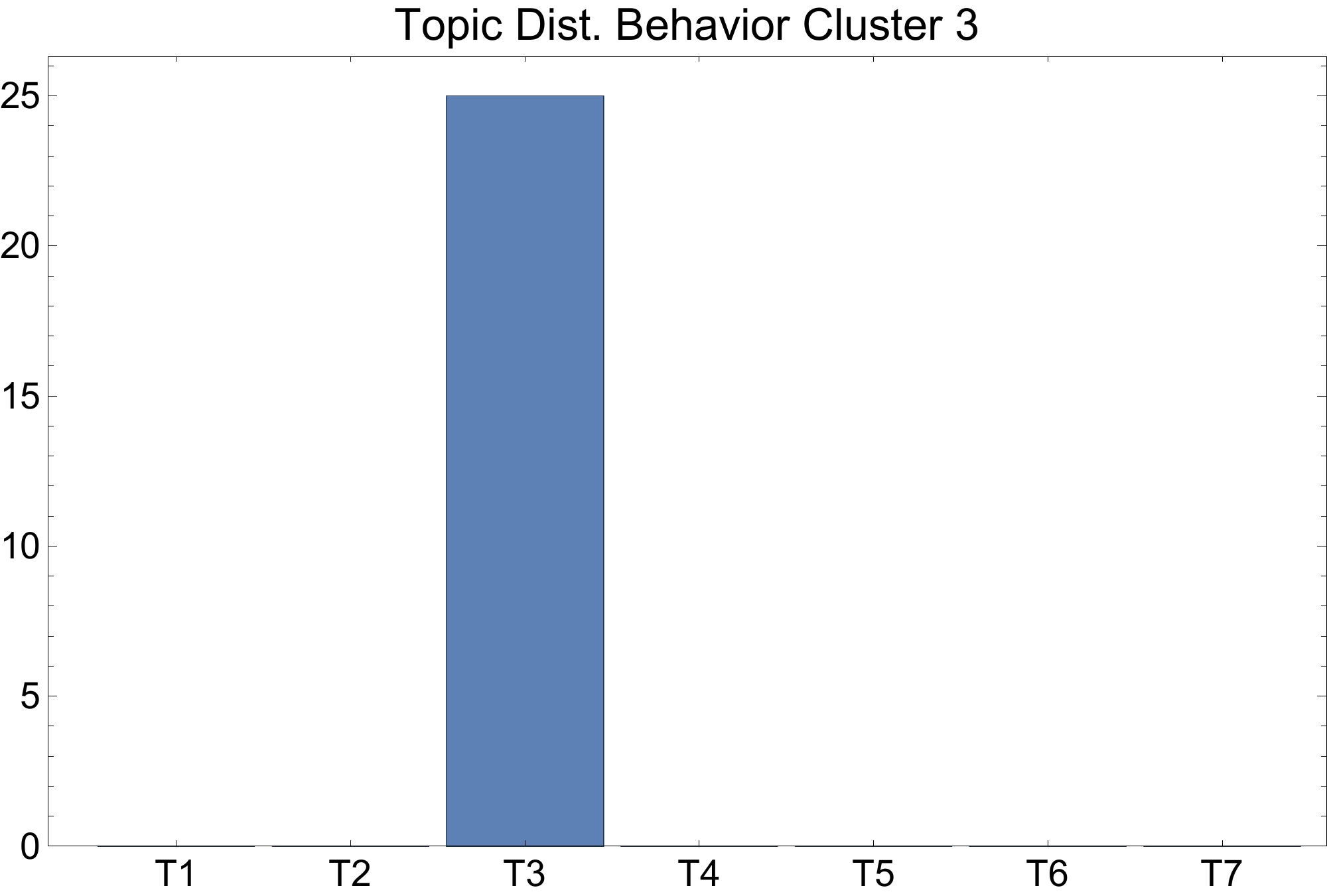}}\quad
\subfloat[Post-Election $C_4$]{\includegraphics[width=0.45\columnwidth]{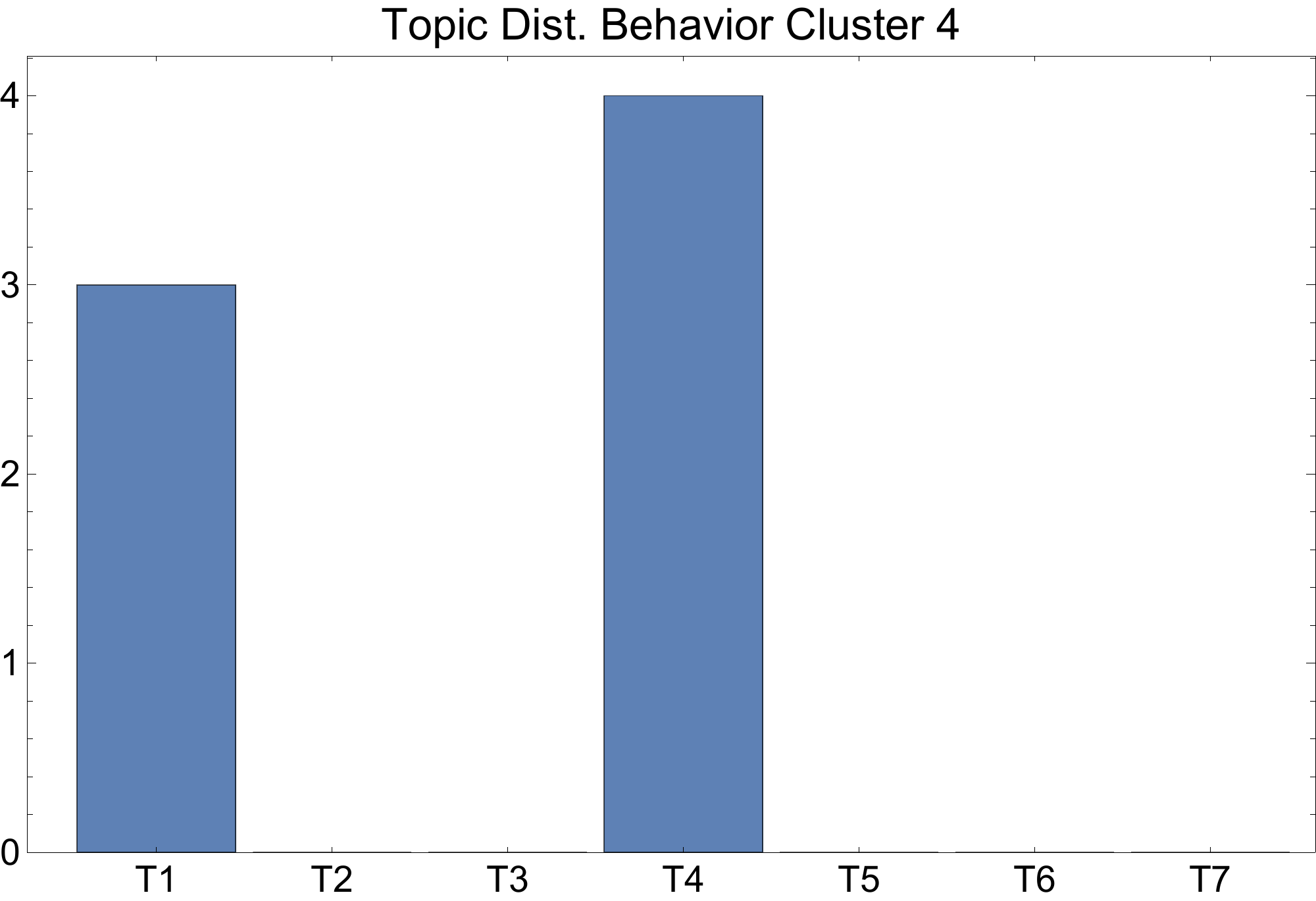}}\\
\caption{A comparison of the spectrum-based user clustering and the topic-based user clustering shows that user groups with well defined spectra have lower topic diversity.}
\label{fig:Comparison}
\end{figure}

We compare the spectrum-based cluster in the pre- and post-election time periods with the topic-based clusters. This is shown in \cref{fig:Comparison}. 

\begin{figure}[htbp]
\centering
\includegraphics[width=0.65\columnwidth]{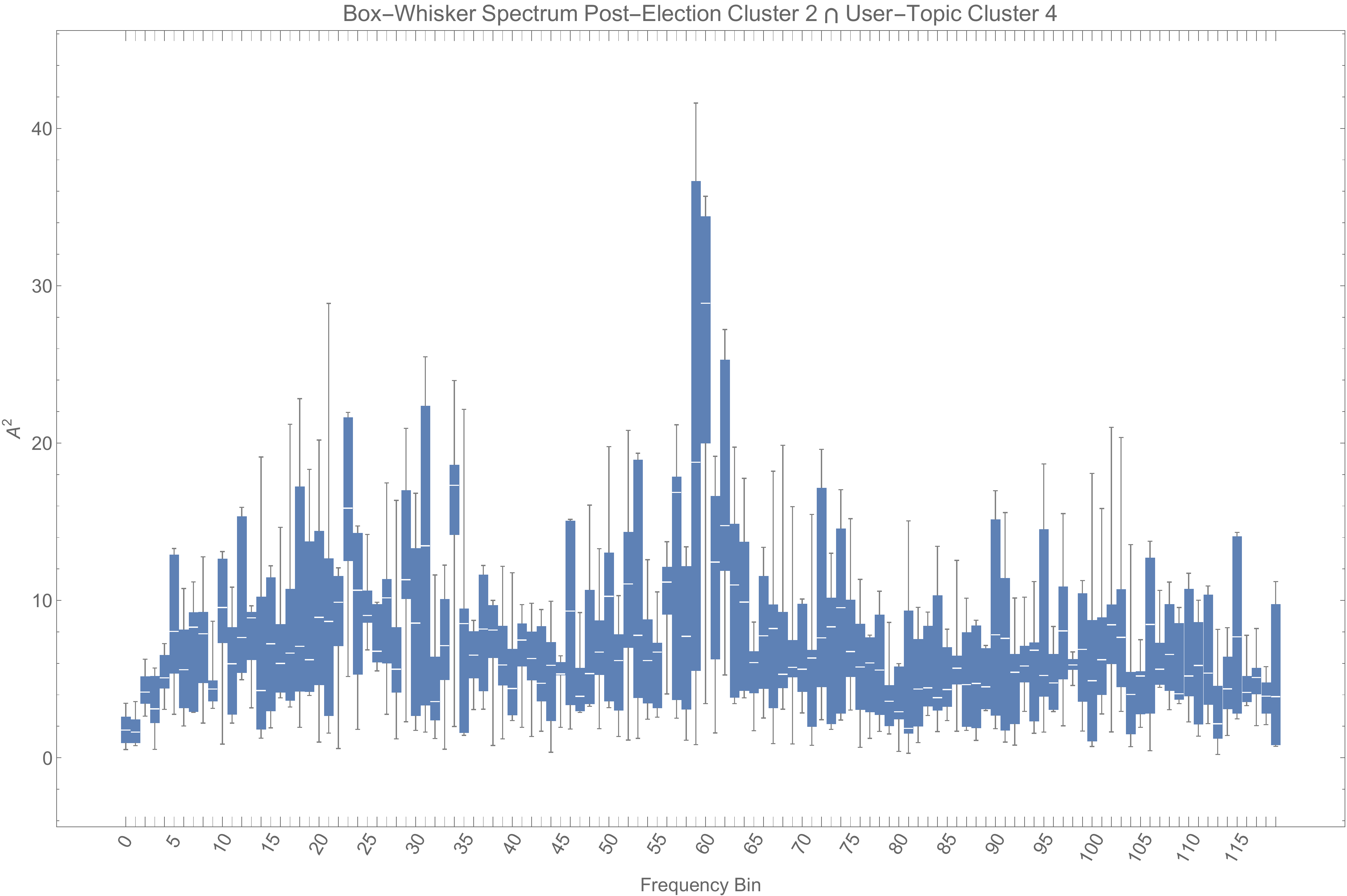}
\caption{The spectral box-and-whiskers chart that results from intersecting a spectral cluster and a user-topic cluster. The dominant frequency is now clearer.}
\label{fig:SubCluster}
\end{figure}

We noted in \cref{sec:Spectral} that clusters 3 and 4 in both the pre- and post-election periods had the clearest defining spectral characteristics. It is not surprising, therefore, that these groups of users show the least topical diversity. Members of pre-election (spectral) cluster 3 can be found only in pre-election user-topic cluster 5 (the news cluster), while members of pre-election (spectral) cluster 4 can be found only in pre-election user-topic cluster 2. In a similar pattern, members of post-election (spectral) cluster 3 are members of post-election user-topic cluster 3 (a news cluster). Members of post-election (spectral) cluster 4 are members of post-election user-topic clusters 1 (holiday) and 4 (Trump-related). Other spectral clusters have users that are more spread out, corresponding to less well-defined spectral properties. It is possible to create sub-clusterings of the users corresponding to both their topics and their spectra. This helps to decrease noise in the spectral clusters. We illustrate this using post-election spectral cluster 2 and post-election user-topic cluster 4. The smaller cluster contains only the users \texttt{hashed\_26}, \texttt{Jenn\_Abrams}, \texttt{Pamela\_Moore13}, \texttt{TEN\_GOP},\texttt{TheFoundingSon}, and \texttt{todayinsyria}, but has a clear dominating frequency corresponding to a 4 day period. This is shown in \cref{fig:SubCluster}.
Exhaustive sub-clustering using both frequency and user-topic information is left for future work. 

\section{Conclusion}\label{sec:Conclusion}
In this paper, we have analyzed the IRA data provided by Twitter Inc. from a dynamical systems perspective. We proposed a mathematical model to describe the per-day tweet output of a subset of the users and showed that this model could be validated using spectral analysis. This analysis allowed us to cluster user behavior in a way that was distinct from prior language-based work. We also developed a novel strategy assignment that allowed us to label each tweet as being \textit{original} to the IRA, \textit{amplifying} of external information or \textit{spreading} existing IRA information. We showed that a subset of the spectral clusters had interpretable strategic behaviors. We also showed the existence of a substantial behavioral shift within the English data set from originating information to amplifying external information just after the 2016 US presidential election. Finally, we compared the spectrally-derived user clusters to clusters derived from topics discussed and showed that more cohesive spectral clusters (based on the distinctiveness of frequency peaks) also have greater topical cohesion. We used this observation to illustrate the utility of sub-clustering using both dynamic behavior and linguistic features. 

In future work, we may investigate joint clustering based on dynamic behavior and linguistic features. Additionally, it would be very instructive to conduct a very large scale analysis on general Twitter data to determine whether there are hidden attractors or fractal properties within the per-day (or per-week) time series. While this data set \textit{seems} large (9M+ tweets), in reality many of the users exhibited bursty behavior making it practically impossible to reconstruct an attracting set. In the case of this data set, the presence of dominant frequencies strongly suggests that chaotic behavior is not present. However, this fact may be an important feature in differentiating bots from humans, who might exhibit far more organic (and therefore complex) tweeting patterns. In any future investigation with even more data the application of more sophisticated analytical methods using wavelet analysis or (at a minimum) spectrograms might be more instructive in understanding bursty data. 

\section*{Acknowledgement}
Portions of CG's work were supported by the National Science Foundation under grant IIS-1909255. SR gratefully acknowledges support from the Center for Security Research and Education at The Pennsylvania State University.

\appendix

\section{Alternative Spectrum and Fit Example}\label{sec:Kansas}
\begin{figure}[htbp]
\centering
\subfloat[Pre-Election Spectra]{\includegraphics[width=0.45\columnwidth]{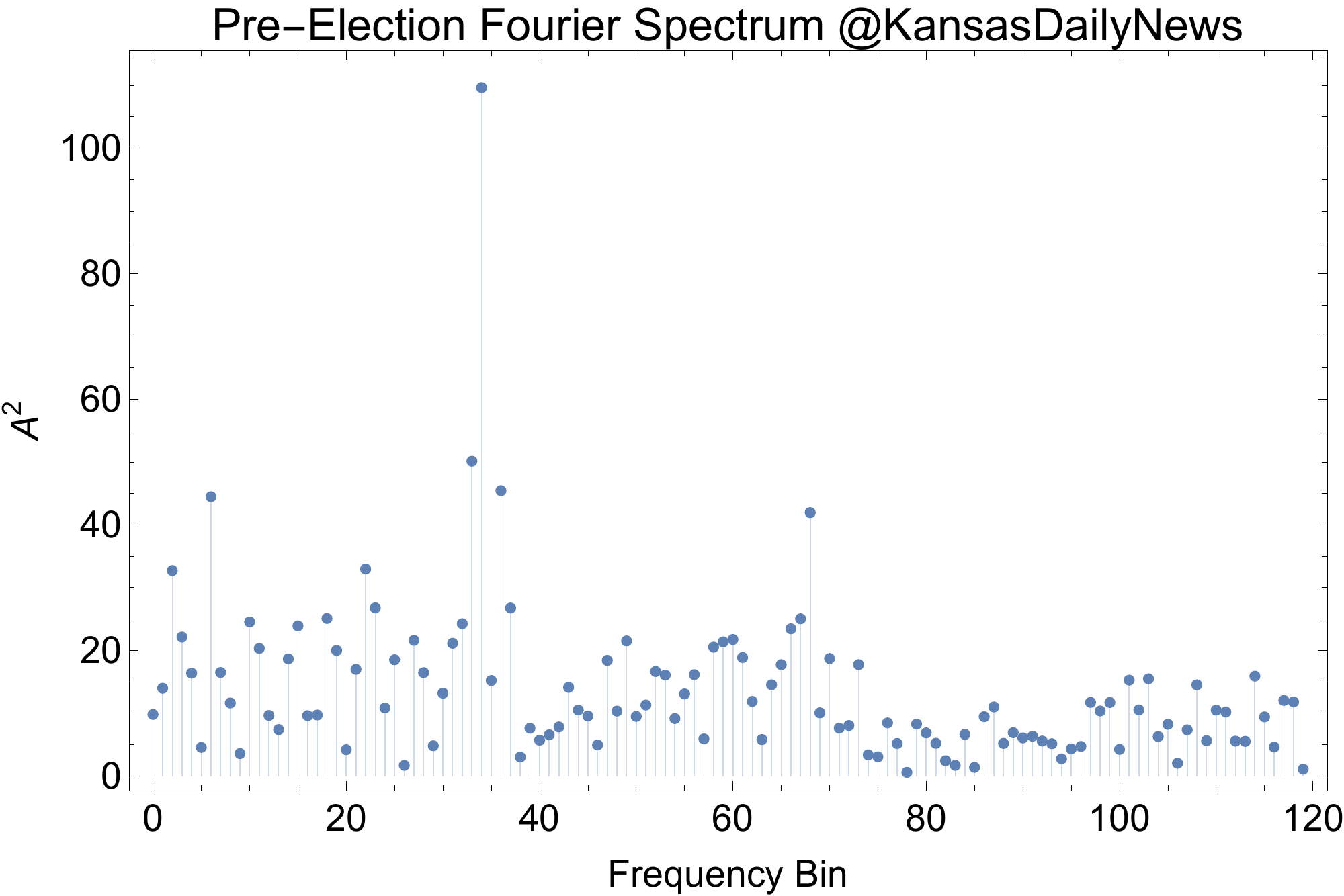}}\quad
\subfloat[Post-Election Spectra]{\includegraphics[width=0.45\columnwidth]{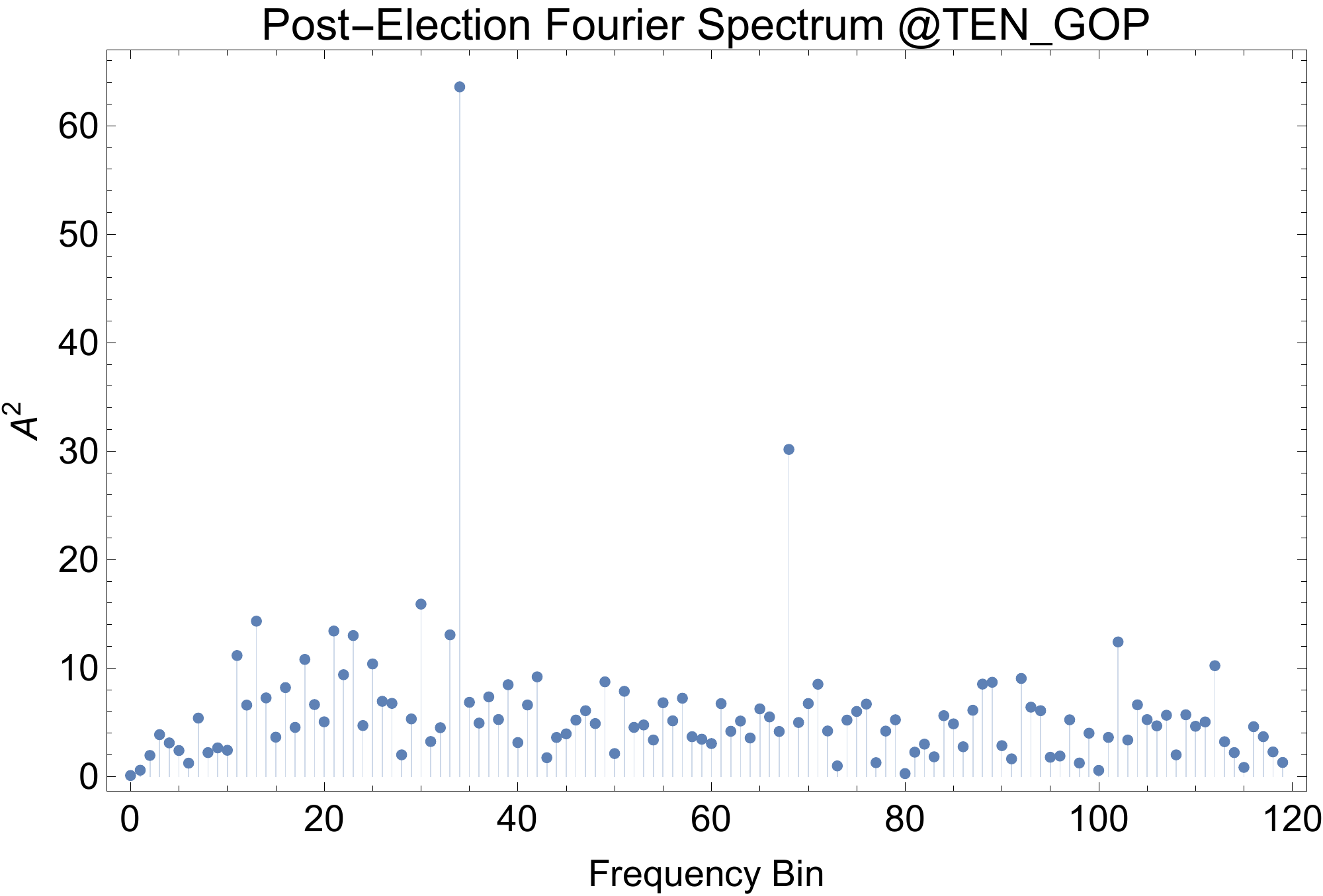}}\\
\subfloat[Pre-Election Fit]{\includegraphics[width=0.45\columnwidth]{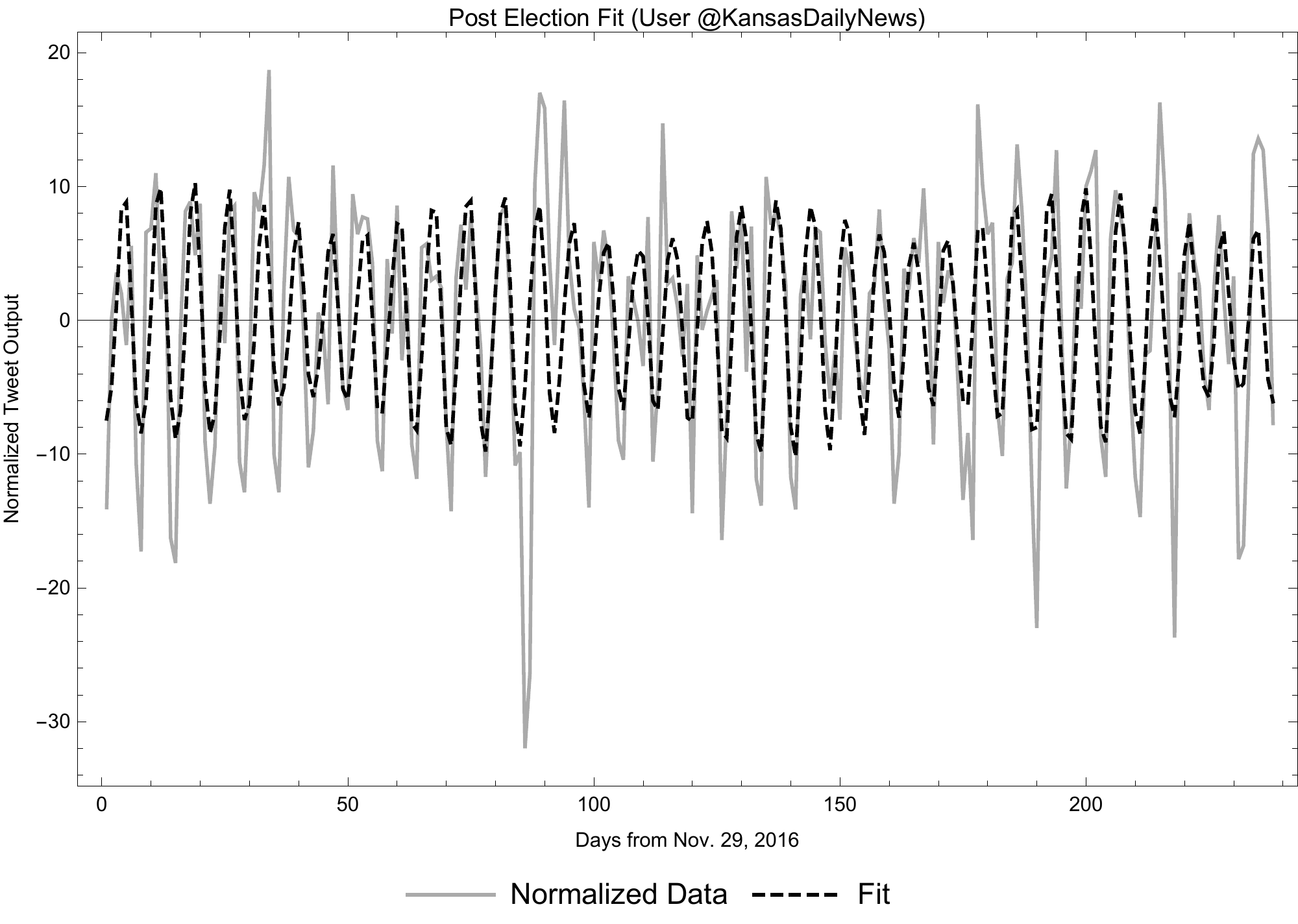}}\quad
\subfloat[Post-Election Fit]{\includegraphics[width=0.45\columnwidth]{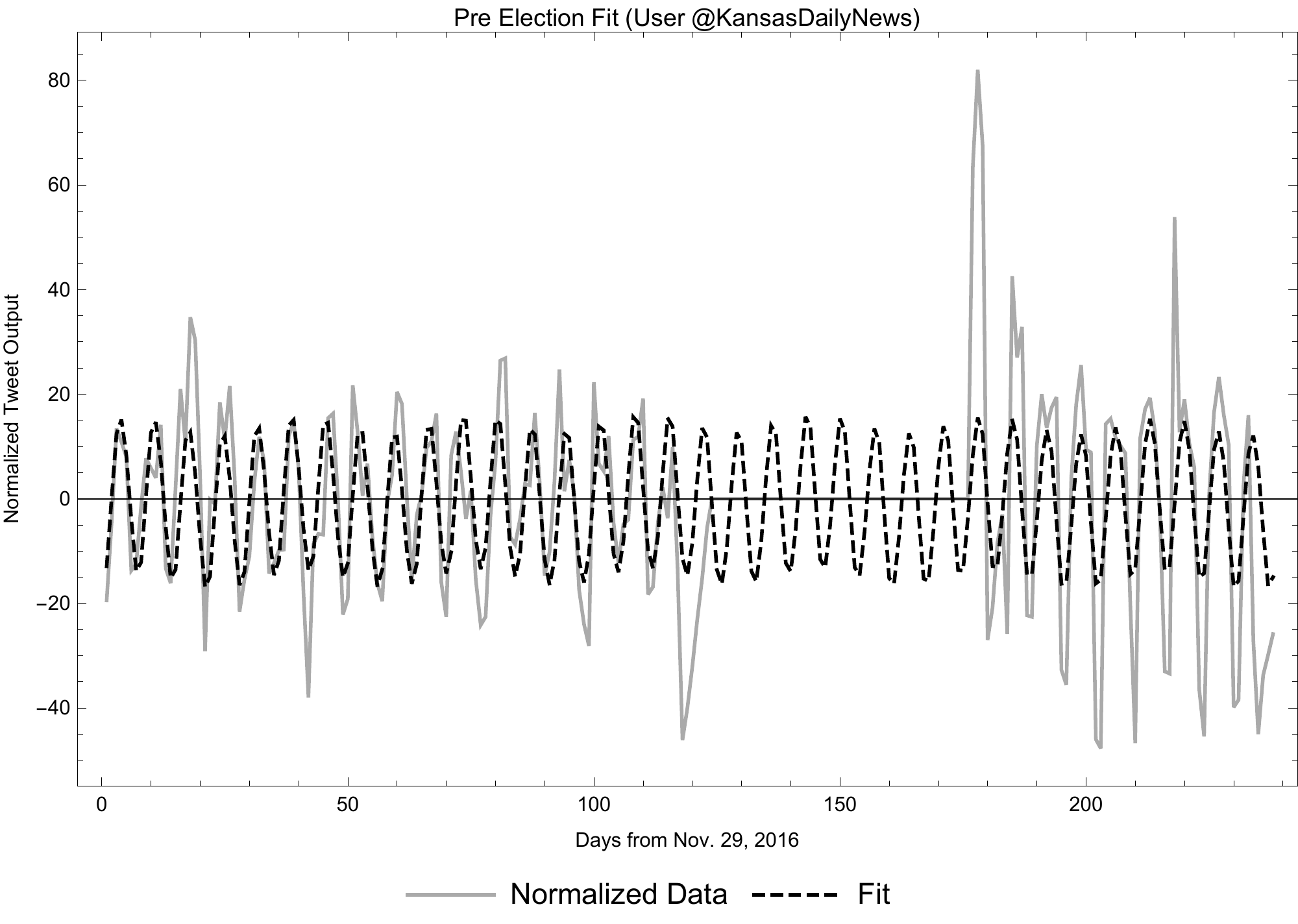}}
\caption{(Top) Fourier spectra for \texttt{KansasDailyNews} before and after the 2016 US Presidential Election. (Bottom) Corresponding fit using a 3 term Fourier sequence.}
\label{fig:KansasDaily-Spectra}
\end{figure}

In this appendix, we provide an alternate spectrum and fit example from a \texttt{KansasDailyNews} (see Figure \ref{fig:KansasDaily-Spectra}). This user has a 7 day period and is substantially less noisy that many other users. 

\section{Spectrum Clustered User IDs for each Time Period}\label{sec:UserId's}
The following user names are associated to the two periods of investigation. We have clustered them according to their primary frequencies of posting.\\
\vspace*{\parsep}

\noindent\textbf{Period 1 (March 8, 2016 - November 8, 2016):}\\
\small
\noindent\textit{Cluster 1:}
\begin{flushleft}
\texttt{TrayneshaCole}, \texttt{BleepThePolice}, \texttt{WarfareWW}, \texttt{Jenn\_Abrams}, \texttt{Blk\_Voice}, \texttt{blackmattersus}, \texttt{BlackToLive}, \texttt{hashed\_68}, \texttt{todayinsyria}, \texttt{TheFoundingSon}
\end{flushleft}

\noindent\textit{Cluster 2:} 
\begin{flushleft}
\texttt{USA\_Gunslinger}, \texttt{TEN\_GOP}, \texttt{gloed\_up}, \texttt{Crystal1Johnson}, 
\end{flushleft}
 
\noindent\textit{Cluster 3:} 
\begin{flushleft}
\texttt{DailySanJose}, \texttt{hashed\_37}, \texttt{NewspeakDaily}, \texttt{RoomOfRumor}, \texttt{hashed\_57}, \texttt{OaklandOnline}, \texttt{ScreamyMonkey}, \texttt{KansasDailyNews}, 
\end{flushleft} 

\noindent\textit{Cluster 4:}
\begin{flushleft}
\texttt{DanaGeezus}, \texttt{GiselleEvns}
\end{flushleft}

\normalsize\noindent\textbf{Period 2 (November 29, 2016 - July 31, 2017)}\\

\small
\noindent\textit{Cluster 1:}
\begin{flushleft}
\texttt{hashed\_19}, \texttt{hashed\_39}, \texttt{DetroitDailyNew}, \texttt{DailySanJose}, \texttt{hashed\_55}, \texttt{hashed\_54}, \texttt{hashed\_12}, \texttt{hashed\_44}, \texttt{hashed\_1}, \texttt{hashed\_32}, \texttt{hashed\_16}, \texttt{hashed\_18}, \texttt{hashed\_36}, \texttt{hashed\_63}, \texttt{hashed\_4}, \texttt{hashed\_33}, \texttt{hashed\_28}, \texttt{hashed\_46}, \texttt{hashed\_77}, \texttt{hashed\_35}, \texttt{hashed\_41}, \texttt{hashed\_5}, \texttt{hashed\_78}, \texttt{hashed\_25}, \texttt{hashed\_38}, \texttt{hashed\_2}, \texttt{hashed\_61}, \texttt{hashed\_60}, \texttt{hashed\_21}, \texttt{hashed\_51}, \texttt{hashed\_66}, \texttt{hashed\_17}, \texttt{hashed\_34}, \texttt{hashed\_56}, \texttt{hashed\_62}, \texttt{hashed\_40}, \texttt{hashed\_24}, \texttt{hashed\_9}, \texttt{hashed\_47}, \texttt{hashed\_45}, \texttt{hashed\_14}, \texttt{hashed\_74}, \texttt{hashed\_73}, \texttt{hashed\_3}, \texttt{hashed\_58}, \texttt{hashed\_13}, \texttt{hashed\_31}, \texttt{hashed\_71}, \texttt{hashed\_67}, \texttt{hashed\_53}, \texttt{hashed\_42}, \texttt{hashed\_76}, \texttt{hashed\_64}, \texttt{hashed\_48}, \texttt{hashed\_27}, \texttt{hashed\_59}, \texttt{hashed\_8}, \texttt{hashed\_15}, \texttt{hashed\_43}, \texttt{hashed\_10}, \texttt{hashed\_50}, \texttt{hashed\_7}, \texttt{hashed\_75}, \texttt{hashed\_65}, \texttt{hashed\_69}
\end{flushleft}

\noindent\textit{Cluster 2:}
\begin{flushleft}
\texttt{TrayneshaCole}, \texttt{BleepThePolice}, \texttt{hashed\_11}, \texttt{hashed\_52}, \texttt{TEN\_GOP}, \texttt{Jenn\_Abrams}, \texttt{hashed\_29}, \texttt{ElPasoTopNews}, \texttt{hashed\_49}, \texttt{gloed\_up}, \texttt{blackmattersus}, \texttt{Crystal1Johnson}, \texttt{BlackToLive}, \texttt{Pamela\_Moore13}, \texttt{hashed\_30}, \texttt{hashed\_26}, \texttt{todayinsyria}, \texttt{BlackNewsOutlet}, \texttt{TheFoundingSon}, \texttt{hashed\_72}
\end{flushleft}

\noindent\textit{Cluster 3:}
\begin{flushleft}
\texttt{hashed\_20}, \texttt{Seattle\_Post}, \texttt{DallasTopNews}, \texttt{DailyLosAngeles}, \texttt{PhiladelphiaON}, \texttt{NewOrleansON}, \texttt{TodayNYCity}, \texttt{TodayCincinnati}, \texttt{TodayMiami}, \texttt{PhoenixDailyNew}, \texttt{ChicagoDailyNew}, \texttt{OnlineCleveland}, \texttt{Baltimore0nline}, \texttt{Atlanta\_Online}, \texttt{TodayBostonMA}, \texttt{TodayPittsburgh}, \texttt{DailySanFran}, \texttt{Politweecs}, \texttt{DailySanDiego}, \texttt{WashingtOnline}, \texttt{OnlineMemphis}, \texttt{HoustonTopNews}, \texttt{StLouisOnline}, \texttt{KansasDailyNews}, \texttt{SanAntoTopNews}
\end{flushleft}

\noindent\textit{Cluster 4:}
\begin{flushleft}
\texttt{hashed\_70}, \texttt{DanaGeezus}, \texttt{GiselleEvns}, \texttt{ChrixMorgan}, \texttt{hashed\_22}, \texttt{hashed\_23}, \texttt{hashed\_6}
\end{flushleft}

\normalsize

\section{Topic Clustered Users for each Time Period}\label{sec:TopicCluster}
Here user names are clustered according to the topic about which they tweet during the two time periods.\\
\vspace*{\parsep}

\noindent\textbf{Period 1 (March 8, 2016 - November 8, 2016):}\\
\small
\noindent\textit{Cluster 1}
\begin{flushleft}
\texttt{Blk\_Voice}, \texttt{Crystal1Johnson}, \texttt{blackmattersus}
\end{flushleft}

\noindent\textit{Cluster 2}
\begin{flushleft}
\texttt{DanaGeezus}, \texttt{GiselleEvns}
\end{flushleft}

\noindent\textit{Cluster 3}
\begin{flushleft}
\texttt{Jenn\_Abrams}, \texttt{TEN\_GOP}, \texttt{TheFoundingSon}, \texttt{USA\_Gunslinger}
\end{flushleft}

\noindent\textit{Cluster 4}
\begin{flushleft}
\texttt{BlackToLive}, \texttt{BleepThePolice}, \texttt{TrayneshaCole}, \texttt{gloed\_up}
\end{flushleft}

\noindent\textit{Cluster 5}
\begin{flushleft}
\texttt{DailySanJose}, \texttt{hashed\_37}, \texttt{KansasDailyNews}, \texttt{hashed\_57}, \texttt{NewspeakDaily}, \texttt{OaklandOnline}, \texttt{hashed\_68}, \texttt{RoomOfRumor}, \texttt{ScreamyMonkey}, \texttt{WarfareWW}, \texttt{todayinsyria}
\end{flushleft}

\normalsize\noindent\textbf{Period 2 (November 29, 2016 - July 31, 2017)}\\

\small
\noindent\textit{Cluster 1}
\begin{flushleft}
\texttt{ChrixMorgan}, \texttt{DanaGeezus}, \texttt{GiselleEvns}
\end{flushleft}

\noindent\textit{Cluster 2}
\begin{flushleft}
\texttt{hashed\_5}, \texttt{hashed\_9}, \texttt{hashed\_7}, \texttt{hashed\_8}, \texttt{hashed\_10}, \texttt{hashed\_18}, \texttt{hashed\_21}, \texttt{hashed\_25}, \texttt{hashed\_27}, \texttt{hashed\_33}, \texttt{hashed\_35}, \texttt{hashed\_39}, \texttt{hashed\_41}, \texttt{hashed\_43}, \texttt{hashed\_45}, \texttt{hashed\_51}, \texttt{hashed\_56}, \texttt{hashed\_62}, \texttt{hashed\_69}, \texttt{hashed\_71}, \texttt{hashed\_76}, \texttt{hashed\_78}, \texttt{hashed\_1}, \texttt{hashed\_53}, \texttt{hashed\_55}
\end{flushleft}

\noindent\textit{Cluster 3}
\begin{flushleft}
\texttt{Atlanta\_Online}, \texttt{Baltimore0nline}, \texttt{BlackNewsOutlet}, \texttt{BleepThePolice}, \texttt{ChicagoDailyNew}, \texttt{Crystal1Johnson}, \texttt{DailyLosAngeles}, \texttt{hashed\_20}, \texttt{DailySanDiego}, \texttt{DailySanFran}, \texttt{DailySanJose}, \texttt{DallasTopNews}, \texttt{DetroitDailyNew}, \texttt{ElPasoTopNews}, \texttt{HoustonTopNews}, \texttt{KansasDailyNews}, \texttt{NewOrleansON}, \texttt{OnlineCleveland}, \texttt{OnlineMemphis}, \texttt{PhiladelphiaON}, \texttt{PhoenixDailyNew}, \texttt{Politweecs}, \texttt{SanAntoTopNews}, \texttt{Seattle\_Post}, \texttt{hashed\_72}, \texttt{StLouisOnline}, \texttt{TodayBostonMA}, \texttt{TodayCincinnati}, \texttt{TodayMiami}, \texttt{TodayNYCity}, \texttt{TodayPittsburgh}, \texttt{TrayneshaCole}, \texttt{WashingtOnline}, \texttt{blackmattersus}, \texttt{gloed\_up}
\end{flushleft}

\noindent\textit{Cluster 4}
\begin{flushleft}
\texttt{hashed\_4}, \texttt{hashed\_6}, \texttt{hashed\_16}, \texttt{hashed\_17}, \texttt{hashed\_19}, \texttt{hashed\_23}, \texttt{hashed\_24}, \texttt{hashed\_26}, \texttt{hashed\_36}, \texttt{hashed\_38}, \texttt{hashed\_44}, \texttt{Jenn\_Abrams}, \texttt{hashed\_47}, \texttt{hashed\_50}, \texttt{Pamela\_Moore13}, \texttt{hashed\_60}, \texttt{hashed\_63}, \texttt{hashed\_70}, \texttt{TEN\_GOP}, \texttt{TheFoundingSon}, \texttt{hashed\_75}, \texttt{hashed\_22}, \texttt{hashed\_28}, \texttt{hashed\_32}, \texttt{todayinsyria}
\end{flushleft}

\noindent\textit{Cluster 5}
\begin{flushleft}
\texttt{hashed\_2}, \texttt{hashed\_46}, \texttt{hashed\_49}, \texttt{hashed\_52}, \texttt{hashed\_14}, \texttt{hashed\_29}, \texttt{hashed\_30}
\end{flushleft}

\noindent\textit{Cluster 6}
\begin{flushleft}
\texttt{BlackToLive}, \texttt{hashed\_11}, \texttt{hashed\_42}, \texttt{hashed\_61}, \texttt{hashed\_64}, \texttt{hashed\_12}, \texttt{hashed\_54}
\end{flushleft}

\noindent\textit{Cluster 7}
\begin{flushleft}
\texttt{hashed\_3}, \texttt{hashed\_13}, \texttt{hashed\_15}, \texttt{hashed\_31}, \texttt{hashed\_40}, \texttt{hashed\_58}, \texttt{hashed\_59}, \texttt{hashed\_65}, \texttt{hashed\_67}, \texttt{hashed\_73}, \texttt{hashed\_77}, \texttt{hashed\_34}, \texttt{hashed\_48}, \texttt{hashed\_66}, \texttt{hashed\_74}
\end{flushleft}

\section{Hashed Usernames}
\tiny
\noindent\textbf{hashed\_1}\\ 71db45e4783602194ced17d50c4070036dfc754e9557fe918b39e58e8bf7d65b 

\noindent\textbf{hashed\_2}\\  1cd9dcae39203ce9df705a6ff0cb4b1961effb11c9959e48055606a108fedd44 

\noindent\textbf{hashed\_3}\\  f0a624171ca1c8d3177521f1362bf9eb78f1015ec3085a1eefdfa3ff9dbf73cb 

\noindent\textbf{hashed\_4}\\  cfb0d237487ebe3520cb83bd82df903c9c7afd5a60acba8a3576bc3ca9346467 

\noindent\textbf{hashed\_5}\\  efbb959a308f1bd1199a15221511bdd9e8945a3a02da83ce9d6585224733a8f6 

\noindent\textbf{hashed\_6}\\  8da5d23207d302647d84c4f38e4c9c48b199d9aa1cb0549940629a68b50b16cf 

\noindent\textbf{hashed\_7}\\  70007b94f796e3b39c18da7711327730b9e58032323c6e7b8438fc9c6acf835c 

\noindent\textbf{hashed\_8}\\  d225180304e4fafe5af936306b254e2cf5f2a950d6947f94c6ab44f7d51e2606 

\noindent\textbf{hashed\_9}\\  e35cc49080427faaf1f73e41752c5fea35a96f6c243863cd60b019661950c27e 

\noindent\textbf{hashed\_10}\\  8e58ab0f46d273103d9e71aa92cdaffb6e330ec7d15ae5fa79c825e2d6f291bf 

\noindent\textbf{hashed\_11}\\  d3f6e58202566043d909d57584be4811916859c8a09ebbe3b5239f72596d9acf 

\noindent\textbf{hashed\_12}\\  c465bceee4e65cc392661fafdecabc98450eb1f0b67e66a5a0d69be631f34ca5 

\noindent\textbf{hashed\_13}\\  14882528f53ed4b8fbadce206679b89258e363212b5a63f31777f4e93e1a5ccf 

\noindent\textbf{hashed\_14}\\  ef983249ef6ed5de427c4dc19ad6d966c6cf572c2505e44142e7e7261f917ae6 

\noindent\textbf{hashed\_15}\\  3da8a7609e8e8bbc93fb2051c4d6568ad5436d629d003735f217ecd3acd1bb7e 

\noindent\textbf{hashed\_16}\\  0994abf9fb8fe1bf699d0e101e9603b30f369e94a0eec1109e2f654b1a9c5caf 

\noindent\textbf{hashed\_17}\\  80f92e973543912617d5c7ec3bc42bd455b6970b7a52416f398ce923389d3a85 

\noindent\textbf{hashed\_18}\\  3f22ec97e6f7c68c28e59390710135dbfd067aa024eeb34f13dbb905b4dde4a5 

\noindent\textbf{hashed\_19}\\  a95a911dd6ae864c48ed062cdbe75e5c28dbe0cf57c6db3fa9dd53b654ee9845 

\noindent\textbf{hashed\_20}\\  8a73099860c2a4790ae6899381c0c327462e626c02ce0d3c899ce6058135786a 

\noindent\textbf{hashed\_21}\\  88669ad69e40d7c199af91e8107f1e0e7988d377d2e41fff77182bf42f31a1ed 

\noindent\textbf{hashed\_22}\\  99299a75c967aa491b27d86a60c2ffe0a6156d1b0dcb4487917f48e7e1c7ef78 

\noindent\textbf{hashed\_23}\\  2aae433624dcea729731ff8e44b0bdedc56e7cc9eea9bedd14bb264b07eedd5c 

\noindent\textbf{hashed\_24}\\  815561925624b72a95321d039bf8bdf0da5620f8475e62ce2dee37beb71abcab 

\noindent\textbf{hashed\_25}\\  40bd0ff013b85c7646ca07ad238bc4dc865ce2cc87034af6e7884e69481f6422 

\noindent\textbf{hashed\_26}\\  dce04b9807276c27a04ad521e37f061bcdfc8bc46cc6dc3d47128b263bd61124 

\noindent\textbf{hashed\_27}\\  a4df865b62425d5eb91b169b0bbb873434078a44b193174962a65408ed66d793 

\noindent\textbf{hashed\_28}\\  5ad1a4ad9a7d67c045ab752c7a946a9d847d5d7e03684174835269a193c2820e 

\noindent\textbf{hashed\_29}\\  f656411ab6d63685f868421c4b381779ce97b3470f701edf9ca9ebb5a0226e3d 

\noindent\textbf{hashed\_30}\\  21c35c77715644e10675df5bb88f02d40412785ace111edda73b40553db68bed 

\noindent\textbf{hashed\_31}\\  0dea399346f2f1a94e11fbc84897a35f1ccd3fcde6ea5f0e207ba0b296412e47 

\noindent\textbf{hashed\_32}\\  396b0b6914e425ca53c49b7f1d6cfe9757eb7368c04c657a09d83b19a8f1251e 

\noindent\textbf{hashed\_33}\\  beb24c4d724cb8a986f15f48e11895e7e3484598a54af5a0fef356ceb3641193 

\noindent\textbf{hashed\_34}\\  be762d9f6e6423c2b11320bd7b4d6abad8aed0f8a8a54f86350d0a584117eb25 

\noindent\textbf{hashed\_35}\\  e1c4cd815926b92691660738cd682335520a24fb08765655c261b07ef591b653 

\noindent\textbf{hashed\_36}\\  5788d958456ab731e042dfff6d344829c158409c1c0911d79a5de12b0a72ced1 

\noindent\textbf{hashed\_37}\\  fea5e166786f00c893ee247a87ab6047e3ed8727dbe4b5a67b64484dc777b070 

\noindent\textbf{hashed\_38}\\  ec2109adb67d2a24091026d5d9aab64dadca1fdb2f7355473c9a82f988a9c2a0 

\noindent\textbf{hashed\_39}\\  6791992ed85cd2e739f57d9e4f1a4d1ac119577cdece986e1e049ade25c2d793 

\noindent\textbf{hashed\_40}\\  8a3c0cdcac971eaf9f7802a5688faea3aadd850e5d2e214941c2ccbdcaf4ac32 

\noindent\textbf{hashed\_41}\\  5f452585194266b94e60c99e1dd07f808ceb2b26b10cfb939c2e5ad47888be1c 

\noindent\textbf{hashed\_42}\\  3c46bd68165bb49be66c657de7f09b1cbd130e65934abd29f5aec28e2a86d43d 

\noindent\textbf{hashed\_43}\\  e579b30026998948100e534d656b6f27770e519fab233bccaf8b6c18b4a862c8 

\noindent\textbf{hashed\_44}\\  0e45d2af1bbca512ab1d8fa1c2216c038f9627bb773ab4cbe351404ee7a2c93e 

\noindent\textbf{hashed\_45}\\  e8e7c2bfdb7272a4a7db5dfcbb01627222d934054e99469baba6a2fe5edb9f92 

\noindent\textbf{hashed\_46}\\  afe3c426e91f3f0e507f9cf7beceeddb2537f56e4fd3b2aab12f87fc6eee3edd 

\noindent\textbf{hashed\_47}\\  aa80df35407d74929dc17b058bdc679ae7f995c237f1cfd09915d970d72f72c1 

\noindent\textbf{hashed\_48}\\  3434652527f405b7fd0a49064a0760dc11c1eaf80a42b7c8afc800d6a3612ce2 

\noindent\textbf{hashed\_49}\\  6183ad40df50200d0001e2b2e0acfcdb1d5f71ddee9495874a3b1fa2b00b6e25 

\noindent\textbf{hashed\_50}\\  93a00b4e2d4198865a35726aab0d5c1a4205301471411a85054b714b3b89c4a9 

\noindent\textbf{hashed\_51}\\  fb6f6cc1a31b47965d9894b89d576add7a679a77e9022aad33d6c0c2310cd789 

\noindent\textbf{hashed\_52}\\  6ca21c26de8159386b71428b8bdadf732bd854614dadea9086770b71bbe1df6a 

\noindent\textbf{hashed\_53}\\  77755623608ac46aacd68d114868ffebad0b2cbd4d29f0fe99143d93d51864ad 

\noindent\textbf{hashed\_54}\\  0a0af8893cdc8454338447004eeaf65ee2934977c71ec48d3bb17e6723e1daef 

\noindent\textbf{hashed\_55}\\  cfe9315d7429aedb7f843a165185d91f0ae819e54026d8a22a4f6e539f5e9132 

\noindent\textbf{hashed\_56}\\  274d180d1de8828a8e3e7f62eec224ba9e3d04481e5d9eb29bf1cab882004a1b 

\noindent\textbf{hashed\_57}\\  080da9e6b0c293d8e7abe6115605790ec8bc9a5c75a2b3e60539e67e3f100875 

\noindent\textbf{hashed\_58}\\  f78a269be68e66b4a8c9bb5adbe49f617b6b3845134de4b9d8afe7a350bfb2cd 

\noindent\textbf{hashed\_59}\\  1b9adccac803d2fc7633cef02e9e6c3d64b0af50c7550a1f3d2b84c54e765cbd 

\noindent\textbf{hashed\_60}\\  63df8aabe784aebc5674b36a34c14dae77bfaa89e1d9bafa5a485ed32f6e2834 

\noindent\textbf{hashed\_61}\\  cd0e6a3af6160a511a2cea1bda54113ea9ba8ba27b70575e4d3612c6074c342b 

\noindent\textbf{hashed\_62}\\  11b2a99966def750bb3e9a4b52d46826cec464aefefecfb792c3b0da7e2d2be7 

\noindent\textbf{hashed\_63}\\  f8d82815f7d03fec70e6e8f50f859b84714024b1024979f418110feb8db9d798 

\noindent\textbf{hashed\_64}\\  c567f55d5430039ab5b53980a337cd6e3399e842f186e3930a92d7f2e11d27d6 

\noindent\textbf{hashed\_65}\\  2d60b8bc4ff97649d507084f47f8e916263eac2d7e60ab7576d7c0722c39f88b 

\noindent\textbf{hashed\_66}\\  366e32f7e0e3b8eac2fff4f7049e1ebd0bdc4ddfdf0c4334edac18eda8b846e6 

\noindent\textbf{hashed\_67}\\  96d4f0c1f4b4ed7638342a48c66c401263da4b151fdffcb98d7b775c996c3fb6 

\noindent\textbf{hashed\_68}\\  9ebdcf10ebedc9abf33a34e07792e18230ecd26cea77ab0f3fcdfac11fe2116f 

\noindent\textbf{hashed\_69}\\  eafd6b68b8855efe312fbb60d5fde0edb8632e9c0364f5044d999b0ee58837a8 

\noindent\textbf{hashed\_70}\\  a396435b41bb21c018281c802df9fa8d1a1eaa9b5b139c72bc1114ca62871b9c 

\noindent\textbf{hashed\_71}\\  21ede93ad40450cf820f249bde68d4622b61894a6597de8c6290f034acc8f059 

\noindent\textbf{hashed\_72}\\  aba127107267464101355d3465e86aafe41aaa31395bf82b3eb2fb801efac56f 

\noindent\textbf{hashed\_73}\\  8773aaa8badf887f0702db4c5c8fe1e45ff82e35ce867539fd26e9dbf804b27f 

\noindent\textbf{hashed\_74}\\  190df5183c66fcdf2efe5a11eb35cc827a5f0726f4788b3ee5d3aadaeeb28dbf 

\noindent\textbf{hashed\_75}\\  74c5eb30de3d6691e150879ee8528463c70b0feaefed9766fcbbcc8b2458221a 

\noindent\textbf{hashed\_76}\\  83216c1bdaf0245f9ac5b98a8c4b3cf2a1634b74d8b38dce5641f292a56c10d6 

\noindent\textbf{hashed\_77}\\  85fdb02196bb8c19d03a262e72132e5c8ec70bde25eaea167f012f0bc1becbf8 

\noindent\textbf{hashed\_78}\\  c7b1bacb73f0f3025dc09452d99b1abb22a5ce7aa0a649330f3820de07f2ecde 

\normalsize

\bibliographystyle{apsrev4-1}
\bibliography{RussianTwitterPaper-2}

\begin{thebibliography}{42}%
\makeatletter
\providecommand \@ifxundefined [1]{%
 \@ifx{#1\undefined}
}%
\providecommand \@ifnum [1]{%
 \ifnum #1\expandafter \@firstoftwo
 \else \expandafter \@secondoftwo
 \fi
}%
\providecommand \@ifx [1]{%
 \ifx #1\expandafter \@firstoftwo
 \else \expandafter \@secondoftwo
 \fi
}%
\providecommand \natexlab [1]{#1}%
\providecommand \enquote  [1]{``#1''}%
\providecommand \bibnamefont  [1]{#1}%
\providecommand \bibfnamefont [1]{#1}%
\providecommand \citenamefont [1]{#1}%
\providecommand \href@noop [0]{\@secondoftwo}%
\providecommand \href [0]{\begingroup \@sanitize@url \@href}%
\providecommand \@href[1]{\@@startlink{#1}\@@href}%
\providecommand \@@href[1]{\endgroup#1\@@endlink}%
\providecommand \@sanitize@url [0]{\catcode `\\12\catcode `\$12\catcode
  `\&12\catcode `\#12\catcode `\^12\catcode `\_12\catcode `\%12\relax}%
\providecommand \@@startlink[1]{}%
\providecommand \@@endlink[0]{}%
\providecommand \url  [0]{\begingroup\@sanitize@url \@url }%
\providecommand \@url [1]{\endgroup\@href {#1}{\urlprefix }}%
\providecommand \urlprefix  [0]{URL }%
\providecommand \Eprint [0]{\href }%
\providecommand \doibase [0]{http://dx.doi.org/}%
\providecommand \selectlanguage [0]{\@gobble}%
\providecommand \bibinfo  [0]{\@secondoftwo}%
\providecommand \bibfield  [0]{\@secondoftwo}%
\providecommand \translation [1]{[#1]}%
\providecommand \BibitemOpen [0]{}%
\providecommand \bibitemStop [0]{}%
\providecommand \bibitemNoStop [0]{.\EOS\space}%
\providecommand \EOS [0]{\spacefactor3000\relax}%
\providecommand \BibitemShut  [1]{\csname bibitem#1\endcsname}%
\let\auto@bib@innerbib\@empty
\bibitem [{\citenamefont {Silva}(2017)}]{S17}%
  \BibitemOpen
  \bibfield  {author} {\bibinfo {author} {\bibfnamefont {D.}~\bibnamefont
  {Silva}},\ }\href@noop {} {\enquote {\bibinfo {title} {Twitter releases
  massive data trove on russian, iranian foreign influence campaigns},}\
  }\bibinfo {howpublished}
  {\url{https://www.nbcnews.com/tech/social-media/twitter-releases-massive-data-trove-russian/-/iranian-foreign-influence-campaigns-n921146}}
  (\bibinfo {year} {2017})\BibitemShut {NoStop}%
\bibitem [{\citenamefont {TwitterInc.}(2017)}]{T17}%
  \BibitemOpen
  \bibfield  {author} {\bibinfo {author} {\bibnamefont {TwitterInc.}},\
  }\href@noop {} {\enquote {\bibinfo {title} {{Information operations}},}\
  }\bibinfo {howpublished}
  {\url{https://about.twitter.com/en_us/values/elections-integrity.html}}
  (\bibinfo {year} {2017})\BibitemShut {NoStop}%
\bibitem [{\citenamefont {Mak}(2018)}]{M18}%
  \BibitemOpen
  \bibfield  {author} {\bibinfo {author} {\bibfnamefont {T.}~\bibnamefont
  {Mak}},\ }\href@noop {} {\enquote {\bibinfo {title} {{Russia's Divisive
  Twitter Campaign Took A Rare Consistent Stance: Pro-Gun}},}\ }\bibinfo
  {howpublished}
  {\url{https://www.npr.org/2018/09/21/648803459/russias-2016-twitter-campaign-was-strongly-pro/-/gun-with-echoes-of-the-nra}}
  (\bibinfo {year} {2018})\BibitemShut {NoStop}%
\bibitem [{\citenamefont {Griffin}\ and\ \citenamefont {Bickel}(2018)}]{GB18}%
  \BibitemOpen
  \bibfield  {author} {\bibinfo {author} {\bibfnamefont {C.}~\bibnamefont
  {Griffin}}\ and\ \bibinfo {author} {\bibfnamefont {B.}~\bibnamefont
  {Bickel}},\ }\href@noop {} {\enquote {\bibinfo {title} {Unsupervised machine
  learning of open source russian twitter data reveals global scope and
  operational characteristics},}\ } (\bibinfo {year} {2018}),\ \Eprint
  {http://arxiv.org/abs/1810.01466} {arXiv:1810.01466 [cs.SI]} \BibitemShut
  {NoStop}%
\bibitem [{\citenamefont {Review}(2018)}]{R18}%
  \BibitemOpen
  \bibfield  {author} {\bibinfo {author} {\bibfnamefont {M.~T.}\ \bibnamefont
  {Review}},\ }\href@noop {} {\enquote {\bibinfo {title} {Data mining has
  revealed previously unknown russian twitter troll campaigns},}\ }\bibinfo
  {howpublished}
  {\url{https://www.technologyreview.com/s/612252/data-mining-has-revealed-previously-unknown/-/russian-twitter-troll-campaigns/}}
  (\bibinfo {year} {2018})\BibitemShut {NoStop}%
\bibitem [{\citenamefont {Popken}(2018)}]{P18}%
  \BibitemOpen
  \bibfield  {author} {\bibinfo {author} {\bibfnamefont {B.}~\bibnamefont
  {Popken}},\ }\href@noop {} {\bibfield  {journal} {\bibinfo  {journal} {NBC
  News}\ }\textbf {\bibinfo {volume}
  {\url{https://www.nbcnews.com/tech/social-media/now-available-more-200-000-deleted-russian-troll/-/tweets-n844731}}}
  (\bibinfo {year} {2018})}\BibitemShut {NoStop}%
\bibitem [{\citenamefont {Kellner}\ \emph {et~al.}(2019)\citenamefont
  {Kellner}, \citenamefont {Rangosch}, \citenamefont {Wressnegger},\ and\
  \citenamefont {Rieck}}]{KRWR19}%
  \BibitemOpen
  \bibfield  {author} {\bibinfo {author} {\bibfnamefont {A.}~\bibnamefont
  {Kellner}}, \bibinfo {author} {\bibfnamefont {L.}~\bibnamefont {Rangosch}},
  \bibinfo {author} {\bibfnamefont {C.}~\bibnamefont {Wressnegger}}, \ and\
  \bibinfo {author} {\bibfnamefont {K.}~\bibnamefont {Rieck}},\ }\href@noop {}
  {\enquote {\bibinfo {title} {Political elections under (social) fire?
  analysis and detection of propaganda on twitter},}\ } (\bibinfo {year}
  {2019}),\ \Eprint {http://arxiv.org/abs/1912.04143} {arXiv:1912.04143
  [cs.CR]} \BibitemShut {NoStop}%
\bibitem [{\citenamefont {Im}\ \emph {et~al.}(2019)\citenamefont {Im},
  \citenamefont {Chandrasekharan}, \citenamefont {Sargent}, \citenamefont
  {Lighthammer}, \citenamefont {Denby}, \citenamefont {Bhargava}, \citenamefont
  {Hemphill}, \citenamefont {Jurgens},\ and\ \citenamefont {Gilbert}}]{ICSL19}%
  \BibitemOpen
  \bibfield  {author} {\bibinfo {author} {\bibfnamefont {J.}~\bibnamefont
  {Im}}, \bibinfo {author} {\bibfnamefont {E.}~\bibnamefont {Chandrasekharan}},
  \bibinfo {author} {\bibfnamefont {J.}~\bibnamefont {Sargent}}, \bibinfo
  {author} {\bibfnamefont {P.}~\bibnamefont {Lighthammer}}, \bibinfo {author}
  {\bibfnamefont {T.}~\bibnamefont {Denby}}, \bibinfo {author} {\bibfnamefont
  {A.}~\bibnamefont {Bhargava}}, \bibinfo {author} {\bibfnamefont
  {L.}~\bibnamefont {Hemphill}}, \bibinfo {author} {\bibfnamefont
  {D.}~\bibnamefont {Jurgens}}, \ and\ \bibinfo {author} {\bibfnamefont
  {E.}~\bibnamefont {Gilbert}},\ }\href@noop {} {\enquote {\bibinfo {title}
  {Still out there: Modeling and identifying russian troll accounts on
  twitter},}\ } (\bibinfo {year} {2019}),\ \Eprint
  {http://arxiv.org/abs/1901.11162} {arXiv:1901.11162} \BibitemShut {NoStop}%
\bibitem [{\citenamefont {Cable}\ and\ \citenamefont {Hugh}(2019)}]{CH19}%
  \BibitemOpen
  \bibfield  {author} {\bibinfo {author} {\bibfnamefont {J.}~\bibnamefont
  {Cable}}\ and\ \bibinfo {author} {\bibfnamefont {G.}~\bibnamefont {Hugh}},\
  }\href@noop {} {\emph {\bibinfo {title} {{Bots in the Net: Applying Machine
  Learningto Identify Social Media Trolls}}}},\ \bibinfo {type} {Tech. Rep.}\
  (\bibinfo  {institution} {Stanford University},\ \bibinfo {year}
  {2019})\BibitemShut {NoStop}%
\bibitem [{\citenamefont {{Lim}}\ \emph {et~al.}(2019)\citenamefont {{Lim}},
  \citenamefont {{Liu}},\ and\ \citenamefont {{Zhou}}}]{LLZ19}%
  \BibitemOpen
  \bibfield  {author} {\bibinfo {author} {\bibfnamefont {J.}~\bibnamefont
  {{Lim}}}, \bibinfo {author} {\bibfnamefont {Z.}~\bibnamefont {{Liu}}}, \ and\
  \bibinfo {author} {\bibfnamefont {L.}~\bibnamefont {{Zhou}}},\ }in\ \href
  {\doibase 10.1109/ISI.2019.8823421} {\emph {\bibinfo {booktitle} {2019 IEEE
  International Conference on Intelligence and Security Informatics (ISI)}}}\
  (\bibinfo {year} {2019})\ pp.\ \bibinfo {pages} {203--205}\BibitemShut
  {NoStop}%
\bibitem [{\citenamefont {Mueller}(2019)}]{M19}%
  \BibitemOpen
  \bibfield  {author} {\bibinfo {author} {\bibfnamefont {R.~S.}\ \bibnamefont
  {Mueller}},\ }\href@noop {} {\emph {\bibinfo {title} {{Report On The
  Investigation Into Russian Interference In The 2016 Presidential
  Election}}}},\ \bibinfo {type} {Tech. Rep.}\ (\bibinfo  {institution}
  {Special Counsel's Office, United States Department of Justice},\ \bibinfo
  {year} {2019})\BibitemShut {NoStop}%
\bibitem [{\citenamefont {Chen}\ \emph {et~al.}(2017)\citenamefont {Chen},
  \citenamefont {Tanash}, \citenamefont {Stoll},\ and\ \citenamefont
  {Subramanian}}]{chen2017hunting}%
  \BibitemOpen
  \bibfield  {author} {\bibinfo {author} {\bibfnamefont {Z.}~\bibnamefont
  {Chen}}, \bibinfo {author} {\bibfnamefont {R.~S.}\ \bibnamefont {Tanash}},
  \bibinfo {author} {\bibfnamefont {R.}~\bibnamefont {Stoll}}, \ and\ \bibinfo
  {author} {\bibfnamefont {D.}~\bibnamefont {Subramanian}},\ }in\ \href@noop {}
  {\emph {\bibinfo {booktitle} {International Conference on Social
  Informatics}}}\ (\bibinfo {organization} {Springer},\ \bibinfo {year}
  {2017})\ pp.\ \bibinfo {pages} {501--510}\BibitemShut {NoStop}%
\bibitem [{\citenamefont {Beskow}\ and\ \citenamefont
  {Carley}(2018)}]{beskow2018bot}%
  \BibitemOpen
  \bibfield  {author} {\bibinfo {author} {\bibfnamefont {D.~M.}\ \bibnamefont
  {Beskow}}\ and\ \bibinfo {author} {\bibfnamefont {K.~M.}\ \bibnamefont
  {Carley}},\ }in\ \href@noop {} {\emph {\bibinfo {booktitle} {SBP-BRiMS:
  International Conference on Social Computing, Behavioral-Cultural Modeling
  and Prediction and Behavior Representation in Modeling and Simulation}}}\
  (\bibinfo {year} {2018})\BibitemShut {NoStop}%
\bibitem [{\citenamefont {{Chavoshi}}\ \emph {et~al.}(2016)\citenamefont
  {{Chavoshi}}, \citenamefont {{Hamooni}},\ and\ \citenamefont
  {{Mueen}}}]{7837909}%
  \BibitemOpen
  \bibfield  {author} {\bibinfo {author} {\bibfnamefont {N.}~\bibnamefont
  {{Chavoshi}}}, \bibinfo {author} {\bibfnamefont {H.}~\bibnamefont
  {{Hamooni}}}, \ and\ \bibinfo {author} {\bibfnamefont {A.}~\bibnamefont
  {{Mueen}}},\ }in\ \href {\doibase 10.1109/ICDM.2016.0096} {\emph {\bibinfo
  {booktitle} {2016 IEEE 16th International Conference on Data Mining
  (ICDM)}}}\ (\bibinfo {year} {2016})\ pp.\ \bibinfo {pages}
  {817--822}\BibitemShut {NoStop}%
\bibitem [{\citenamefont {Radziwill}\ and\ \citenamefont
  {Benton}(2016)}]{Radziwill2016BotON}%
  \BibitemOpen
  \bibfield  {author} {\bibinfo {author} {\bibfnamefont {N.~M.}\ \bibnamefont
  {Radziwill}}\ and\ \bibinfo {author} {\bibfnamefont {M.~C.}\ \bibnamefont
  {Benton}},\ }\href@noop {} {\bibfield  {journal} {\bibinfo  {journal}
  {ArXiv}\ }\textbf {\bibinfo {volume} {abs/1605.06555}} (\bibinfo {year}
  {2016})}\BibitemShut {NoStop}%
\bibitem [{\citenamefont {{Khan}}\ and\ \citenamefont {{Thakur}}(2018)}]{KT18}%
  \BibitemOpen
  \bibfield  {author} {\bibinfo {author} {\bibfnamefont {Y.}~\bibnamefont
  {{Khan}}}\ and\ \bibinfo {author} {\bibfnamefont {S.}~\bibnamefont
  {{Thakur}}},\ }in\ \href {\doibase 10.1109/ICONIC.2018.8601294} {\emph
  {\bibinfo {booktitle} {2018 International Conference on Intelligent and
  Innovative Computing Applications (ICONIC)}}}\ (\bibinfo {year} {2018})\ pp.\
  \bibinfo {pages} {1--5}\BibitemShut {NoStop}%
\bibitem [{\citenamefont {Ozbay}\ and\ \citenamefont {Alatas}(2020)}]{OA20}%
  \BibitemOpen
  \bibfield  {author} {\bibinfo {author} {\bibfnamefont {F.~A.}\ \bibnamefont
  {Ozbay}}\ and\ \bibinfo {author} {\bibfnamefont {B.}~\bibnamefont {Alatas}},\
  }\href {\doibase https://doi.org/10.1016/j.physa.2019.123174} {\bibfield
  {journal} {\bibinfo  {journal} {Physica A: Statistical Mechanics and its
  Applications}\ }\textbf {\bibinfo {volume} {540}},\ \bibinfo {pages} {123174}
  (\bibinfo {year} {2020})}\BibitemShut {NoStop}%
\bibitem [{\citenamefont {{Watson}}(2015)}]{7363824}%
  \BibitemOpen
  \bibfield  {author} {\bibinfo {author} {\bibfnamefont {M.~C.}\ \bibnamefont
  {{Watson}}},\ }in\ \href {\doibase 10.1109/BigData.2015.7363824} {\emph
  {\bibinfo {booktitle} {2015 IEEE International Conference on Big Data (Big
  Data)}}}\ (\bibinfo {year} {2015})\ pp.\ \bibinfo {pages}
  {793--800}\BibitemShut {NoStop}%
\bibitem [{\citenamefont {Gao}\ \emph {et~al.}(2016)\citenamefont {Gao},
  \citenamefont {Shen}, \citenamefont {Liu},\ and\ \citenamefont
  {Cheng}}]{Gao}%
  \BibitemOpen
  \bibfield  {author} {\bibinfo {author} {\bibfnamefont {J.}~\bibnamefont
  {Gao}}, \bibinfo {author} {\bibfnamefont {H.}~\bibnamefont {Shen}}, \bibinfo
  {author} {\bibfnamefont {S.}~\bibnamefont {Liu}}, \ and\ \bibinfo {author}
  {\bibfnamefont {X.}~\bibnamefont {Cheng}},\ }in\ \href {\doibase
  10.1145/2872518.2889389} {\emph {\bibinfo {booktitle} {Proceedings of the
  25th International Conference Companion on World Wide Web}}},\ \bibinfo
  {series and number} {WWW '16 Companion}\ (\bibinfo  {publisher}
  {International World Wide Web Conferences Steering Committee},\ \bibinfo
  {address} {Republic and Canton of Geneva, CHE},\ \bibinfo {year} {2016})\
  pp.\ \bibinfo {pages} {33--34}\BibitemShut {NoStop}%
\bibitem [{\citenamefont {Stillgherrian}(2020)}]{stilgherrian}%
  \BibitemOpen
  \bibfield  {author} {\bibinfo {author} {\bibnamefont {Stillgherrian}},\
  }\href@noop {} {\enquote {\bibinfo {title} {{Twitter bots and trolls promote
  conspiracy theories about Australian bushfires}},}\ }\bibinfo {howpublished}
  {\url{https://www.zdnet.com/article/twitter-bots-and-trolls-promote-conspiracy/-/theories-about-australian-bushfires/}}
  (\bibinfo {year} {2020})\BibitemShut {NoStop}%
\bibitem [{\citenamefont {Manning}\ and\ \citenamefont
  {Sch\"{u}tze}(1999)}]{MS99}%
  \BibitemOpen
  \bibfield  {author} {\bibinfo {author} {\bibfnamefont {C.~D.}\ \bibnamefont
  {Manning}}\ and\ \bibinfo {author} {\bibfnamefont {H.}~\bibnamefont
  {Sch\"{u}tze}},\ }\href@noop {} {\emph {\bibinfo {title} {{Foundations of
  Statistical Natural Language Processing}}}}\ (\bibinfo  {publisher} {MIT
  Press},\ \bibinfo {year} {1999})\BibitemShut {NoStop}%
\bibitem [{\citenamefont {Inc.}(2019)}]{TwitterHowTo}%
  \BibitemOpen
  \bibfield  {author} {\bibinfo {author} {\bibfnamefont {T.}~\bibnamefont
  {Inc.}},\ }\href@noop {} {\enquote {\bibinfo {title} {{Getting Started:
  Twitter Help Center}},}\ }\bibinfo {howpublished}
  {\url{https://help.twitter.com/en/twitter-guide}} (\bibinfo {year}
  {2019})\BibitemShut {NoStop}%
\bibitem [{\citenamefont {Griffin}\ and\ \citenamefont
  {Belmonte}(2017)}]{GB17}%
  \BibitemOpen
  \bibfield  {author} {\bibinfo {author} {\bibfnamefont {C.}~\bibnamefont
  {Griffin}}\ and\ \bibinfo {author} {\bibfnamefont {A.}~\bibnamefont
  {Belmonte}},\ }\href@noop {} {\bibfield  {journal} {\bibinfo  {journal}
  {Physical Review E}\ }\textbf {\bibinfo {volume} {95}},\ \bibinfo {pages}
  {052309} (\bibinfo {year} {2017})}\BibitemShut {NoStop}%
\bibitem [{\citenamefont {Badawy}\ \emph {et~al.}(2018)\citenamefont {Badawy},
  \citenamefont {Ferrara},\ and\ \citenamefont {Lerman}}]{Lerman}%
  \BibitemOpen
  \bibfield  {author} {\bibinfo {author} {\bibfnamefont {A.}~\bibnamefont
  {Badawy}}, \bibinfo {author} {\bibfnamefont {E.}~\bibnamefont {Ferrara}}, \
  and\ \bibinfo {author} {\bibfnamefont {K.}~\bibnamefont {Lerman}},\ }\href
  {http://arxiv.org/abs/1802.04291} {\bibfield  {journal} {\bibinfo  {journal}
  {CoRR}\ }\textbf {\bibinfo {volume} {abs/1802.04291}} (\bibinfo {year}
  {2018})},\ \Eprint {http://arxiv.org/abs/1802.04291} {arXiv:1802.04291}
  \BibitemShut {NoStop}%
\bibitem [{\citenamefont {Cleary}(2019)}]{Armchair1}%
  \BibitemOpen
  \bibfield  {author} {\bibinfo {author} {\bibfnamefont {G.}~\bibnamefont
  {Cleary}},\ }\href@noop {} {\enquote {\bibinfo {title} {{Twitterbots: Anatomy
  of a Propaganda Campaign}},}\ }\bibinfo {howpublished}
  {\url{https://www.symantec.com/blogs/threat-intelligence/twitterbots-propaganda-disinformation}}
  (\bibinfo {year} {2019})\BibitemShut {NoStop}%
\bibitem [{\citenamefont {Lab}(2018)}]{Armchair2}%
  \BibitemOpen
  \bibfield  {author} {\bibinfo {author} {\bibfnamefont {A.~C. D. F.~R.}\
  \bibnamefont {Lab}},\ }\href@noop {} {\enquote {\bibinfo {title}
  {{TrollTracker: Twitter Troll Farm Archives}},}\ }\bibinfo {howpublished}
  {\url{https://medium.com/dfrlab/trolltracker-twitter-troll-farm-archives/-/8d5dd61c486b}}
  (\bibinfo {year} {2018})\BibitemShut {NoStop}%
\bibitem [{\citenamefont {Thompson}(2018)}]{Armchair3}%
  \BibitemOpen
  \bibfield  {author} {\bibinfo {author} {\bibfnamefont {A.}~\bibnamefont
  {Thompson}},\ }\href@noop {} {\enquote {\bibinfo {title} {{This is How We
  Troll}},}\ }\bibinfo {howpublished}
  {\url{https://www.ceros.com/originals/russian-tweet-bot-data/}} (\bibinfo
  {year} {2018})\BibitemShut {NoStop}%
\bibitem [{\citenamefont {Di~Lorenzo}(2013)}]{D13}%
  \BibitemOpen
  \bibfield  {author} {\bibinfo {author} {\bibfnamefont {R.}~\bibnamefont
  {Di~Lorenzo}},\ }in\ \href@noop {} {\emph {\bibinfo {booktitle} {Basic
  Technical Analysis of Financial Markets}}}\ (\bibinfo  {publisher}
  {Springer},\ \bibinfo {year} {2013})\ pp.\ \bibinfo {pages}
  {189--220}\BibitemShut {NoStop}%
\bibitem [{\citenamefont {Friedlander}\ \emph {et~al.}(2003)\citenamefont
  {Friedlander}, \citenamefont {Griffin}, \citenamefont {Jacobson},
  \citenamefont {Phoha},\ and\ \citenamefont {Brooks}}]{FGNP+03}%
  \BibitemOpen
  \bibfield  {author} {\bibinfo {author} {\bibfnamefont {D.}~\bibnamefont
  {Friedlander}}, \bibinfo {author} {\bibfnamefont {C.}~\bibnamefont
  {Griffin}}, \bibinfo {author} {\bibfnamefont {N.}~\bibnamefont {Jacobson}},
  \bibinfo {author} {\bibfnamefont {S.}~\bibnamefont {Phoha}}, \ and\ \bibinfo
  {author} {\bibfnamefont {R.~R.}\ \bibnamefont {Brooks}},\ }\href@noop {}
  {\bibfield  {journal} {\bibinfo  {journal} {EURASIP J. Applied Signal
  Processing}\ }\textbf {\bibinfo {volume} {4}},\ \bibinfo {pages} {371}
  (\bibinfo {year} {2003})}\BibitemShut {NoStop}%
\bibitem [{\citenamefont {Jolliffe}(2011)}]{J11}%
  \BibitemOpen
  \bibfield  {author} {\bibinfo {author} {\bibfnamefont {I.}~\bibnamefont
  {Jolliffe}},\ }\href@noop {} {\emph {\bibinfo {title} {Principal component
  analysis}}}\ (\bibinfo  {publisher} {Springer},\ \bibinfo {year}
  {2011})\BibitemShut {NoStop}%
\bibitem [{\citenamefont {Kaufman}\ \emph {et~al.}(1987)\citenamefont
  {Kaufman}, \citenamefont {Rousseeuw},\ and\ \citenamefont {Dodge}}]{KRD87}%
  \BibitemOpen
  \bibfield  {author} {\bibinfo {author} {\bibfnamefont {L.}~\bibnamefont
  {Kaufman}}, \bibinfo {author} {\bibfnamefont {P.}~\bibnamefont {Rousseeuw}},
  \ and\ \bibinfo {author} {\bibfnamefont {Y.}~\bibnamefont {Dodge}},\
  }\href@noop {} {\enquote {\bibinfo {title} {Clustering by means of medoids in
  statistical data analysis based on the},}\ } (\bibinfo {year}
  {1987})\BibitemShut {NoStop}%
\bibitem [{\citenamefont {Meves}\ and\ \citenamefont {Allgemein}(2016)}]{MA16}%
  \BibitemOpen
  \bibfield  {author} {\bibinfo {author} {\bibfnamefont {N.}~\bibnamefont
  {Meves}}\ and\ \bibinfo {author} {\bibfnamefont {K.}~\bibnamefont
  {Allgemein}},\ }\href@noop {} {\enquote {\bibinfo {title} {{Der Bot Boost
  \#merkelmussbleiben}},}\ }\bibinfo {howpublished}
  {{https://www.wahl.de/aktuell/2016/08/05/social-bot-bundesregierung/}}
  (\bibinfo {year} {2016})\BibitemShut {NoStop}%
\bibitem [{\citenamefont {Tenenbaum}\ \emph {et~al.}(2000)\citenamefont
  {Tenenbaum}, \citenamefont {Silva},\ and\ \citenamefont {Langford}}]{TSL00}%
  \BibitemOpen
  \bibfield  {author} {\bibinfo {author} {\bibfnamefont {J.~B.}\ \bibnamefont
  {Tenenbaum}}, \bibinfo {author} {\bibfnamefont {V.~d.}\ \bibnamefont
  {Silva}}, \ and\ \bibinfo {author} {\bibfnamefont {J.~C.}\ \bibnamefont
  {Langford}},\ }\href {\doibase 10.1126/science.290.5500.2319} {\bibfield
  {journal} {\bibinfo  {journal} {Science}\ }\textbf {\bibinfo {volume}
  {290}},\ \bibinfo {pages} {2319} (\bibinfo {year} {2000})},\ \Eprint
  {http://arxiv.org/abs/http://science.sciencemag.org/content/290/5500/-/2319.full.pdf}
  {http://science.sciencemag.org/content/290/5500/-/2319.full.pdf} \BibitemShut
  {NoStop}%
\bibitem [{\citenamefont {Belkin}\ and\ \citenamefont {Niyogi}(2002)}]{BN02}%
  \BibitemOpen
  \bibfield  {author} {\bibinfo {author} {\bibfnamefont {M.}~\bibnamefont
  {Belkin}}\ and\ \bibinfo {author} {\bibfnamefont {P.}~\bibnamefont
  {Niyogi}},\ }in\ \href@noop {} {\emph {\bibinfo {booktitle} {Advances in
  neural information processing systems}}}\ (\bibinfo {year} {2002})\ pp.\
  \bibinfo {pages} {929--936}\BibitemShut {NoStop}%
\bibitem [{\citenamefont {Newman}(2006)}]{N06}%
  \BibitemOpen
  \bibfield  {author} {\bibinfo {author} {\bibfnamefont {M.~E.~J.}\
  \bibnamefont {Newman}},\ }\href {\doibase 10.1073/pnas.0601602103} {\bibfield
   {journal} {\bibinfo  {journal} {Proceedings of the National Academy of
  Sciences}\ }\textbf {\bibinfo {volume} {103}},\ \bibinfo {pages} {8577}
  (\bibinfo {year} {2006})},\ \Eprint
  {http://arxiv.org/abs/http://www.pnas.org/content/103/23/8577.full.pdf}
  {http://www.pnas.org/content/103/23/8577.full.pdf} \BibitemShut {NoStop}%
\bibitem [{\citenamefont {Newman}(2016)}]{N16}%
  \BibitemOpen
  \bibfield  {author} {\bibinfo {author} {\bibfnamefont {M.~E.~J.}\
  \bibnamefont {Newman}},\ }\href {\doibase 10.1103/PhysRevE.94.052315}
  {\bibfield  {journal} {\bibinfo  {journal} {Phys. Rev. E}\ }\textbf {\bibinfo
  {volume} {94}},\ \bibinfo {pages} {052315} (\bibinfo {year}
  {2016})}\BibitemShut {NoStop}%
\bibitem [{\citenamefont {Newman}(2013)}]{N13}%
  \BibitemOpen
  \bibfield  {author} {\bibinfo {author} {\bibfnamefont {M.~E.~J.}\
  \bibnamefont {Newman}},\ }\href {\doibase 10.1103/PhysRevE.88.042822}
  {\bibfield  {journal} {\bibinfo  {journal} {Phys. Rev. E}\ }\textbf {\bibinfo
  {volume} {88}},\ \bibinfo {pages} {042822} (\bibinfo {year}
  {2013})}\BibitemShut {NoStop}%
\bibitem [{\citenamefont {Zipf}(1932)}]{Z32}%
  \BibitemOpen
  \bibfield  {author} {\bibinfo {author} {\bibfnamefont {G.~K.}\ \bibnamefont
  {Zipf}},\ }\href@noop {} {\emph {\bibinfo {title} {{Selected Studies of the
  Principle of Relative Frequency in Language}}}}\ (\bibinfo  {publisher}
  {Harvard University Press},\ \bibinfo {year} {1932})\BibitemShut {NoStop}%
\bibitem [{\citenamefont {Mandelbrot}(1965)}]{M65}%
  \BibitemOpen
  \bibfield  {author} {\bibinfo {author} {\bibfnamefont {B.}~\bibnamefont
  {Mandelbrot}},\ }in\ \href@noop {} {\emph {\bibinfo {booktitle} {Language}}}\
  (\bibinfo  {publisher} {Penguin Books},\ \bibinfo {year} {1965})\BibitemShut
  {NoStop}%
\bibitem [{\citenamefont {Hart}\ \emph {et~al.}(2000)\citenamefont {Hart},
  \citenamefont {Stork},\ and\ \citenamefont {Duda}}]{HSD00}%
  \BibitemOpen
  \bibfield  {author} {\bibinfo {author} {\bibfnamefont {P.~E.}\ \bibnamefont
  {Hart}}, \bibinfo {author} {\bibfnamefont {D.~G.}\ \bibnamefont {Stork}}, \
  and\ \bibinfo {author} {\bibfnamefont {R.~O.}\ \bibnamefont {Duda}},\
  }\href@noop {} {\emph {\bibinfo {title} {Pattern Classification}}}\ (\bibinfo
   {publisher} {Wiley-Interscience},\ \bibinfo {year} {2000})\BibitemShut
  {NoStop}%
\bibitem [{\citenamefont {Albert}\ and\ \citenamefont
  {Barab\'asi}(2000)}]{AB00}%
  \BibitemOpen
  \bibfield  {author} {\bibinfo {author} {\bibfnamefont {R.}~\bibnamefont
  {Albert}}\ and\ \bibinfo {author} {\bibfnamefont {A.-L.}\ \bibnamefont
  {Barab\'asi}},\ }\href {\doibase 10.1103/PhysRevLett.85.5234} {\bibfield
  {journal} {\bibinfo  {journal} {Phys. Rev. Lett.}\ }\textbf {\bibinfo
  {volume} {85}},\ \bibinfo {pages} {5234} (\bibinfo {year}
  {2000})}\BibitemShut {NoStop}%
\bibitem [{\citenamefont {Borg}\ and\ \citenamefont {Groenen}(1997)}]{BG97}%
  \BibitemOpen
  \bibfield  {author} {\bibinfo {author} {\bibfnamefont {I.}~\bibnamefont
  {Borg}}\ and\ \bibinfo {author} {\bibfnamefont {P.}~\bibnamefont {Groenen}},\
  }\href@noop {} {\emph {\bibinfo {title} {Modern Multidimensional Scaling:
  theory and applications,}}}\ (\bibinfo  {publisher} {Springer-Verlag},\
  \bibinfo {address} {New York, NY},\ \bibinfo {year} {1997})\BibitemShut
  {NoStop}%
\end{thebibliography}%

\end{document}